\documentclass[10pt,a4paper,reqno,article]{amsart}
\usepackage{mathrsfs}
\usepackage{stmaryrd}
\usepackage{amsthm,amstext}
\usepackage{amssymb,amsmath}
\usepackage{amsfonts}
\usepackage{MnSymbol,slashed}
\usepackage{fancyhdr}
\usepackage{graphicx}
\usepackage{pdfpages}
\usepackage{float}
\usepackage{subfig}
\usepackage{simplewick}
\usepackage{graphics,psfrag}
\usepackage{pstricks}
\usepackage{latexsym}
\usepackage{changepage}
\usepackage{caption}
\usepackage{paralist}
\usepackage{tikzit}

\tikzstyle{boundary vertex}=[inner sep=0mm, minimum size=1mm, shape=circle, draw=black, fill=black]
\tikzstyle{grey_dot}=[fill={rgb,255: red,191; green,191; blue,191}, draw={rgb,255: red,191; green,191; blue,191}, shape=circle, minimum size=1mm, inner sep=0mm]

\tikzstyle{arrow}=[->]
\tikzstyle{red_arrow}=[->, draw=red]
\tikzstyle{cyan_arrow}=[->, draw=cyan]
\tikzstyle{red_dash}=[-, dashed, draw=red]
\tikzstyle{grey dash}=[-, fill=none, draw={rgb,255: red,191; green,191; blue,191}, dashed]
\tikzstyle{dashed arrow}=[->, dashed]

\tikzstyle{gate}=[shape=rectangle, text height=1.5ex, text depth=0.25ex, yshift=0.5mm, fill=white, draw=black, minimum height=5mm, yshift=-0.5mm, minimum width=5mm, font={\small}, tikzit category=circuit]
\tikzstyle{big gate}=[shape=rectangle, text height=1.5ex, text depth=0.25ex, yshift=0.5mm, fill=white, draw=black, minimum height=10mm, yshift=-0.5mm, minimum width=5mm, font={\small}, tikzit category=circuit]
\tikzstyle{Z dot}=[inner sep=0mm, minimum size=2mm, shape=circle, draw=black, fill={rgb,255: red,221; green,255; blue,221}, tikzit category=zx]
\tikzstyle{Z phase dot}=[minimum size=5mm, font={\footnotesize\boldmath}, shape=rectangle, rounded corners=2mm, inner sep=0.2mm, outer sep=-2mm, scale=0.8, tikzit shape=circle, draw=black, fill={rgb,255: red,221; green,255; blue,221}, tikzit draw=blue, tikzit category=zx]
\tikzstyle{X dot}=[Z dot, shape=circle, draw=black, fill={rgb,255: red,255; green,136; blue,136}, tikzit category=zx]
\tikzstyle{X phase dot}=[Z phase dot, tikzit shape=circle, tikzit draw=blue, fill={rgb,255: red,255; green,136; blue,136}, font={\footnotesize\boldmath}, tikzit category=zx]
\tikzstyle{hadamard}=[fill=yellow, draw=black, shape=rectangle, inner sep=0.6mm, minimum height=1.5mm, minimum width=1.5mm, tikzit category=zx]
\tikzstyle{paulibox}=[fill={rgb,255: red,221; green,221; blue,255}, draw=black, shape=rectangle, inner sep=0.6mm, minimum height=5mm, minimum width=5mm, font={\footnotesize}, text height=1.5ex, text depth=0.25ex, tikzit category=zx]
\tikzstyle{vertex}=[inner sep=0mm, minimum size=1mm, shape=circle, draw=black, fill=black, tikzit category=misc]
\tikzstyle{vertex set}=[inner sep=0mm, minimum size=1mm, shape=circle, draw=black, fill=white, font={\footnotesize\boldmath}, tikzit category=misc]
\tikzstyle{small black dot}=[fill=black, draw=black, shape=circle, inner sep=0pt, minimum width=1.2mm, tikzit category=circuit]
\tikzstyle{cnot ctrl}=[fill=black, draw=black, shape=circle, inner sep=0pt, minimum width=1.2mm, tikzit category=circuit]
\tikzstyle{cnot targ}=[fill=white, draw=white, shape=circle, tikzit category=circuit, label={center:$\oplus$}, inner sep=0pt, minimum width=2.1mm, tikzit fill={rgb,255: red,102; green,204; blue,255}, tikzit draw=black]
\tikzstyle{ket}=[fill=white, draw=black, shape=regular polygon, regular polygon sides=3, regular polygon rotate=-30, scale=0.7, inner sep=1pt, tikzit category=circuit, tikzit shape=rectangle, tikzit fill=green]
\tikzstyle{bra}=[fill=white, draw=black, shape=regular polygon, regular polygon sides=3, regular polygon rotate=30, scale=0.7, inner sep=1pt, tikzit category=circuit, tikzit shape=rectangle, tikzit fill=red]
\tikzstyle{scalar}=[shape=rectangle, text height=1.5ex, text depth=0.25ex, yshift=0.5mm, fill=white, draw=black, minimum height=5mm, yshift=-0.5mm, minimum width=5mm, font={\small}]
\tikzstyle{clabel}=[fill=white, draw=none, shape=rectangle, tikzit fill={rgb,255: red,56; green,255; blue,242}, font={\footnotesize}, inner sep=1pt, tikzit category=labels]
\tikzstyle{empty diagram}=[draw={gray!40!white}, dashed, shape=rectangle, minimum width=1cm, minimum height=1cm, tikzit category=misc]

\tikzstyle{hadamard edge}=[-, dashed, dash pattern=on 2pt off 0.5pt, thick, draw={rgb,255: red,68; green,136; blue,255}]
\tikzstyle{box edge}=[-, dashed, dash pattern=on 2pt off 0.5pt, thick, draw={rgb,255: red,203; green,192; blue,225}]
\tikzstyle{brace edge}=[-, tikzit draw=blue, decorate, decoration={brace,amplitude=1mm,raise=-1mm}]
\tikzstyle{diredge}=[->]
\tikzstyle{double edge}=[-, double, shorten <=-1mm, shorten >=-1mm, double distance=2pt]
\tikzstyle{gray edge}=[-, {gray!60!white}]
\tikzstyle{pointer edge}=[->, very thick, gray]
\tikzstyle{boldedge}=[-, line width=1.6pt, shorten <=-0.17mm, shorten >=-0.17mm]
\allowdisplaybreaks[1]

\theoremstyle{plain}
\newtheorem{theorem}{Theorem}[section] 

\newtheorem{lemma}[theorem]{Lemma}
\newtheorem{corollary}[theorem]{Corollary}
\newtheorem{proposition}[theorem]{Proposition}
\newtheorem{definition}[theorem]{Definition}
\newtheorem{example}[theorem]{Example}
\newtheorem{remark}[theorem]{Remark}
\newcommand{\CC}{{\hbox{{$\mathcal C$}}}}

\newcommand{\CG}{\hbox{{$\mathcal G$}}}

\newcommand{\CH}{\hbox{{$\mathcal H$}}}

\newcommand{\CM}{\hbox{{$\mathcal M$}}}
\newcommand{\CT}{\hbox{{$\mathcal T$}}}
\newcommand{\CL}{\hbox{{$\mathcal L$}}}

\newcommand{\CR}{\hbox{{$\mathcal R$}}}

\newcommand{\CO}{\hbox{{$\mathcal O$}}}

\newcommand{\CV}{{\hbox{{$\mathcal V$}}}}

\newcommand{\CZ}{\hbox{{$\mathcal Z$}}}
\renewcommand{\O}{\mathbb{O}}  
\newcommand{\C}{\mathbb{C}}
\newcommand{\R}{\mathbb{R}}
\newcommand{\F}{\mathbb{F}}
\newcommand{\Z}{\mathbb{Z}}

\newcommand{\isom}{{\cong}}
\newcommand{\eps}{{\epsilon}}
\newcommand{\tens}{\mathop{{\otimes}}}
\newcommand{\la}{{\triangleright}}
\newcommand{\ra}{{\triangleleft}}

\newcommand{\id}{\mathrm{id}}

\renewcommand{\>}{\rangle}
\newcommand{\End}{\mathrm{ End}}
\newcommand{\Tr}{\mathrm{ Tr}}

\def\rcross{{\triangleright\!\!\!<}}
\def\lcross{{>\!\!\!\triangleleft}}

\def\lcocross{{>\!\!\blacktriangleleft}}
\def\cobicross{{\triangleright\!\blacktriangleleft}}

\def\dcross{{\bowtie}}

\newcommand{\vac}{{|{\rm vac}\>}}

\begin{document}
\author{Alexander Cowtan${}^\dagger$}
\address{Department of Computer Science,  University of Oxford\\
\& Cambridge Quantum Computing}
\author{Shahn Majid${}^{*}$}
\address{School of Mathematical Sciences, Queen Mary University of London\\
\& Cambridge Quantum Computing${}$}
\email{${}^\dagger$akcowtan@gmail.com,${}^*$\\s.majid@qmul.ac.uk}

\title{Algebraic aspects of boundaries in the Kitaev quantum double model}
	\begin{abstract} 
	We provide a systematic treatment of boundaries based on subgroups $K\subseteq G$ with the Kitaev quantum double $D(G)$ model in the bulk. The boundary sites are representations of a $*$-subalgebra $\Xi\subseteq D(G)$ and we explicate its structure as a strong $*$-quasi-Hopf algebra dependent on a choice of transversal $R$. We provide decomposition formulae for irreducible representations of $D(G)$ pulled back to $\Xi$. We also provide explicitly the monoidal equivalence of the category of $\Xi$-modules and the category of $G$-graded $K$-bimodules and use this to prove that different choices of $R$ are related by Drinfeld cochain twists. Examples include $S_{n-1}\subset S_n$ and an example related to the octonions where $\Xi$ is also a Hopf quasigroup. As an application of our treatment, we study patches with boundaries based on $K=G$ horizontally and $K=\{e\}$ vertically and show how these could be used in a quantum computer using the technique of lattice surgery.
	\end{abstract}
\maketitle

\section{Introduction}

The Kitaev model is defined for a finite group $G$ \cite{Kit} with quasiparticles given by representations of the quantum double $D(G)$, and their dynamics described by intertwiners. In quantum computing, the quasiparticles correspond to measurement outcomes at sites on a lattice, and their dynamics correspond to linear maps on the data, with the aim of performing fault-tolerant quantum computation. The lattice can be any ciliated ribbon graph embedded on a surface \cite{Meu}, although throughout we will assume a square lattice on the plane for convenience. The Kitaev model generalises to replace $G$ by a finite-dimensional semisimple Hopf algebra, as well as aspects that work of a general finite-dimensional Hopf algebra. We refer to \cite{CowMa} for details of the relevant algebraic aspects of this theory, which applies in the bulk of the Kitaev model. We now extend this work with a study of the algebraic structure that underlies an approach to the treatment of boundaries.

The treatment of boundaries here originates in a more categorical point of view. In the original Kitaev model the relevant category that defines the `topological order' in condensed matter terms\cite{LK}  is the category ${}_{D(G)}\mathcal{M}$ of $D(G)$-modules, which one can think of as an instance of the `dual' or `centre' $\CZ(\CC)$ construction\cite{Ma:rep}, where  $\CC=\CM^G$ is the category of $G$-graded vector spaces.  Levin-Wen `string-net' models \cite{LW} are a sort of generalisation of Kitaev models specified now by a unitary fusion category $\mathcal{C}$ with topological order $\CZ(\mathcal{C})$, meaning that at every site on the lattice one has an object in $\CZ(\mathcal{C})$, and now on a trivalent lattice. Computations correspond to morphisms in the same category. 
A so-called gapped boundary condition of a string-net model preserves a finite energy gap between the vacuum and the lowest excited state(s), which is independent of system size. Such boundary conditions are defined by module categories of the fusion category $\CC$. By definition,  a (right) $\CC$-module means\cite{Os,KK} a category $\CV$ equipped with a bifunctor $\CV \times \CC \rightarrow \CV$ obeying coherence equations which are a polarised version of the properties of $\tens: \CC\times\CC\to \CC$ (in the same way that a right module of an algebra obeys a polarised version of the axioms for the product). One can also see a string-net model as a discretised quantum field theory \cite{Kir2, Meu}, and indeed boundaries of a conformal field theory can also be similarly defined by module categories \cite{FS}.  For our purposes, we care about \textit{indecomposable} module categories, that is module categories which are not equivalent to a direct sum of other module categories. Excitations on the boundary with condition $\mathcal{V}$ are then given by functors $F \in \mathrm{End}_\CC(\mathcal{V})$ that commute with the $\CC$ action\cite{KK}, beyond the vacuum state which is the identity functor $\mathrm{id}_{\mathcal{V}}$. More than just the boundary conditions above, we care about these excitations, and so $\mathrm{End}_\CC(\mathcal{V})$ is the category of interest. 

The Kitaev model is not exactly a string-net model (the lattice in our case will not even be trivalent) but closely related. In particular, it can be shown that indecomposable module categories for $\CC=\CM^G$, the category of $G$-graded vector spaces, are\cite{Os2} classified by subgroups $K\subseteq G$ and cocycles $\alpha\in H^2(K,\C^\times)$. We will stick to the trivial $\alpha$ case here, and the upshot is that the boundary conditions in the regular  Kitaev model should be given by $\CV={}_K\CM^G$ the $G$-graded $K$-modules where $x\in K$ itself has grade $|x|=x\in G$. Then the excitations are governed by objects of $\mathrm{End}_\CC(\CV) \simeq {}_K\CM_K^G$, the category of $G$-graded bimodules over $K$. This is necessarily equivalent, by Tannaka-Krein reconstruction\cite{Ma:tan} to the category of modules ${}_{\Xi(R,K)}\mathcal{M}$ of a certain quasi-Hopf algebra $\Xi(R,K)$. Here $R\subseteq G$ is a choice of transversal so that every element of $G$ factorises uniquely as $RK$, but the algebra of $\Xi(R,K)$ depends only on the choice of subgroup $K$ and not on the transversal $R$. This is the algebra which we use to define measurement protocols on the boundaries of the Kitaev model. One also has that $\CZ({}_\Xi\CM)\simeq\CZ(\CM^G)\simeq{}_{D(G)}\CM$ as braided monoidal categories. 

Categorical aspects will be deferred to Section~\ref{sec:cat_just}, our main focus prior to that being on a full understanding of the algebra $\Xi$, its properties and aspects of the physics. In fact, lattice boundaries of Kitaev models based on subgroups have been defined and characterised previously, see \cite{BSW, Bom}, with \cite{CCW} giving an overview for computational purposes, and we build on these works. We begin in Section~\ref{sec:bulk} with a recap of  the algebras and actions involved in the bulk of the lattice model, then in Section~\ref{sec:gap} we accommodate the boundary conditions in a manner which works with features important for quantum computation, such as sites, quasiparticle projectors and ribbon operators. These sections mostly cover well-trodden ground, although we correct errors and clarify some algebraic subtleties which appear to have gone unnoticed in previous works. In particular, we obtain formulae for the decomposition of bulk irreducible representations of $D(G)$ into $\Xi$-representations which we believe to be new. Key to our results here is an observation that in fact $\Xi(R,K)\subseteq D(G)$ as algebras, which gives a much more direct route than previously to an adjunction between $\Xi(R,K)$-modules and $D(G)$-modules describing how excitations pass between the bulk and boundary. This is important for the physical picture\cite{CCW} and previously was attributed to an adjunction between ${}_{D(G)}\CM$ and ${}_K\CM_K^G$ in \cite{PS2}.

In Section~\ref{sec:patches}, as an application of our explicit description of boundaries, we generalise the quantum computational model called \textit{lattice surgery} \cite{HFDM,Cow2} to the nonabelian group case. We find that for every finite group $G$ one can simulate the group algebra $\C G$ and its dual $\C(G)$ on a lattice patch with `rough' and `smooth' boundaries. This is an alternative model of fault-tolerant computation to the well-known method of braiding anyons or defects \cite{Kit,FMMC}, although we do not know whether there are choices of group such that lattice surgery is natively universal without state distillation.

In Section~\ref{sec:quasi}, we look at $\Xi(R,K)$ as a quasi-Hopf algebra in somewhat more detail than we have found elsewhere. As well as the quasi-bialgebra structure, we provide and verify the antipode for any choice of transversal $R$ for which right-inversion is bijective. This case is in line with \cite{Nat}, but we will also consider antipodes more generally.  We then show that an obvious $*$-algebra structure on $\Xi$ meets all the axioms of a strong $*$-quasi-Hopf algebra in the sense of \cite{BegMa:bar} coming out of the theory of bar categories. The key ingredient here is a somewhat nontrivial map that relates the complex conjugate $\Xi$-module to $V\tens W$ to those of $W$ and $V$. We also give an extended series of examples, including one related to the octonions. 

Lastly, in Section~\ref{sec:cat_just}, we connect the algebraic notions up to the abstract description of boundaries conditions via module categories and use this to obtain more results about $\Xi(R,K)$. We first calculate the relevant categorical equivalence  ${}_K\CM_K^G \simeq {}_{\Xi(R,K)}\mathcal{M}$ concretely, deriving the quasi-bialgebra structure of $\Xi(R,K)$ precisely such that this works. 
Since the left hand side is independent of $R$, we deduce by Tannaka-Krein arguments that changing $R$ changes $\Xi(R,K)$ by a Drinfeld cochain twist and we find this cochain as a main result of the section. This is important as Drinfeld twists do not change the category of modules up to equivalence, so such aspects of the physics do not depend on $R$. Twisting arguments then imply that we have an antipode more generally for any $R$. We also look at $\CV = {}_K\CM^G$ as a module category for $\CC=\CM^G$. Section~\ref{sec:rem} provides some concluding remarks relating to generalisations of the boundaries to models based on other Hopf algebras \cite{BMCA}.

\subsection*{Acknowledgements}
The first author thanks Stefano Gogioso for useful discussions regarding nonabelian lattice surgery as a model for computation. Thanks also to Paddy Gray \& Kathryn Pennel for their hospitality while some of this paper was written and to Simon Harrison for the Wolfson Harrison UK Research Council Quantum Foundation Scholarship, which made this work possible. The second author was on sabbatical at Cambridge Quantum Computing and we thank members of the team there. 

\section{Preliminaries: recap of the Kitaev model in the bulk}\label{sec:bulk}
We begin with the model in the bulk. This is a largely a recap of eg. \cite{Kit, CowMa}. 

\subsection{Quantum double}\label{sec:double}Let $G$ be a finite group with identity $e$, then $\C G$ is the group Hopf algebra with basis $G$. Multiplication is extended linearly, and $\C G$ has comultiplication $\Delta h = h \otimes h$ and counit $\eps h = 1$ on basis elements $h\in G$. The antipode is given by $Sh = h^{-1}$. $\C G$ is a Hopf $*$-algebra with  $h^* = h^{-1}$ extended antilinearly. Its dual Hopf algebra $\C(G)$ of functions on $G$ has basis of $\delta$-functions $\{\delta_g\}$ with $\Delta\delta_g=\sum_h \delta_h\tens\delta_{h^{-1}g}$, $\eps \delta_g=\delta_{g,e}$ and $S\delta_g=\delta_{g^{-1}}$ for the Hopf algebra structure, and $\delta_g^* = \delta_{g}$ for all $g\in G$. The normalised integral elements \textit{in} $\C G$ and $\C(G)$ are 
\[ \Lambda_{\C G}={1\over |G|}\sum_{h\in G} h\in \C G,\quad \Lambda_{\C(G)}=\delta_e\in \C(G).\]
The integrals \textit{on} $\C G$ and $\C(G)$ are
\[ \int h = \delta_{h,e}, \quad \int \delta_g = 1\]
normalised so that $\int 1 = 1$ for $\C G$ and $\int 1 = |G|$ for $\C(G)$.

For the Drinfeld double we have $D(G)=\C(G)\lcross \C G$ as in \cite{Ma:book}, with $\C G$ and $\C(G)$ sub-Hopf algebras and the cross relations $ h\delta_g =\delta_{hgh^{-1}} h$ (a semidirect product). The Hopf algebra antipode is $S(\delta_gh)=\delta_{h^{-1}g^{-1}h} h^{-1}$, and over $\C$ we have a Hopf $*$-algebra with $(\delta_g h)^* = \delta_{h^{-1}gh} h^{-1}$. There is also a quasitriangular structure which in subalgebra notation is 
\begin{equation}\label{RDG} \CR=\sum_{h\in G} \delta_h\tens h\in D(G) \otimes D(G).\end{equation}
If we want to be totally explicit we can build $D(G)$ on either the vector space $\C(G)\tens \C G$ or on the vector space $\C G\tens\C(G)$. In fact the latter is more natural but we follow the conventions in \cite{Ma:book,CowMa} and use the former. Then one can say the above more explicitly as \[(\delta_g\tens h)(\delta_f\tens k)=\delta_g\delta_{hfh^{-1}}\tens hk=\delta_{g,hfh^{-1}}\delta_g\tens hk,\quad S(\delta_g\tens h)=\delta_{h^{-1}g^{-1}h} \tens h^{-1}\]
etc. for the operations on the underlying vector space. 

As a semidirect product, irreducible representations of $D(G)$ are given by standard theory as labelled by pairs $(\CC,\pi)$ consisting of an orbit under the action (i.e. by a conjugacy class $\CC\subset G$ in this case) and an irrep $\pi$ of the isotropy subgroup, in our case  
\[ G^{c_0}=\{n\in G\ |\ nc_0 n^{-1}=c_0\}\]
of a fixed element $c_0\in\CC$, i.e. the centraliser $C_G(c_0)$. The choice of $c_0$ does not change the isotropy group up to isomorphism but does change how it sits inside $G$. We also fix data $q_c\in G$ for each $c\in \CC$ such that $c=q_cc_0q_c^{-1}$ with $q_{c_0}=e$ and define from this a cocycle $\zeta_c(h)=q^{-1}_{hch^{-1}}hq_c$ as a map $\zeta: \CC\times G\to G^{c_0}$. The associated irreducible representation is then 
\[ W_{\CC,\pi}=\C \CC\tens W_\pi,\quad \delta_g.(c\tens w)=\delta_{g,c}c\tens w,\quad  h.(c\tens w)=hch^{-1}\tens \zeta_c(h).w \]
for all $w\in W_\pi$, the carrier space of $\pi$. This constructs all irreps of $D(G)$ and, over $\C$, these are unitary in a Hopf $*$-algebra sense if $\pi$ is unitary. Moreover, $D(G)$ is semisimple and hence has a block decomposition $D(G)\isom\oplus_{\CC,\pi} \End(W_{\CC,\pi})$ given by a complete orthogonal set of self-adjoint central idempotents 
\begin{equation}\label{Dproj}P_{(\CC,\pi)}={{\rm dim}(W_\pi)\over |G^{c_0}|}\sum_{c\in \CC}\sum_{n\in G^{c_0}}\Tr_\pi(n^{-1})\delta_{c}\tens q_c nq_c^{-1}.\end{equation}
We refer to \cite{CowMa} for more details and proofs. Acting on a state, this will become a projection operator that determines if a quasiparticle of type $\CC,\pi$ is present. Chargeons are quasiparticles with $\CC=\{e\}$ and $\pi$ an irrep of $G$, and fluxions are quasiparticles with $\CC$ a conjugacy class and $\pi=1$, the trivial representation.

\subsection{Bulk lattice model}\label{sec:lattice}
Having established the prerequisite algebra, we move on to the lattice model itself. This first part is largely a recap of \cite{Kit, CowMa} and we use the notations of the latter. Let $\Sigma = \Sigma(V, E, P)$ be a square lattice viewed as a directed graph with its usual (cartesian) orientation, vertices $V$, directed edges $E$ and faces $P$. The Hilbert space $\CH$ will be a tensor product of vector spaces with one copy of $\C G$ at each arrow in $E$. We have group elements for the basis of each copy. Next, to each adjacent pair of vertex $v$ and face $p$ we associate a site $s = (v, p)$, or equivalently a line (the `cilium') from $p$ to $v$. We then define an action of $\C G$ and $\C(G)$ at each site by
\[ \includegraphics[scale=0.7]{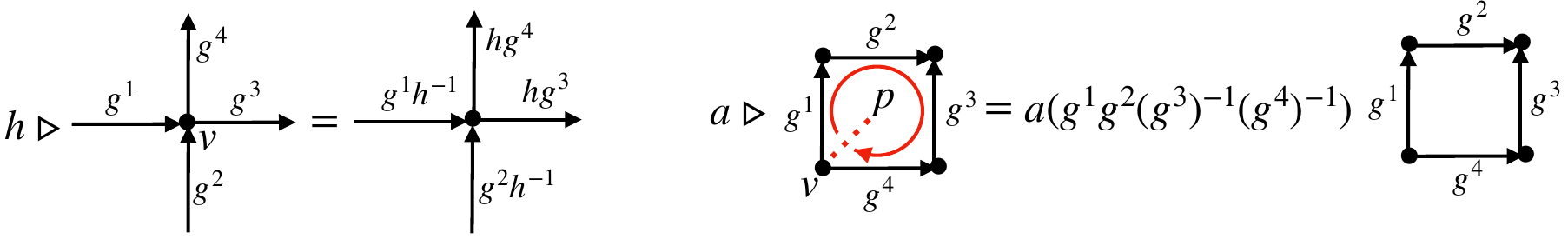}\]
Here $h\in \C G$, $a\in \C(G)$ and $g^1,\cdots,g^4$ denote independent elements of $G$ (not powers). Observe that the vertex action is invariant under the location of $p$ relative to its adjacent $v$, so the red dashed line has been omitted.

\begin{lemma}\label{lemDGrep} \cite{Kit,CowMa} $h\la$ and $a\la$ for all $h\in G$ and $a\in \C(G)$ define a representation of $D(G)$ on $\CH$ associated to each site $(v,p)$. 
\end{lemma}

We next define  
\[ A(v):=\Lambda_{\C G}\la={1\over |G|}\sum_{h\in G}h\la,\quad B(p):=\Lambda_{\C(G)}\la=\delta_e\la\]
where $\delta_{e}(g^1g^2g^3g^4)=1$ iff $g^1g^2g^3g^4=e$, which is iff $(g^4)^{-1}=g^1g^2g^3$, which is iff $g^4g^1g^2g^3=e$. Hence $\delta_{e}(g^1g^2g^3g^4)=\delta_{e}(g^4g^1g^2g^3)$ is invariant under cyclic rotations, hence $\Lambda_{\C(G)}\la$ computed at site $(v,p)$ does not depend on the location of $v$ on the boundary of $p$. Moreover,
\[ A(v)B(p)=|G|^{-1}\sum_h h\delta_e\la=|G|^{-1}\sum_h \delta_{heh^{-1}}h\la=|G|^{-1}\sum_h \delta_{e}h\la=B(p)A(v)\]  
if $v$ is a vertex on the boundary of $p$ by Lemma~\ref{lemDGrep}, and more trivially if not. We also have the rest of
\[ A(v)^2=A(v),\quad B(p)^2=B(p),\quad [A(v),A(v')]=[B(p),B(p')]=[A(v),B(p)]=0\]
for all $v\ne v'$ and $p\ne p'$, as easily checked. We then define the Hamiltonian
\[ H=\sum_v (1-A(v)) + \sum_p (1-B(p))\]
and the space of vacuum states
\[ \CH_{\rm vac}=\{|\psi\>\in\CH\ |\ A(v)|\psi\>=B(p)|\psi\>=|\psi\>,\quad \forall v,p\}.\]

Quasiparticles in Kitaev models are labelled by representations of $D(G)$ occupying a given site $(v,p)$, which take the system out of the vacuum. Detection of a quasiparticle is via a {\em projective measurement} of the operator $\sum_{\CC, \pi} p_{\CC,\pi} P_{\mathcal{C}, \pi}$ acting at each site on the lattice for distinct coefficients $p_{\CC,\pi} \in \R$. By definition, this is a process which yields the classical value $p_{\CC,\pi}$ with a probability given by the likelihood of the state prior to the measurement being in the subspace in the image of $P_{\mathcal{C},\pi}$, and in so doing performs the corresponding action of the projector $P_{\mathcal{C}, \pi}$ at the site. The projector $P_{e,1}$ corresponds to the vacuum quasiparticle.

In computing terms, this system of measurements encodes a logical Hilbert subspace, which we will always take to be the vacuum space $\CH_{\rm vac}$, within the larger physical Hilbert space given by the lattice; this subspace is dependent on the topology of the surface that the lattice is embedded in, but not the size of the lattice. For example, there is a convenient closed-form expression for the dimension of $\CH_{\rm vac}$ when $\Sigma$ occupies a closed, orientable surface \cite{Cui}. Computation can then be performed on states in the logical subspace in a fault-tolerant manner, with unwanted excitations constituting detectable errors.

In the interest of brevity, we forgo a detailed exposition of such measurements, ribbon operators and fault-tolerant quantum computation on the lattice. The interested reader can learn about these in e.g. \cite{Kit,Bom,CCW,CowMa}. We do give a brief recap of ribbon operators, although without much rigour, as these will be useful later.

\begin{figure}
\[ \includegraphics[scale=0.8]{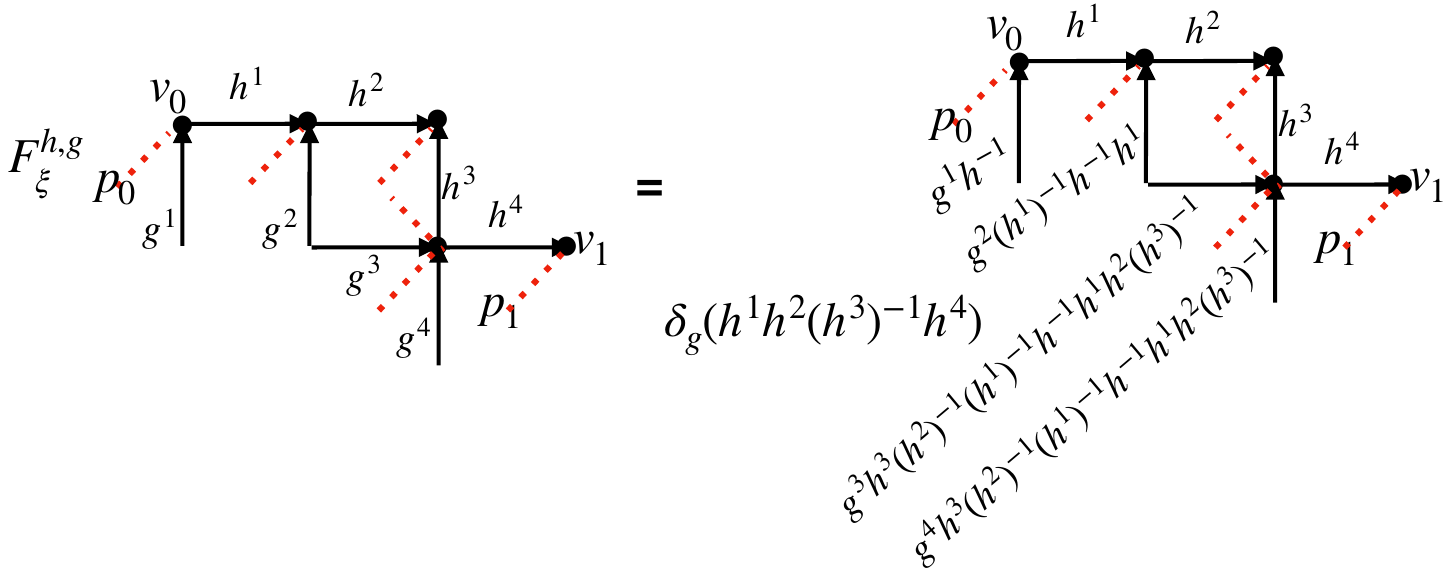}\]
\caption{\label{figribbon} Example of a ribbon operator for a ribbon $\xi$ from $s_0=(v_0,p_0)$ to $s_1=(v_1,p_1)$.} 
\end{figure}

\begin{definition}\rm \label{def:ribbon}
A ribbon $\xi$ is a strip of face width that connects two sites $s_0 = (v_0,p_0)$ and $s_1 = (v_1,p_1)$ on the lattice. A ribbon operator $F^{h,g}_\xi$ acts on the vector spaces associated to the edges along the path of the ribbon, as shown in Fig~\ref{figribbon}. We call this basis of ribbon operators labelled by $h$ and $g$ the \textit{group basis}.
\end{definition}

\begin{lemma}\label{lem:concat}
If $\xi'$ is a ribbon concatenated with $\xi$, then the associated ribbon operators in the group basis satisfy
\[F_{\xi'\circ\xi}^{h,g}=\sum_{f\in G}F_{\xi'}^{f^{-1}hf,f^{-1}g}\circ F_\xi^{h,f}, \quad F^{h,g}_\xi \circ F^{h',g'}_\xi=\delta_{g,g'}F_\xi^{hh',g}.\]
\end{lemma}
The first identity shows the role of the comultiplication of $D(G)^*$, 
\[\Delta(h\delta_g) = \sum_{f\in G} h\delta_f\otimes f^{-1}hf\delta_{f^{-1}g}.\]
using subalgebra notation, while the second identity implies that
\[(F_\xi^{h,g})^\dagger = F_\xi^{h^{-1},g}.\]

\begin{lemma}\label{ribcom}\cite{Kit} Let $\xi$ be a ribbon with the orientation as shown in Figure~\ref{figribbon} between sites $s_0=(v_0,p_0)$ and $s_1=(v_1,p_1)$. Then
\[ [F_\xi^{h,g},f\la_v]=0,\quad [F_\xi^{h,g},\delta_e\la_p]=0,\]
for all $v \notin \{v_0, v_1\}$ and $p \notin \{p_0, p_1\}$.
\[ f\la_{s_0}\circ F_\xi^{h,g}=F_\xi^{fhf^{-1},fg} \circ f\la_{s_0},\quad \delta_f\la_{s_0}\circ F_\xi^{h,g}=F_\xi^{h,g} \circ\delta_{h^{-1}f}\la_{s_0},\]
\[ f\la_{s_1}\circ F_\xi^{h,g}=F_\xi^{h,gf^{-1}} \circ f\la_{s_1},\quad \delta_f\la_{s_1}\circ F_\xi^{h,g}=F_\xi^{h,g}\circ  \delta_{fg^{-1}hg}\la_{s_1}\]
for all ribbons where $s_0,s_1$ are disjoint, i.e. when $s_0$ and $s_1$ share neither vertices or faces. The subscript notation $f\la_v$ means the local action of $f\in \C G$ at vertex $v$, and the dual for $\delta_f\la_s$ at a site $s$.
\end{lemma}

We call the above lemma the \textit{equivariance property} of ribbon operators. Such ribbon operators may be deformed according to a sort of discrete isotopy, so long as the endpoints remain the same. We formalised ribbon operators as left and right module maps in \cite{CowMa}, but skim over any further details here. The physical interpretation of ribbon operators is that they create, move and annihilate quasiparticles.

\begin{lemma}\cite{Kit}\label{lem:ribs_only}
Let $s_0$, $s_1$ be two sites on the lattice. The only operators in ${\rm End}(\CH)$ which change the states at these sites, and therefore create quasiparticles and change the distribution of measurement outcomes, but leave the state in vacuum elsewhere, are ribbon operators.
\end{lemma}
This lemma is somewhat hard to prove rigorously but a proof was sketched in \cite{CowMa}. Next, there is an alternate basis for these ribbon operators in which the physical interpretation becomes more obvious. The \textit{quasiparticle basis} has elements 
\begin{equation}F_\xi^{'\CC,\pi;u,v} = \sum_{n\in G^{c_0}}  \pi(n^{-1})_{ji}  F_\xi^{c, q_c n q_d^{-1}},\end{equation}
where $\CC$ is a conjugacy class, $\pi$ is an irrep of the associated isotropy subgroup $G^{c_0}$ and $u = (c,i)$, $v = (d,j)$ label basis elements of $W_{\CC,\pi}$ in which $c,d \in \CC$ and $i,j$ label a basis of $W_\pi$. This amounts to a nonabelian Fourier transform of the space of ribbons (that is, the Peter-Weyl isomorphism of $D(G)$) and has inverse
\begin{equation}F_\xi^{h,g} = \sum_{\CC,\pi\in \hat{G^{c_0}}}\sum_{c\in\CC}\delta_{h,gcg^{-1}} \sum_{i,j = 0}^{{\rm dim}(W_\pi)}\pi(q^{-1}_{gcg^{-1}}g q_c)_{ij}F_\xi^{'\CC,\pi;a,b},\end{equation}
where $a = (gcg^{-1},i)$ and $b=(c,j)$. This reduces in the chargeon sector to the special cases
\begin{equation}\label{chargeon_ribbons}F_\xi^{'e,\pi;i,j} = \sum_{n\in G}\pi(n^{-1})_{ji}F_\xi^{e,n}\end{equation}
and
\begin{equation}F_\xi^{e,g} = \sum_{\pi\in \hat{G}}\sum_{i,j = 0}^{{\rm dim}(W_\pi)}\pi(g)_{ij}F_\xi^{'e,\pi;i,j}\end{equation}
Meanwhile, in the fluxion sector we have
\begin{equation}\label{fluxion_ribbons}F_\xi^{'\CC,1;c,d}=\sum_{n\in G^{c_0}}F_\xi^{c,q_c nq_d^{-1}}\end{equation}
but there is no inverse in the fluxion sector. This is because the chargeon sector corresponds to the irreps of $\C G$, itself a semisimple algebra; the fluxion sector has no such correspondence.

If $G$ is Abelian then $\pi$ are 1-dimensional and we do not have to worry about the indices for the basis of $W_\pi$; this then looks like a more usual Fourier transform.

\begin{lemma}\label{lem:quasi_basis}
If $\xi'$ is a ribbon concatenated with $\xi$, then the associated ribbon operators in the quasiparticle basis satisfy
\[ F_{\xi'\circ\xi}^{'\CC,\pi;u,v}=\sum_w F_{\xi'}^{'\CC,\pi;w,v}\circ F_\xi^{'\CC,\pi;u,w}\]
and are such that the nonabelian Fourier transform takes convolution to multiplication and vice versa, as it does in the abelian case.
\end{lemma}

In particular, we have the \textit{ribbon trace operators}, $W^{\CC,\pi}_\xi := \sum_u F_\xi^{'\CC,\pi;u,u}$. Such ribbon trace operators create exactly quasiparticles of the type $\CC,\pi$ from the vacuum, meaning that
\[P_{(\CC,\pi)}\la_{s_0}W^{\CC,\pi}_\xi\vac = W^{\CC,\pi}_\xi\vac = W^{\CC,\pi}_\xi\vac\ra_{s_1}P_{(\CC,\pi)}.\]
We refer to \cite{CowMa} for more details and proofs of the above. 

\begin{example}\rm \label{exDS3} Our go-to example for our expositions will be $G=S_3$ generated by transpositions $u=(12), v=(23)$ with $w=(13)=uvu=vuv$. There are then 8 irreducible representations of $D(S_3)$ according to the choices $\CC_0=\{e\}$, $\CC_1=\{u,v,w\}$, $\CC_2=\{uv,vu\}$ for which we pick representatives $c_0=e$, $q_e=e$, $c_1=u$, $q_u=e$, $q_v=w$, $q_w=v$ and $c_2=uv$ with $q_{uv}=e,q_{vu}=v$   (with the $c_i$ in the role of $c_0$ in the general theory). Here $G^{c_0}=S_3$ with 3 representations $\pi=$ trivial, sign and $W_2$ the 2-dimensional one given by (say) $\pi(u)=\sigma_3, \pi(v)=(\sqrt{3}\sigma_1-\sigma_3)/2$, $G^{c_1}=\{e,u\}=\Z_2$ with $\pi(u)=\pm1$ and $G^{c_2}=\{e,uv,vu\}=\Z_3$ with $\pi(uv)=1,\omega,\omega^2$ for $\omega=e^{2\pi\imath\over 3}$.  See \cite{CowMa} for details and calculations of the associated projectors and some $W_\xi^{\CC,\pi}$ operators. 
\end{example}

\section{Gapped Boundaries}\label{sec:gap}
While $D(G)$ is the relevant algebra for the bulk of the model, our focus is on the boundaries. For these, we require a different class of algebras.

\subsection{The boundary subalgebra $\Xi(R,K)$}\label{sec:xi}

Let $K\subseteq G$ be a subgroup of a finite group $G$ and $G/K=\{gK\ |\ g\in G\}$ be the set of left cosets. It is not necessary in this section, but convenient,  to fix a representative $r$ for each coset and let $R\subseteq G$ be the set of these, so there is a bijection between $R$ and $G/K$  whereby $r\leftrightarrow rK$. We assume that $e\in R$ and call such a subset (or section of the map $G\to G/K$) a {\em transversal}.  Every element of $G$ factorises uniquely as $rx$ for $r\in R$ and $x\in K$, giving a coordinatisation of $G$ which we will use. Next, as we quotiented by $K$ from the right, we still have an action of $K$ from the left on $G/K$, which we denote $\la$. By the above bijection, this equivalently means an action $\la:K\times R\to R$ on $R$ which in terms of the factorisation is determined by $xry=(x\la r)y'$, where we refactorise $xry$ in the form $RK$ for some $y'\in R$. There is much more information in this factorisation, as will see in Section~\ref{sec:quasi}, but  this action is all we need for now. Also note that we have chosen to work with left cosets so as to be consistent with the literature \cite{CCW,BSW}, but one could equally choose a right coset factorisation to build a class of algebras similar to those in \cite{KM2}. We consider the algebra $\C(G/K)\lcross \C K$ as the cross product by the above action. Using our coordinatisation, this becomes the following algebra.

\begin{definition}\label{defXi}  $\Xi(R,K)=\C(R)\lcross \C K$ is generated by $\C(R)$ and $\C K$ with cross relations $x\delta_r=\delta_{x\la r} x$. Over $\C$, this is a $*$-algebra with $(\delta_r x)^*=x^{-1}\delta_r=\delta_{x^{-1}\la r}x^{-1}$.
\end{definition}
If we choose a different transversal $R$ then the algebra does not change up to an isomorphism which maps the $\delta$-functions between the corresponding choices of representative. Of relevance to the applications, we also have:
\begin{lemma} $\Xi(R,K)$ has the `integral element'
\[\Lambda:=\Lambda_{\C(R)} \otimes \Lambda_{\C K} = \delta_e \frac{1}{|K|}\sum_{x\in K}x\]
characterised by $\xi\Lambda=\eps(\xi)\Lambda=\Lambda\xi$ for all $\xi\in \Xi$, and $\eps(\Lambda)=1$.
\end{lemma}
\proof We check that
\begin{align*}
\xi\Lambda& = (\delta_s y)(\delta_e\frac{1}{|K|}\sum_{x\in K}x) = \delta_{s,y\la e}\delta_s\frac{1}{|K|}\sum_{x\in K}yx= \delta_{s,e}\delta_e \frac{1}{|K|}\sum_{x\in K}x\\ 
&= \eps(\xi)\Lambda = \frac{1}{|K|}\sum_{x\in K}\delta_{e,x\la y}\delta_e xy = \frac{1}{|K|}\sum_{x\in K}\delta_{e,y}\delta_e x = \Lambda\xi. 
\end{align*}
And clearly, $\eps(\Lambda) = \delta_{e,e} {|K|\over |K|} = 1$. 
\endproof

As a cross product algebra, we can take the same approach as with $D(G)$ to the classification of its irreps:
\begin{lemma} Irreps of  $\Xi(R,K)$  are classified by pairs $(\CO,\rho)$ where  $\CO\subseteq R$ is an orbit under the action $\la$ and $\rho$ is an irrep of the isotropy group $K^{r_0}:=\{x\in K\ |\ x\la r_0=r_0\}$. Here we fix a base point $r_0\in \CO$ as well as $\kappa: \CO\to K $  a choice of lift such that 
\[ \kappa_r\la r_0 = r,\quad\forall r\in \CO,\quad \kappa_{r_0}=e.\]
Then 
\[ V_{\CO,\rho}=\C \CO\tens V_\rho,\quad \delta_r(s\tens v)=\delta_{r,s}s\tens v,\quad x.(s\tens v)=x\la s\tens\zeta_s(x).v,\quad \zeta_s(x)=\kappa^{-1}_{x\la s}x\kappa_s\]
for $v\in V_\rho$, the carrier space for $\rho$, and
\[ \zeta: \CO\times K\to K^{r_0},\quad \zeta_r(x)=\kappa_{x\la r}^{-1}x\kappa_r.\]
\end{lemma}
\proof One can check that $\zeta_r(x)$ lives in $K^{r_0}$,  
\[ \zeta_r(x)\la r_0=(\kappa_{x\la r}^{-1}x\kappa_r)\la r_0=\kappa_{x\la r}^{-1}\la(x\la r)=\kappa_{x\la r}^{-1}\la(\kappa_{x\la r}\la r_0)=r_0\]
and the cocycle property
\[ \zeta_r(xy)=\kappa^{-1}_{x\la y\la r}x \kappa_{y\la r}\kappa^{-1}_{y\la r}y \kappa_r=\zeta_{y\la r}(x)\zeta_r(y),\]
from which it is easy to see that $V_{\CO,\rho}$ is a representation,
\[ x.(y.(s\tens v))=x.(y\la s\tens \zeta_s(y). v)=x\la(y\la s)\tens\zeta_{y\la s}(x)\zeta_s(y).v=xy\la s\tens\zeta_s(xy).v=(xy).(s\tens v),\]
\[ x.(\delta_r.(s\tens v))=\delta_{r,s}x\la s\tens \zeta_s(x). v= \delta_{x\la r,x\la s}x\la s\tens\zeta_s(x).v=\delta_{x\la r}.(x.(s\tens v)).\]
One can show that  $V_{\CO,\pi}$ are irreducible and do not depend up to isomorphism on the choice of $r_0$ or $\kappa_r$.\endproof

In the $*$-algebra case as here, we obtain a unitary representation if $\rho$ is unitary. One can also show that all irreps can be obtained this way. In fact the algebra $\Xi(R,K)$ is semisimple and has a block associated to the $V_{\CO,\pi}$.

\begin{lemma}\label{Xiproj} $\Xi(R,K)$ has a complete orthogonal set of central  idempotents  
\[ P_{(\CO,\rho)}={\dim V_\rho\over |K^{r_0}|}\sum_{r\in\CO}\sum_{n\in K^{r_0}} \Tr_{\rho}(n^{-1})\delta_r\tens \kappa_r n \kappa_r^{-1}.\]
\end{lemma}
\proof The proofs are similar to those for $D(G)$ in \cite{CowMa}. That we have a projection is 
\begin{align*}P_{(\CO,\rho)}^2&={\dim(V_\rho)^2\over |K^{r_0}|^2}\sum_{m,n\in K^{r_0}}\Tr_\rho(m^{-1})\Tr_\rho(n^{-1})\sum_{r,s\in \CO}(\delta_r\tens \kappa_rm\kappa_r^{-1})(\delta_s\tens\kappa_sn\kappa_s^{-1})\\
&={\dim(V_\rho)^2\over |K^{r_0}|^2}\sum_{m,n\in K^{r_0}}\Tr_\rho(m^{-1})\Tr_\rho(n^{-1})\sum_{r,s\in \CO}\delta_r\delta_{r,s}\tens \kappa_rm\kappa_r^{-1}\kappa_s n\kappa_s^{-1}\\
&={\dim(V_\rho)^2\over |K^{r_0}|^2}\sum_{m,m'\in K^{r_0}}\Tr_\rho(m^{-1})\Tr_\rho(m m'{}^{-1})\sum_{r\in \CO}\delta_r\tens \kappa_rm'\kappa_r^{-1}= P_{(\CO,\rho)}
\end{align*}
where we used $r=\kappa_r m\kappa_r^{-1}\la s$ iff $s=\kappa_r m^{-1}\kappa_r^{-1}\la r=\kappa_r m^{-1}\la r_0=\kappa_r\la r_0=r$. We then changed $mn=m'$ as a new variable and used the orthogonality formula for characters on $K^{r_0}$. Similarly, for different projectors to be orthogonal.  The sum of projectors is 1 since
\begin{align*}\sum_{\CO,\rho}P_{(\CO,\rho)}=\sum_{\CO, r\in \CC}\delta_r\tens \kappa_r\sum_{\rho\in \hat{K^{r_0}}} \left({\dim V_\rho\over |K^{r_0}|}\sum_{n\in K^{r_0}} \Tr_{\rho}(n^{-1}) n\right) \kappa_r^{-1}=\sum_{\CO,r\in\CO}\delta_r\tens 1=1,
\end{align*}
where the bracketed expression is the projector $P_\rho$ for $\rho$ in the group algebra of $K^{r_0}$, and these sum to 1 by the Peter-Weyl decomposition of the latter. \endproof

\begin{remark}\rm
In the previous literature, the irreps have been described using double cosets and representatives thereof \cite{CCW}. In fact a double coset in ${}_KG_K$ is an orbit for the left action of $K$ on $G/K$ and hence has the form $\CO K$ corresponding to an orbit $\CO\subset R$ in our approach. We will say more about this later, in Proposition~\ref{prop:mon_equiv}.
\end{remark}

An important question for the physics is how representations on the bulk relate to those on the boundary. This is afforded by functors in the two directions. Here we give a much more direct approach to this issue as follows.
\begin{proposition}\label{Xisub} There is an inclusion of algebras $i:\Xi(R,K)\hookrightarrow D(G)$
\[ i(x)=x,\quad i(\delta_r)=\sum_{x\in K} \delta_{rx}.\]
 The pull-back or restriction of a $D(G)$-module $W$ to a $\Xi$-module $i^*(W)$ is simply for $\xi\in \Xi$ to act by $i(\xi)$. Going the other way, the induction functor sends a $\Xi$-module $V$ to a $D(G)$-module $D(G)\tens_\Xi V$, where $\xi\in \Xi$ right acts on $D(G)$ by right multiplication by $i(\xi)$. These two functors are adjoint.
 \end{proposition}
 \proof We just need to check that $i$ respects the relations of $\Xi$. Thus, 
 \begin{align*} i(\delta_r)i(\delta_s)&=\sum_{x,y\in K}\delta_{rx}\delta_{sy}=\sum_{x\in K}\delta_{r,s}\delta_{rx}=i(\delta_r\delta_s),
 \\ i(x)i(\delta_r)&=\sum_{y\in K}x\delta_{ry}=\sum_{y\in K}\delta_{xryx^{-1}}x=\sum_{y\in K}\delta_{(x\la r)x'yx^{-1}}x=\sum_{y'\in K}\delta_{(x\la r)y'}x=i(\delta_{x\la r} x),\end{align*}
as required. For the first line, we used the unique factorisation $G=RK$ to break down the $\delta$-functions. For the second line, we use this in the form $xr=(x\la r)x'$ for some $x'\in K$ and then changed variables from $y$ to $y'=x'yx^{-1}$. The rest follows as for any algebra inclusion. \endproof

In fact, $\Xi$ is a quasi-bialgebra and at least when $(\ )^R$ is bijective a quasi-Hopf algebra, as we see in Section~\ref{sec:quasi}. In the latter case, it  has a quantum double $D(\Xi)$ which contains $\Xi$ as a sub-quasi Hopf algebra. Moreover, it can be shown that $D(\Xi)$ is a `Drinfeld cochain twist' of $D(G)$, which implies it has the same algebra as $D(G)$. We do not need details, but this is the abstract reason for the above inclusion. (An explicit proof of this twisting result in the usual Hopf algebra case with $R$ a group is in \cite{BGM}.) Meanwhile, the statement that the two functors in the lemma are adjoint is that 
\[ \hom_{D(G)}(D(G)\tens_\Xi V,W))=\hom_\Xi(V, i^*(W))\]
for all $\Xi$-modules $V$ and all $D(G)$-modules $W$. These functors do not take irreps to irreps and of particular interest are the multiplicities for the decompositions back into irreps, i.e. if  $V_i, W_a$ are respective irreps and $D(G)\tens_\Xi V_i=\oplus_{a} n^i{}_a W_a$ then 
\[ {\rm dim}(\hom_{D(G)}(D(G)\tens_\Xi V_i,W_a))={\rm dim}(\hom_\Xi(V_i,i^*(W_a)))\]
and hence $i^*(W_a)=\oplus_i n^i_a V_i$. This explains one of the observations in \cite{CCW}. It remains to give a formula for these multiplicities, but here we were not able to reproduce the formula in \cite{CCW}. Our approach goes via a general lemma as follows. 

\begin{lemma}\label{lemfrobn} Let  $i:A\hookrightarrow B$ be an inclusion of finite-dimensional semisimple algebras and $\int$ the unique symmetric special Frobenius linear form on $B$ such that $\int 1=1$. Let $V_i$ be an irrep of $A$ and $W_a$ an irrep of $B$. Then the multiplicity $V_i$ in the pull-back $i^*(W_a)$ (which is the same as the multiplicity of $W_a$ in $B\tens_A V_i$) is  given by 
\[ n^i{}_a={\dim(B)\over\dim(V_i)\dim(W_a)}\int i(P_i)P_a,\]
where $P_i\in A$ and $P_a\in B$ are the associated central idempotents. Moreover, $i(P_i)P_a  =0$ if and only if $n^i_a = 0$.
\end{lemma}
\proof Recall that a linear map $\int:B\to \C$ is Frobenius if the bilinear form $(b,c):=\int bc$ is nondegenerate, and is symmetric if this bilinear form is symmetric. Also, let $g=g^1\tens g^2\in B\tens B$ (in a notation with the sum of such terms understood) be the associated `metric' such that $(\int b g^1 )g^2=b=g^1\int g^2b$ for all $b$ (it is the inverse matrix in a basis of the algebra). We say that the Frobenius form is special if the algebra product $\cdot$ obeys $\cdot(g)=1$. It is well-known that there is a unique symmetric special Frobenius form up to scale, given by the trace in the left regular representation, see \cite{MaRie:spi} for a recent study.  In our case, over $\C$, we also know that a finite-dimensional semisimple algebra $B$ is a direct sum of matrix algebras ${\rm End}(W_a)$ associated to the irreps $W_a$ of $B$. Then
\begin{align*} \int i(P_i)P_a&={1\over\dim(B)}\sum_{\alpha,\beta}\<f^\alpha\tens e_\beta,i(P_i)P_a (e_\alpha\tens f^\beta)\>\\
&={1\over\dim(B)}\sum_{\alpha}\dim(W_a)\<f^\alpha, i(P_i)e_\alpha\>={\dim(W_a)\dim(V_i)\over\dim(B)} n^i{}_a.
\end{align*}
where $\{e_\alpha\}$ is a basis of $W_a$ and $\{f^\beta\}$ is a dual basis, and $P_a$ acts as the identity on $\End(W_a)$ and zero on the other blocks. We then used that if  $i^*(W_a)=\oplus_i {n^i{}_a}V_i$ as $A$-modules, then $i(P_i)$ just picks out the $V_i$ components where $P_i$ acts as the identity. 

For the last part, the forward direction is immediate given the first part of the lemma. For the other direction, suppose 
$n^i_a = 0$ so that $i^*(W_a)=\oplus_j n^j_aV_j$ with $j\ne a$ running over the other irreps of $A$. Now, we can view $P_{a}\in W_{a}\tens W_{a}^*$ (as the identity element) and left multiplication by $i(P_i)$ is the same as $P_i$ acting on $P_{a}$ viewed as an element of $i^*(W_{a})\tens W_{a}^*$, which is therefore zero.\endproof

We apply Lemma~\ref{lemfrobn} in our case of $A=\Xi$ and $B=D(G)$, where \[ \dim(V_i)=|\CO|\dim(V_\rho), \quad \dim(W_a)=|\CC|\dim(W_\pi)\]
with $i=(\CC,\rho)$ as described above and $a=(\CC,\pi)$ as described in Section~\ref{sec:bulk}.

\begin{proposition}\label{nformula} For the inclusion $i:\Xi\hookrightarrow D(G)$ in Proposition~\ref{Xisub}, the multiplicities for restriction and induction as above are given by
\[ n^{(\CO,\rho)}_{(\CC,\pi)}= {|G|\over |\CO| |\CC|  |K^{r_0}| |G^{c_0}|} \sum_{{r\in \CO,  c\in \CC\atop
   r^{-1}c\in K}} |K^{r,c}|\sum_{\tau\in \hat{K^{r,c}} } n_{\tau,\tilde\rho|_{K^{r,c}}} n_{\tau, \tilde\pi|_{K^{r,c}}},\quad K^{r,c}=K^r\cap G^c,\]
where $\tilde \pi(m)=\pi(q_c^{-1}mq_c)$ and $\tilde\rho(m)=\rho(\kappa_r^{-1}m\kappa_r)$ are the corresponding representation of $K^r,G^c$ decomposing as $K^{r,c}$ representations as
\[ \tilde\rho|_{K^{r,c}}\isom\oplus_\tau n_{\tau,\tilde\rho|_{K^{r,c}}}\tau,\quad  \tilde\pi|_{K^{r,c}}\isom\oplus_\tau n_{\tau,\tilde\pi|_{K^{r,c}}}\tau.\]
\end{proposition}
\proof We include the projector from Lemma~\ref{Xiproj} as 
\[ i(P_{(\CO,\rho)})={{\rm dim}(V_\rho)\over |K^{r_0}|}\sum_{r\in \CO, x\in K}\sum_{m\in K^{r_0}}\Tr_\rho(m^{-1})\delta_{rx}\tens \kappa_r m\kappa_r^{-1}\]
and multiply this by $P_{(\CC,\pi)}$ from (\ref{Dproj}). In the latter, we write $c=sy$ for the factorisation of $c$. Then when we multiply these out, for $(\delta_{rx}\tens \kappa_r m \kappa_r^{-1})(\delta_{c}\tens q_c n q_c^{-1})$ we will need $\kappa_r m\kappa_r^{-1}\la s=r$ or equivalently $s=\kappa_r m^{-1}\kappa_r^{-1}\la r=r$ so we are actually summing not over $c$ but over $y\in K$ such that $ry\in \CC$. Also then $x$ is uniquely determined in terms of $y$.
Hence
\[ i(P_{(\CO,\rho)})P_{(\CC,\pi)}={{\rm dim}(V_\rho){\rm dim}(W_\pi)\over |K^{r_0}| |G^{c_0}|}\sum_{m\in K^{r_0}, n\in G^{c_0}}\sum_{r\in \CO,  y\in K | ry\in\CC}  \Tr_\rho(m^{-1})\Tr_\pi(n^{-1})   \delta_{rx}\tens \kappa_r m\kappa_r^{-1} q_c nq_c^{-1}\]
Now we apply the integral of $D(G)$, $\int\delta_g\tens h=\delta_{h,e}$ which requires 
\[ n=q_c^{-1}\kappa_r m^{-1}\kappa_r^{-1}q_c\]
and  $x=y$ for $n\in G^{c_0}$ given that $c=ry$. We refer to this condition on $y$ as $(\star)$. Remembering that $\int$ is normalised so that $\int 1=|G|$, we have from the lemma
\begin{align*}n^{(\CO,\rho)}_{(\CC,\pi)}&={|G|\over\dim(V_i)\dim(W_a)}\int i(P_{(\CO,\rho)})P_{(\CC,\pi)}\\
&={|G|\over |\CO| |\CC| |K^{r_0}| |G^{c_0}|}\sum_{m\in K^{r_0}}\sum_{{r\in \CO,  y\in K\atop  (*),   ry\in\CC}}  \Tr_\rho(m^{-1})\Tr_\pi(q_{ry}^{-1}\kappa_r m\kappa_r^{-1}q_{ry}) \\
&={|G|\over |\CO| |\CC| |K^{r_0}| |G^{c_0}|}\sum_{m\in K^{r_0}}\sum_{{r\in \CO,  c\in \CC\atop
   r^{-1}c\in K}}\sum_{m'\in  K^r\cap G^c} \Tr_\rho(\kappa_r^{-1}m'{}^{-1}\kappa_r)\Tr_\pi(q_{c}^{-1} m q_{c}),
\end{align*}
where we compute in $G$ and view $(\star)$ as $m':=\kappa_r m \kappa_r^{-1}\in G^c$.  We then use the group orthogonality formula
\[ \sum_{m\in K^{r,c}}\Tr_{\tau}(m^{-1})\Tr_{\tau'}(m)=\delta_{\tau,\tau'}|K^{r,c}| \]
for any irreps $\tau,\tau'$ of the group 
\[ K^{r,c}:=K^r\cap G^c=\{x\in K\ |\ x\la r=r,\quad x c x^{-1}=c\}\]
to  obtain the formula stated. \endproof

This simplifies in four (overlapping) special cases as follows.

\noindent{(i) $V_i$ trivial: }
\[ n^{(\{e\},1)}_{(\CC,\pi)}={|G|\over |\CC||K||G^{c_0}|}\sum_{c\in \CC\cap K}\sum_{m\in K\cap G^c}\Tr_\pi(q_c^{-1}mq_c)={|G| \over |\CC| |K||G^{c_0}|}\sum_{c\in \CC\cap K} |K^c|  n_{1,\tilde\pi}\]
as $\rho=1$ implies $\tilde\rho=1$ and forces $\tau=1$. Here $K^c$ is the centraliser of $c\in K$. If  $n_{1,\tilde\pi}$ is independent of the choice of $c$ then we can simplify this further
as 
\[ n^{(\{e\},1)}_{(\CC,\pi)}={|G| |(\CC\cap K)/K|\over |\CC| |G^{c_0}|} n_{1,\pi|_{K^{c_0}}}\]
using the orbit-counting lemma, where $K$ acts on $\CC\cap K$ by conjugation.

\noindent{(ii) $W_a$ trivial:}
\[ n^{(\CO,\rho)}_{(\{e\},1)}= {|G|\over |\CO||K^{r_0}||G|}\sum_{r\in \CO\cap K}\sum_{m\in K^{r_0}}\Tr_\rho(m^{-1})=\begin{cases} 1 & {\rm if\ }\CO, \rho\ {\rm trivial}\\ 0 & {\rm else}\end{cases} \]
as $\CO\cap K=\{e\}$ if $\CO=\{e\}$ (but is otherwise empty) and in this case only $r=e$ contributes. This is consistent with the fact that if $W_a$ is the trivial representation of $D(G)$ then its pull back is also trivial and hence contains only the trivial representation of $\Xi$. 

\noindent{(iii) Fluxion sector:}
\[ n^{(\CO,1)}_{(\CC,1)}= {|G|\over |\CO||\CC||K^{r_0}| |G^{c_0}|} \sum_{{r\in \CO,  c\in \CC\atop
   r^{-1}c\in K}} |K^r\cap G^c|.\]

\noindent{(iv) Chargeon sector: }
\[ n^{(\{e\},\rho)}_{(\{e\},\pi)}=  n_{\rho, \pi|_{K}},\]
where $\rho,\pi$ are arbitrary irreps of $K,G$ respectively and only $r=c=e$ are allowed so  $K^{r,c}=K$, and then only $\tau=\rho$ contributes.

\begin{example}\label{exS3n}\rm (i) We take $G=S_3$, $K=\{e,u\}=\Z_2$, where $u=(12)$. Here $G/K$ consists of
\[ G/K=\{\{e, u\}, \{w, uv\}, \{v, vu\}\}\]
and our standard choice of $R$ will be $R=\{e,uv, vu\}$, where we take one from each coset (but any other transversal will have the same irreps and their decompositions).  This leads to 3 irreps of $\Xi(R,K)$ as follows. In $R$, we have two orbits $\CO_0=\{e\}$, $\CO_1=\{uv,vu\}$ and we choose representatives $r_0=e,\kappa_e=e$, $r_1=uv, \kappa_{uv}=e, \kappa_{vu}=u$ since  $u\la (uv)=vu$  for the two cases (here $r_1$ was denoted $r_0$ in the general theory and is the choice for $\CO_1$). We also have $u\la(vu)=uv$. Note that it happens that these orbits are also conjugacy classes but this is an accident of $S_3$ and not true for $S_4$. We have $K^{r_0}=K=\Z_2$ with representations $\rho(u)=\pm1$ and $K^{r_1}=\{e\}$ with only the trivial representation.  

(ii) For $D(S_3)$, we have the 8 irreps in Example~\ref{exDS3} and hence there is a $3\times 8$ table of the $\{n^i{}_a\}$. We can easily compute some of the special cases from the above. For example, the trivial $\pi$ restricted to $K$ is $\rho=1$, the sign representation restricted to $K$ is the $\rho=-1$ representation, the $W_2$ restricted to $K$ is $1\oplus -1$, which gives the upper left $2\times 3$ submatrix for the chargeon sector. Another 6 entries (four new ones) are given from the fluxion formula. We also have $\CC_2\cap K=\emptyset$ so that the latter part of the first two rows is zero by our first special case formula. For $\CC_1,\pm1$ in the first row, we have $\CC_1\cap K=\{u\}$ with trivial action of $K$, so just one orbit. This gives us a nontrivial result in the $+1$ case and 0 in the $-1$ case. The story for $\CC_1,\pm1$ in the second row follows the same derivation, but needs $\tau=-1$ and hence  $\pi=-1$ for the nonzero case. 
In the third row with $\CC_2,\pi$, we have $K^{r}=\{e\}$ so $G'=\{e\}$ and we only have $\tau=1=\rho$  as well as $\tilde\pi=1$ independently of $\pi$ as this is 1-dimensional. So both $n$ factors in the formula in Proposition~\ref{nformula} are 1. In the sum over $r,c$, we need  $c=r$ so we sum over 2 possibilities, giving a nontrivial result as shown. For $\CC_1,\pi$, the first part goes the same way and we similarly have $c$ determined from $r$ in the case of $\CC_1,\pi$, so again two contributions in the sum, giving  the answer shown independently of $\pi$. Finally, for $\CC_0,\pi$ we have $r=\{uv,vu\}$ and $c=e$, and can never meet the condition $r^{-1}c\in K$. So these all have $0$. Thus,  Proposition~\ref{nformula} in this example tells us:
\[\begin{array}{c|c|c|c|c|c|c|c|c} n^i{}_a & \CC_0,1 & \CC_0,{\rm sign} & \CC_0,W_2 & \CC_1, 1& \CC_1,-1 & \CC_2,1& \CC_2,\omega & \CC_2,\omega^2\\
 \hline\
 \CO_0,1&1 & 0 & 1  &1 & 0& 0 &0 &0 \\
\hline
 \CO_0,-1&0 & 1&1& 0& 1&0 &0 &  0\\
 \hline
  \CO_1,1&0 &0&0  & 1& 1 &1  &1  & 1
  \end{array}\]
One can check for consistency that for each $W_a$,  $\dim(W_a)$ is the sum of the dimensions of the $V_i$ that it contains, which determines one row from the other two.  
\end{example}

\subsection{Boundary lattice model}\label{sec:boundary_lat}
Consider a vertex on the lattice $\Sigma$. Fixing a subgroup $K \subseteq G$, we define an action of $\C K$ on $\CH$ by
\begin{equation}\label{actXi0}\input{CaK_vertex_action.tikz}\end{equation}
One can see that this is an action as it is a tensor product of representations on each edge, or simply because it is the restriction to $K$ of the vertex action of $G$ in the bulk. Next, we define the action of $\C (R)$ at a face relative to a cilium,
\begin{equation}\label{actXi}\input{CGK_face_action.tikz}\end{equation}
with a clockwise rotation. That this is indeed an action is also easy to check explicitly, recalling that either $rK = r'K$ when $r= r'$ or $rK \cap r'K = \emptyset$ otherwise, for any $r, r'\in R$. These actions define a representation of $\Xi(R,K)$, which is just the bulk $D(G)$ action restricted to $\Xi(R,K)\subseteq D(G)$ by the inclusion in Proposition~\ref{Xisub}. This says that $x\in K$ acts as in $G$ and  $\C(R)$ acts on faces by the $\C(G)$ action after sending $\delta_r \mapsto \sum_{a\in rK}\delta_a$. To connect the above representation to the topic at hand, we now define what we mean by a boundary.

\subsubsection{Smooth boundaries}
Consider the lattice in the half-plane for simplicity,
\[\input{smooth_halfplane.tikz}\]
where each solid black arrow still carries a copy of $\C G$ and ellipses indicate the lattice extending infinitely. The boundary runs along the left hand side and we refer to the rest of the lattice as the `bulk'. The grey dashed edges and vertices are there to indicate empty space and  the lattice borders the edge with faces; we will call this case a `smooth' boundary. There is a site $s_0$ indicated at the boundary.

There is an action of $\C K$ at the boundary vertex associated to $s_0$, identical to the action of $\C K$ defined above but with the left edge undefined. Similarly, there is an action of $\C(R)$ at the face associated to $s_0$. However, this is more complicated, as the face has three edges undefined and the action must be defined slightly differently from in the bulk:
\[\input{smooth_face_action.tikz}\]
\[\input{smooth_face_actionB.tikz}\]
where the action is given a superscript $\la^b$ to differentiate it from the actions in the bulk. In the first case, we follow the same
clockwise rotation rule but skip over the undefined values on the grey edges, but for the second case we go round round anticlockwise. The resulting rule is then according to whether the cilium is associated to the top or bottom of the edge. It is easy to check that this defines a representation of $\Xi(R,K)$ on $\CH$ associated to each smooth boundary site, such as $s_0$, and that the actions of $\C(R)$ have been chosen such that this holds. A similar principle holds for $\la^b$ in other orientations of the boundary.

The integral actions  at a boundary vertex $v$ and at a face $s_0=(v,p)$ of a smooth boundary are then
\[ A^b_1(v):=\Lambda_{\C K}\la^b_v = {1\over |K|}\sum_k k\la^b_v,\quad B^b_1(p):=\Lambda_{\C(R)}\la^b_{p} = \delta_e\la^b_{p},\]
where the superscript $b$ and subscript $1$ label that these are at a smooth boundary. We have the convenient property that
\[\input{smooth_face_integral.tikz}\]
so both the vertex and face integral actions at a smooth face each depend only on the vertex and face respectively, not the precise cilium, similar to the integral actions.

\begin{remark}\rm
There is similarly an action of $\C(G) \lcross \C K \subseteq D(G)$ on $\CH$ at each site in the next layer into the bulk, where the site has the vertex at the boundary but an internal face. We mention this for completeness, and because using this fact it is easy to show that
\[A_1^b(v)B(p) = B(p)A_1^b(v),\]
where $B(p)$ is the usual integral action in the bulk.
\end{remark}

\begin{remark}\rm
In \cite{BSW} it is claimed that one can similarly introduce actions at smooth boundaries defined not only by $R$ and $K$ but also a 2-cocycle $\alpha$. This makes some sense categorically, as the module categories of $\CM^G$ may also include such a 2-cocycle, which enters by way of a \textit{twisted} group algebra $\C_\alpha K$ \cite{Os2}. However, in Figure 6 of \cite{BSW} one can see that when the cocycle $\alpha$ is introduced all edges on the boundary are assumed to be copies of $\C K$, rather than $\C G$. On closer inspection, it is evident that this means that the action on faces of $\delta_e\in\C(R)$ will always yield 1, and the action of any other basis element of $\C(R)$ will yield 0. Similarly, the action on vertices is defined to still be an action of $\C K$, not $\C_\alpha K$. Thus, the excitations on this boundary are restricted to only the representations of $\C K$, without either $\C(R)$ or $\alpha$ appearing, which appears to defeat the purpose of the definition. It is not obvious to us that a cocycle can be included along these lines in a consistent manner.
\end{remark}

In quantum computational terms, in addition to the measurements in the bulk we now measure the operator $\sum_{\CO,\rho}p_{\CO,\rho}P_{(\CO,\rho)}\la^b$ for distinct coefficients $p_{\CO,\rho} \in \R$ at all sites along the boundary. 

\subsubsection{Rough boundaries}
We now consider the half-plane lattice with a different kind of boundary,
\[\input{rough_halfplane.tikz}\]
This time, there is an action of $\C K$ at the exterior vertex and an action of $\C(R)$ at the face at the boundary with an edge undefined. Again, the former is just the usual action of $\C K$ with three edges undefined, but the action of $\C(R)$ requires more care and is defined as
\[\input{rough_face_action.tikz}\]
\[\input{rough_face_actionB.tikz}\]
\[\input{rough_face_actionC.tikz}\]
\[\input{rough_face_actionD.tikz}\]
All but the second action are just clockwise rotations as in the bulk, but with the greyed-out edge missing from the $\delta$-function. The second action goes counterclockwise in order to have an associated representation of $\Xi(R,K)$ at the bottom left. We have similar actions for other orientations of the lattice. 

\begin{remark}\rm  Although one can check that one has a representation of $\Xi(R,K)$ at each site using these actions and the action of $\C K$ defined before, this requires $g_1$ and $g_2$ on opposite sides of the $\delta$-function, and $g_1$ and $g_3$ on opposite sides, respectively for the last two actions. This means that there is no way to get $\delta_e\la^b$ to always be invariant under choice of site in the face. Indeed, we have not been able to reproduce the implicit claim in \cite{CCW} that $\delta_e\la^b$ at a rough boundary can be defined in a way that depends only on the face. 
\end{remark}

The integral actions at a boundary vertex $v$ and at a site $s_0=(v,p)$ of a rough boundary are then
\[ A_2^b(v):=\Lambda_{\C K}\la^b_v = {1\over |K|}\sum_k k\la^b_v,\quad B_2^b(v,p):=\Lambda_{\C(R)}\la^b_{s_0} = \delta_e\la_{s_0}^b \]
where the superscript $b$ and subscript $2$ label that these are at a rough boundary. In computational terms, we measure the operator $\sum_{\CO,\rho}p_{\CO,\rho}P_{(\CO,\rho)}\la^b$ at each site along the boundary, as with smooth boundaries.

Unlike the smooth boundary case, there is not an action of, say, $\C (R)\lcross \C G$ at each site in the next layer into the bulk, with a boundary face but interior vertex. In particular, we do not have $B_2^b(v,p)A(v) = A(v)B_2^b(v,p)$ in general, but we can still consistently define a Hamiltonian. When the action at $v$ is restricted to $\C K$ we recover an action of $\Xi(R,K)$ again.

As with the bulk, the Hamiltonian incorporating the boundaries uses the actions of the integrals. We can accommodate both rough and smooth boundaries into the Hamiltonian. Let $V,P$ be the set of vertices and faces in the bulk, $S_1$ the set of all sites $(v,p)$ at smooth boundaries, and $S_2$ the same for rough boundaries. Then
\begin{align*}H&=\sum_{v_i\in V} (1-A(v_i)) + \sum_{p_i\in P} (1-B(p_i)) \\
&\quad + \sum_{s_{b_1} \in S_1} ((1 - A_1^b(s_{b_1}) + (1 - B_1^b(s_{b_1}))) + \sum_{s_{b_2} \in S_2} ((1 - A_2^b(s_{b_2}) + (1 - B_2^b(s_{b_2})).\end{align*}

We can pick out two vacuum states immediately:
\begin{equation}\label{eq:vac1}|{\rm vac}_1\> := \prod_{v_i,s_{b_1},s_{b_2}}A(v_i)A_1^b(s_{b_1})A_2^b(s_{b_2})\bigotimes_E e\end{equation}
and
\begin{equation}\label{eq:vac2}|{\rm vac}_2\> := \prod_{p_i,s_{b_1},s_{b_2}}B(p_i)B_1^b(s_{b_1})B_2^b(s_{b_2})\bigotimes_E \sum_{g \in G} g\end{equation}
where the tensor product runs over all edges in the lattice.

\begin{remark}\rm
There is no need for two different boundaries to correspond to the same subgroup $K$, and the Hamiltonian can be defined accordingly. This principle is necessary when performing quantum computation by braiding `defects', i.e. finite holes in the lattice, on the toric code \cite{FMMC}, and also for the lattice surgery in Section~\ref{sec:patches}. We do not write out this Hamiltonian in all its generality here, but its form is obvious.
\end{remark}

\subsection{Quasiparticle condensation}
Quasiparticles on the boundary correspond to irreps of $\Xi(R,K)$. It is immediate from Section~\ref{sec:xi} that when $\CO = \{e\}$, we must have $r_0 = e, K^{r_0} = K$. We may choose the trivial representation of $K$ and then we have $P_{e,1} = \Lambda_{\C(R)} \otimes \Lambda_{\C K}$. We say that this particular measurement outcome corresponds to the absence of nontrivial quasiparticles, as the states yielding this outcome are precisely the locally vacuum states with respect to the Hamiltonian. This set of quasiparticles on the boundary will not in general be the same as quasiparticles defined in the bulk, as ${}_{\Xi(R,K)}\mathcal{M} \not\simeq {}_{D(G)}\mathcal{M}$ for all nontrivial $G$.

Quasiparticles in the bulk can be created from a vacuum and moved using ribbon operators \cite{Kit}, where the ribbon operators are seen as left and right module maps $D(G)^* \rightarrow \mathrm{End}(\CH)$, see \cite{CowMa}. Following \cite{CCW}, we could similarly define a different set of ribbon operators for the boundary excitations, which use $\Xi(R,K)^*$ rather than $D(G)^*$. However, these have limited utility. For completeness we cover them in Appendix~\ref{app:ribbon_ops}. Instead, for our purposes we will keep using the normal ribbon operators.

Such normal ribbon operators can extend to boundaries, still using Definition~\ref{def:ribbon}, so long as none of the edges involved in the definition are greyed-out. When a ribbon operator ends at a boundary site $s$, we are not concerned with equivariance with respect to the actions of $\C(G)$ and $\C G$ at $s$, as in Lemma~\ref{ribcom}. Instead we should calculate equivariance with respect to the actions of $\C(R)$ and $\C K$. We will study the matter in more depth in Section~\ref{sec:quasi}, but note that if $s,t\in R$ then unique factorisation means that $st=(s\cdot t)\tau(s,t)$ for unique elements $s\cdot t\in R$ and $\tau(s,t)\in K$.  Similarly, if $y\in K$ and $r\in R$ then unique factorisation $yr=(y\la r)(y\ra r)$ defines $y\ra r$ to be studied later.

\begin{lemma}\label{boundary_ribcom}
Let $\xi$ be an open ribbon from $s_0$ to $s_1$, where $s_0$ is located at a smooth boundary, for example:
\[\input{smooth_halfplane_ribbon_short.tikz}\]
and where $\xi$ begins at the specified orientation in the example, leading from $s_0$ into the bulk, rather than running along the boundary. Then
\[x\la^b_{s_0}\circ F_\xi^{h,g}=F_\xi^{xhx^{-1},xg} \circ x\la^b_{s_0};\quad \delta_r\la^b_{s_0}\circ F_\xi^{h,g}=F_\xi^{h,g} \circ\delta_{s\cdot(y\la r)}\la^b_{s_0}\]
$\forall x\in K, r\in R, h,g\in G$, and where $sy$ is the unique factorisation of $h^{-1}$.
\end{lemma}
\proof
The first is just the vertex action of $\C G$ restricted to $\C K$, with an edge greyed-out which does not influence the result. For the second, expand $\delta_r\la^b_{s_0}$ and verify explicitly:
\[\input{smooth_halfplane_ribbon_shortA1.tikz}\]
\[\input{smooth_halfplane_ribbon_shortA2.tikz}\]
where we see $(s\cdot(y\la r))K = s(y\la r)\tau(s,y\la r)^{-1}K = s(y\la r)K = s(y\la r)(y\ra r)K = syrK = h^{-1}rK$. We check the other site as well:
\[\input{smooth_halfplane_ribbon_shortB1.tikz}\]
\[\input{smooth_halfplane_ribbon_shortB2.tikz}\]
\endproof

\begin{remark}\rm
One might be surprised that the equivariance property holds for the latter case when $s_0$ is attached to the vertex at the bottom of the face, as in this case $\delta_r\la^b_{s_0}$ confers a $\delta$-function in the counterclockwise direction, different from the bulk. This is because the well-known equivariance properties in the bulk \cite{Kit} are not wholly correct, depending on orientation, as pointed out in \cite[Section~3.3]{YCC}. We accommodated for this by specifying an orientation in Lemma~\ref{ribcom}.
\end{remark}

\begin{remark}\rm\label{rem:rough_ribbon}
We have a similar situation for a rough boundary, albeit we found only one orientation for which the same equivariance property holds, which is:
\[\input{rough_halfplane_ribbon.tikz}\]
In the reverse orientiation, where the ribbon crosses downwards instead, equivariance is similar but with the introduction of an antipode. For other orientations we do not find an equivariance property at all.
\end{remark}

As with the bulk, we can define an excitation space using a ribbon between the two endpoints $s_0$, $s_1$, although more care must be taken in the definition.

\begin{lemma}\label{Ts0s1}
Let $\vac$ be a vacuum state on a half-plane $\Sigma$, where there is one smooth boundary beyond which there are no more edges. Let $\xi$ be a ribbon between two endpoints $s_0, s_1$ where $s_0 = \{v_0,p_0\}$ is on the boundary and $s_1 = \{v_1,p_1\}$ is in the bulk, such that $\xi$ interacts with the boundary only once, when crossing from $s_0$ into the bulk; it cannot cross back and forth multiple times. Let $|\psi^{h,g}\>:=F_\xi^{h,g}\vac$, and $\CT_{\xi}(s_0,s_1)$ be the space with basis $|\psi^{h,g}\>$.

(1)$|\psi^{h,g}\>$ is independent of the choice of ribbon through the bulk between fixed sites $s_0, s_1$, so long as the ribbon still only interacts with the boundary at the chosen location.

(2)$\CT_\xi(s_0,s_1)\subset\CH$ inherits actions at disjoint sites $s_0, s_1$, 
\[ x\la^b_{s_0}|\psi^{h,g}\>=|\psi^{ xhx^{-1},xg}\>,\quad \delta_r\la^b_{s_0}|\psi^{h,g}\>=\delta_{rK,hK}|\psi^{h,g}\>\]
\[ f\la_{s_1}|\psi^{h,g}\>=|\psi^{h,gf^{-1}}\>,\quad \delta_f\la_{s_1}|\psi^{h,g}\>=\delta_{f,g^{-1}h^{-1}g}|\psi^{h,g}\>\]
where we use the isomorphism $|\psi^{h,g}\>\mapsto \delta_hg$ to see the action at $s_0$ as a representation of $\Xi(R,K)$ on $D(G)$. In particular it is the restriction of the left regular representation of $D(G)$ to $\Xi(R,K)$, with inclusion map $i$ from Lemma~\ref{Xisub}. The action at $s_1$ is the right regular representation of $D(G)$, as in the bulk.
\end{lemma}
\proof
(1) is the same as the proof in \cite[Prop.3.10]{CowMa}, with the exception that if the ribbon $\xi'$ crosses the boundary multiple times it will incur an additional energy penalty from the Hamiltonian for each crossing, and thus $\CT_{\xi'}(s_0,s_1) \neq \CT_{\xi}(s_0,s_1)$ in general.

(2) This follows by the commutation rules in Lemma~\ref{boundary_ribcom} and Lemma~\ref{ribcom} respectively, using 
\[x\la^b_{s_0}\vac = \delta_e\la^b_{s_0}\vac = \vac; \quad f\la_{s_1}\vac = \delta_e\la_{s_1}\vac = \vac\]
$\forall x\in K, f \in G$. For the hardest case we have
\begin{align*}\delta_r\la^b_{s_0}F^{h,g}\vac &= F_\xi^{h,g} \circ\delta_{s\cdot(y\la r)}\la^b_{s_0}\vac\\
&= F_\xi^{h,g}\delta_{s\cdot(y\la r)K,K}\vac\\ &= F_\xi^{h,g}\delta_{rK,hK}\vac.
\end{align*}
For the restriction of the action at $s_0$ to $\Xi(R,K)$, we have that 
\[\delta_r\cdot\delta_hg = \delta_{rK,hK}\delta_hg = \sum_{a\in rK}\delta_{a,h}\delta_hg=i(\delta_r)\delta_hg.\]
and $x\cdot \delta_hg = x\delta_hg = i(x)\delta_hg$.
\endproof

In the bulk, the excitation space $\CL(s_0,s_1)$ is totally independent of the ribbon $\xi$ \cite{Kit,CowMa}, but we do not know of a similar property for $\CT_\xi(s_0,s_1)$ when interacting with the boundary without the restrictions stated.

We explained in Section~\ref{sec:xi} how representations of $D(G)$ at sites in the bulk relate to those of $\Xi(R,K)$ in the boundary by functors in both directions. Physically, if we apply ribbon trace operators, that is operators of the form $W_\xi^{\CC,\pi}$, to the vacuum, then in the bulk we create exactly a quasiparticle of type $(\CC,\pi)$ and $(\CC^*,\pi^*)$ at either end. Now let us include a boundary. 

\begin{definition}Given an irrep of $D(G)$ provided by $(\CC,\pi)$, we define the {\em boundary projection} $P_{i^*(\CC,\pi)}\in \Xi(R,K)$ by
\[ P_{i^*(\CC,\pi)}=\sum_{(\CO,\rho)\ |\ n^{(\CO,\rho)}_{(\CC,\pi)}\ne 0} P_{(\CO,\rho)}\]
i.e. we sum over the projectors of all the types of irreps of $\Xi(R,K)$ contained in the restriction of the given $D(G)$ irrep. 
\end{definition}
It is clear that $P_{i^*(\CC,\pi)}$ is a projection as a sum of orthogonal projections. 

\begin{proposition}\label{prop:boundary_traces}
Let $\xi$ be an open ribbon extending from an external site $s_0$ on a smooth boundary with associated algebra $\Xi(R,K)$ to a site $s_1$ in the bulk, for example:
\[\input{smooth_halfplane_ribbon.tikz}\]
Then
\[P_{(\CO,\rho)}\la^b_{s_0}W^{\CC,\pi}_\xi\vac = 0\quad {\rm iff} \quad n^{(\CO,\rho)}_{(\CC,\pi)} = 0.\]
In addition, we have
\[P_{i^*(\CC,\pi)}\la^b_{s_0} W^{\CC,\pi}_\xi\vac = W^{\CC,\pi}_\xi\vac = W^{\CC,\pi}_\xi\vac \ra_{s_1} P_{(\CC,\pi)},\]
where we see the left action at $s_1$ of $P_{(\CC^*,\pi^*)}$ as a right action using the antipode.
\end{proposition}
\proof
Under the isomorphism in Lemma~\ref{Ts0s1} we have that $W^{\CC,\pi}_\xi\vac \mapsto P_{(\CC,\pi)} \in D(G)$. For the first part we therefore have
\[P_{(\CO,\rho)}\la^b_{s_0}W^{\CC,\pi}_\xi\vac \mapsto  i(P_{(\CO,\rho)}) P_{(\CC,\pi)}\]
so the result follows from the last part of Lemma~\ref{lemfrobn}. Since the sum of projectors over the irreps of $\Xi$ is 1, this then implies the second part:
\[W^{\CC,\pi}_\xi\vac = \sum_{\CO,\rho}P_{(\CO,\rho)}\la^b_{s_0}W^{\CC,\pi}_\xi\vac = P_{i^*(\CC,\pi)}\la^b_{s_0}W^{\CC,\pi}_\xi\vac.\]
The action at $s_1$ is the same as for bulk ribbon operators.
\endproof

The physical interpretation is that application of a ribbon trace operator $W_\xi^{\CC,\pi}$ to a vacuum state creates a quasiparticle at $s_0$ of all the types contained in $i^*(\CC,\pi)$, while still creating one of type $(\CC^*,\pi^*)$ at $s_1$; this is called the \textit{condensation} of $({\CC,\pi})$ at the boundary. While we used a smooth boundary in this example, the proposition applies equally to rough boundaries with the specified orientation in Remark~\ref{rem:rough_ribbon} by similar arguments. 

\begin{example}\rm
In the bulk, we take the $D(S_3)$ model. Then by Example~\ref{exDS3}, we have exactly 8 irreps in the bulk. At the boundary, we take $K=\{e,u\} = \Z_2$ with $R = \{e,uv,vu\}$. As per the table in Example~\ref{exS3n} and Proposition~\ref{prop:boundary_traces} above, we then have for example that
\[(P_{\CO_0,-1}+P_{\CO_1,1})\la^b_{s_0}W_\xi^{\CC_1,-1}\vac = W_\xi^{\CC_1,-1}\vac = W_\xi^{\CC_1,-1}\vac \ra_{s_1}P_{\CC_1,-1}.\]
We can see this explicitly. Recall that
\[\Lambda_{\C(R)}\la^b_{s_0}\vac = \Lambda_{\C K}\la^b_{s_0}\vac = \vac.\]
All other vertex and face actions give 0 by orthogonality. Then,
\[P_{\CO_0,-1} = {1\over 2}\delta_e \tens (e-u); \quad P_{\CO_1, 1} = (\delta_{uv} + \delta_{vu})\tens e\]
and
\[W_\xi^{\CC_1,-1} = \sum_{c\in \{u,v,w\}}F_\xi^{c,e}-F_\xi^{c,c}\]
by Lemmas~\ref{Xiproj} and \ref{lem:quasi_basis} respectively. For convenience, we break the calculation up into two parts, one for each projector. Throughout, we will make use of Lemma~\ref{boundary_ribcom} to commute projectors through ribbon operators. First, we have that
\begin{align*}
&P_{\CO_0,-1}\la^b_{s_0}W_\xi^{\CC_1,-1}\vac = {1\over 2}(\delta_e \tens (e - u))\la^b_{s_0} \sum_{c\in \{u,v,w\}}(F_\xi^{c,e}-F_\xi^{c,c})\vac\\
&= {1\over 2}\delta_e\la^b_{s_0}[\sum_{c\in\{u,v,w\}}(F_\xi^{c,e}-F_\xi^{c,c})-(F_\xi^{u,u}-F_\xi^{e,u}+F_\xi^{v,u}-F_\xi^{v,uv}+F_\xi^{w,u}-F_\xi^{w,vu})]\vac\\
&= {1\over 2}[(F_\xi^{u,e}-F_\xi^{u,u})\delta_e\la^b_{s_0}+(F_\xi^{v,e}-F_\xi^{v,v})\delta_{vu}\la^b_{s_0}+(F_\xi^{w,e}-F_\xi^{w,w})\delta_{uv}\la^b_{s_0}\\
&+ (F^{u,e}_\xi-F^{u,u}_\xi)\delta_e\la^b_{s_0} + (F^{v,uv}_\xi-F^{v,u}_\xi)\delta_{vu}\la^b_{s_0} + (F^{w,vu}_\xi-F^{w,u}_\xi)\delta_{uv}\la^b_{s_0}]\vac\\
&= (F_\xi^{u,e}-F_\xi^{u,u})\vac
\end{align*}
where we used the fact that $u = eu, v=vuu, w=uvu$ to factorise these elements in terms of $R,K$. Second,
\begin{align*}
P_{\CO_1,1}\la^b_{s_0}W_\xi^{\CC_1,-1}\vac &= ((\delta_{uv} + \delta_{vu})\tens e)\la^b_{s_0}\sum_{c\in \{u,v,w\}}(F_\xi^{c,e}-F_\xi^{c,c})\vac\\
&= (F_\xi^{v,e}-F_\xi^{v,v}+F_\xi^{w,e}-F_\xi^{w,w})(\delta_e\tens e)\la^b_{s_0}\vac\\
&= (F_\xi^{v,e}-F_\xi^{v,v}+F_\xi^{w,e}-F_\xi^{w,w})\vac.
\end{align*}
The result follows immediately. All other boundary projections of $D(S_3)$ ribbon trace operators can be worked out in a similar way.
\end{example}

\begin{remark}\rm
Proposition~\ref{prop:boundary_traces} does not tell us exactly how \textit{all} ribbon operators in the quasiparticle basis are detected at the boundary, only the ribbon trace operators. We do not know a similar general formula for all ribbon operators.
\end{remark}

Now, consider a lattice in the plane with two boundaries, namely to the left and right,
\[\input{smooth_twobounds.tikz}\]
Recall that a lattice on an infinite plane admits a single ground state $\vac$ as explained in\cite{CowMa}. However, in the present case, we may be able to also use ribbon operators in the quasiparticle basis extending from one boundary, at $s_0$ say, to the other, at $s_1$ say, such that no quasiparticles are detected at either end. These ribbon operators do not form a closed, contractible loop, as all undetectable ones do in the bulk; the corresponding states $|\psi\>$ are ground states and the vacuum has increased degeneracy. We can similarly induce additional degeneracy of excited states. This justifies the term \textit{gapped boundaries}, as the boundaries give rise to additional states with energies that are `gapped'; that is, they have a finite energy difference $\Delta$ (which may be zero) independently of the width of the lattice.

\section{Patches}\label{sec:patches}
For any nontrivial group, $G$ there are always at least two distinct choices of boundary conditions, namely with $K=\{e\}$ and $K=G$ respectively. In these cases, we necessarily have $R=G$ and $R=\{e\}$ respectively.

Considering $K=\{e\}$ on a smooth boundary, we can calculate that $A^b_1(v) = \id$ and $B^b_1(s)g = \delta_{e,g} g$, for $g$ an element corresponding to the single edge associated with the boundary site $s$. This means that after performing the measurements at a boundary, these edges are totally constrained and not part of the large entangled state incorporating the rest of the lattice, and hence do not contribute to the model whatsoever. If we remove these edges then we are left with a rough boundary, in which all edges participate, and therefore we may consider the $K=\{e\}$ case to imply a rough boundary. A similar argument applies for $K=G$ when considered on a rough boundary, which has $A^b_2(v)g = A(v)g = {1\over |G|}\sum_k kg = {1\over |G|}\sum_k k$ for an edge with state $g$ and $B^b_2(s) = \id$. $K=G$ therefore naturally corresponds instead to a smooth boundary, as otherwise the outer edges are totally constrained by the projectors. From now on, we will accordingly use smooth to refer always to the $K=G$ condition, and rough for $K=\{e\}$.

These boundary conditions are convenient because the condensation of bulk excitations to the vacuum at a boundary can be partially worked out in the group basis. For $K=\{e\}$, it is easy to see that the ribbon operators which are undetected at the boundary (and therefore leave the system in a vaccum state) are exactly those of the form $F_\xi^{e,g}$, for all $g\in G$, as any nontrivial $h$ in $F_\xi^{h,g}$ will be detected by the boundary face projectors. This can also be worked out representation-theoretically using Proposition~\ref{nformula}.

\begin{lemma}\label{lem:rough_functor}
Let $K=\{e\}$. Then the multiplicity of an irrep $(\CC,\pi)$ of $D(G)$ with respect to the trivial representation of $\Xi(G,\{e\})$ is
\[n^{(\{e\},1)}_{(\CC,\pi)} = \delta_{\CC,\{e\}}{\rm dim}(W_\pi)\]
\end{lemma}
\proof
Applying Proposition~\ref{nformula} in the case where $V_i$ is trivial, we start with
\[n^{(\{e\},1)}_{(\CC,\pi)}={|G| \over |\CC| |G^{c_0}|}\sum_{c\in \CC\cap \{e\}} |\{e\}^c|  n_{1,\tilde\pi}\]
where $\CC\cap \{e\} = \{e\}$ iff $\CC=\{e\}$, or otherwise $\emptyset$. Also, $\tilde\pi = \oplus_{{\rm dim}(W_\pi)} (\{e\},1)$, and if $\CC = \{e\}$ then $|G^{c_0}| = |G|$.
\endproof
The factor of ${\rm dim}(W_\pi)$ in the r.h.s. implies that there are no other terms in the decomposition of $i^*(\{e\},\pi)$. In physical terms, this means that the trace ribbon operators $W^{e,\pi}_\xi$ are the only undetectable trace ribbon operators, and any ribbon operators which do not lie in the block associated to $(e,\pi)$ are detectable. In fact, in this case we have a further property which is that all ribbon operators in the chargeon sector are undetectable, as by equation~(\ref{chargeon_ribbons}) chargeon sector ribbon operators are Fourier isomorphic to those of the form $F_\xi^{e,g}$ in the group basis.
In the more general case of a rough boundary for an arbitrary choice of $\Xi(R,K)$ the orientation of the ribbon is important for using the representation-theoretic argument. When $K=\{e\}$, for $F^{e,g}_\xi$ one can check that regardless of orientation the rough boundary version of Proposition~\ref{Ts0s1} applies.

The $K=G$ case is slightly more complicated:
\begin{lemma}\label{lem:smooth_functor}
Let $K=G$. Then the multiplicity of an irrep $(\CC,\pi)$ of $D(G)$ with respect to the trivial representation of $\Xi(\{e\},G)$ is
\[n^{(\{e\},1)}_{(\CC,\pi)} = \delta_{\pi,1}\]
\end{lemma}
\proof
We start with 
\[n^{(\{e\},1)}_{(\CC,\pi)}={1 \over |\CC| |G^{c_0}|}\sum_{c\in \CC} |G^c|  n_{1,\tilde\pi}.\]
Now, $K^{r,c} = G^c$ and so $\tilde\pi = \pi$, giving $n_{1,\tilde\pi} = \delta_{1,\pi}$. Then $\sum_{c\in\CC}|G^c| = |\CC||G^{c_0}|$.
\endproof
This means that the only undetectable ribbon operators between smooth boundaries are those in the fluxion sector, i.e. those with assocated irrep $(\CC, 1)$. However, there is no factor of $|\CC|$ on the r.h.s. and so the decomposition of $i^*(\CC,1)$ will generally have additional terms other than just $(\{e\},1)$ in ${}_{\Xi(\{e\},G)}\CM$. As a consequence, a fluxion trace ribbon operator $W^{\CC,1}_\zeta$ between smooth boundaries is undetectable iff its associated conjugacy class is a singlet, say $\CC= \{c_0\}$, and thus $c_0 \in Z(G)$, the centre of $G$.

\begin{definition}\rm
A \textit{patch} is a finite rectangular lattice segment with two opposite smooth sides, each equipped with boundary conditions $K=G$, and two opposite rough sides, each equipped with boundary conditions $K=\{e\}$, for example:
\[\input{patch.tikz}\]
\end{definition}
One can alternatively define patches with additional sides, such as in \cite{Lit1}, or with other boundary conditions which depend on another subgroup $K$ and transversal $R$, but we find this definition convenient. Note that our definition does not put conditions on the size of the lattice; the above diagram is just a conveniently small and yet nontrivial example.

We would like to characterise the vacuum space $\CH_{\rm vac}$ of the patch. To do this, let us begin with $|{\rm vac}_1\>$ from equation~(\ref{eq:vac1}), and denote $|e\>_L := |{\rm vac}_1\>$. This is the \textit{logical zero state} of the patch. We will use this as a reference state to calculate other states in $\CH_{\rm vac}$.

Now, for any other state $|\psi\>$ in $\CH_{\rm vac}$, there must exist some linear map $D \in {\rm End}(\CH_{\rm vac})$ such that $D|e\>_L = |\psi\>$, and thus if we can characterise the algebra of linear maps ${\rm End}(\CH_{\rm vac})$, we automatically characterise $\CH_{\rm vac}$. To help with this, we have the following useful property:
\begin{lemma}\label{lem:rib_move}
Let $F_\xi^{e,g}$ be a ribbon operator for some $g\in G$, with $\xi$ extending from the top rough boundary to the bottom rough boundary. Then the endpoints of $\xi$ may be moved along the rough boundaries with $G=\{e\}$ boundary conditions while leaving the action invariant on any vacuum state.
\end{lemma}
\proof
We explain this on an example patch with initial state $|\psi\> \in \CH_{\rm vac}$ and a ribbon $\xi$,
\[\input{bigger_patch.tikz}\]
\[\input{bigger_patch2.tikz}\]
using the fact that $a = cb$ and $m = lk$ by the definition of $\CH_{\rm vac}$ for the second equality. Thus,  we see that the ribbon through the bulk may be deformed as usual. As the only new component of the proof concerned the endpoints, we see that this property holds regardless of the size of the patch.
\endproof

One can calculate in particular that $F_\xi^{e,g}|e\>_L = \delta_{e,g}|e\>_L$, which we will prove more generally later. The undetectable ribbon operators between the smooth boundaries are of the form
\[W^{\CC,1}_\xi = \sum_{n\in G} F_\zeta^{c_0,n}\]
when $\CC = \{c_0\}$ by Lemma~\ref{lem:smooth_functor}, hence $G^{c_0} = G$. Technically, this lemma only tells us the ribbon trace operators which are undetectable, but in the present case none of the individual component operators are undetectable, only the trace operators. There are thus exactly $|Z(G)|$ orthogonal undetectable ribbon operators between smooth boundaries. These do not play an important role, but we describe them to characterise the operator algebra on $\CH_{\rm vac}$. They obey a similar rule as Lemma~\ref{lem:rib_move}, which one can check in the same way.

In addition to the ribbon operators between sides, we also have undetectable ribbon operators between corners on the lattice. These corners connect smooth and rough boundaries, and thus careful application of specific ribbon operators can avoid detection from either face or vertex measurements,
\[\input{corner_ribbons.tikz}\]
where one can check that these do indeed leave the system in a vacuum using familiar arguments about $B(p)$ and $A(v)$. We could equally define such operators extending from either left corner to either right corner, and they obey the discrete isotopy laws as in the bulk. If we apply $F_\xi^{h,g}$ for any $g\neq e$ then we have $F_\xi^{h,g}|\psi\> =0$ for any $|\psi\>\in \CH_{\rm vac}$, and so these are the only ribbon operators of this form.

\begin{remark}\rm
Corners of boundaries are algebraically interesting themselves, and can be used for quantum computation, but for brevity we skim over them. See e.g. \cite{Bom2,Brown} for details.
\end{remark}

These corner to corner, left to right and top to bottom ribbon operators span ${\rm End}(\CH_{\rm vac})$, the linear maps which leave the system in vacuum. Due to Lemma~\ref{lem:ribs_only}, all other linear maps must decompose into ribbon operators, and these are the only ribbon operators in ${\rm End}(\CH_{\rm vac})$ up to linearity.

As a consequence, we have well-defined patch states $|h\>_L := \sum_gF^{h,g}_\xi|e\>_L$ for each $h\in G$, where $\xi$ is any ribbon extending from the bottom left corner to right. Now, working explicitly on the small patch below, we have
\[\input{wee_patch.tikz}\]
to start with, then:
\[\input{wee_patch2.tikz}\]
It is easy to see that we may always write $|h\>_L$ in this manner, for an arbitrary size of patch. Now, ribbon operators which are undetectable when $\xi$ extends from bottom to top are those of the form $F_\xi^{e,g}$, for example
\[\input{wee_patch3.tikz}\]
and so $F_\xi^{e,g}|h\>_L = \delta_{g,h}|h\>_L$, where again if we take a larger patch all additional terms will clearly cancel. Lastly, undetectable ribbon operators for a ribbon $\zeta$ extending from left to right are exactly those of the form $\sum_{n\in G} F_\zeta^{c_0,n}$ for any $c_0 \in Z(G)$. One can check that $|c_0 h\>_L = \sum_{n\in G} F_\zeta^{c_0,n} |h\>_L$, thus these give us no new states in $\CH_{\rm vac}$.

\begin{lemma}\label{lem:patch_dimension}
For a patch with the $D(G)$ model in the bulk, ${\rm dim}(\CH_{\rm vac}) = |G|$.
\end{lemma}
\proof
By the above characterisation of undetectable ribbon operators, the states $\{|h\>_L\}_{h\in G}$ span ${\rm dim}(\CH_{\rm vac})$. The result then follows from the adjointness of ribbon operators, which means that the states $\{|h\>_L\}_{h\in G}$ are orthogonal. 
\endproof

We can also work out that for $|{\rm vac}_2\>$ from equation~(\ref{eq:vac2}), we have $|{\rm vac}_2\> = \sum_h |h\>_L$. More generally:
\begin{corollary}\label{cor:matrix_basis}
$\CH_{\rm vac}$ has an alternative basis with states $|\pi;i,j\>_L$, where $\pi$ is an irreducible representation of $G$ and $i,j$ are indices such that $0\leq i,j<{\rm dim}(V_\pi)$. We call this the quasiparticle basis of the patch.
\end{corollary}
\proof
First, use the nonabelian Fourier transform on the ribbon operators $F_\xi^{e,g}$, so we have $F_\xi^{'e,\pi;i,j} = \sum_{n\in G}\pi(n^{-1})_{ji}F_\xi^{e,n}$. If we start from the reference state $|1;0,0\>_L := \sum_h |h\>_L = |{\rm vac}_2\>$ and apply these operators with $\xi$ from bottom to top of the patch then we get
\[|\pi;i,j\>_L = F_\xi^{'e,\pi;i,j}|1;0,0\>_L = \sum_{n\in G}\pi(n^{-1})_{ji} |n\>_L\]
which are orthogonal. Now, as $\sum_{\pi\in \hat{G}}\sum_{i,j=0}^{{\rm dim}(V_\pi)} = |G|$ and we know ${\rm dim}(\CH_{\rm vac}) = |G|$ by the previous Lemma~\ref{lem:patch_dimension}, $\{|\pi;i,j\>_L\}_{\pi,i,j}$ forms a basis of ${\rm dim}(\CH_{\rm vac})$.
\endproof

\begin{remark}\rm
Kitaev models are designed in general to be fault tolerant. The minimum number of component Hilbert spaces, that is copies of $\C G$ on edges, for which simultaneous errors will undetectably change the logical state and cause errors in the computation is called the `code distance' $d$ in the language of quantum codes. For the standard method of computation using nonabelian anyons \cite{Kit}, data is encoded using excited states, which are states with nontrivial quasiparticles at certain sites. The code distance can then be extremely small, and constant in the size of the lattice, as the smallest errors need only take the form of ribbon operators winding round a single quasiparticle at a site. This is no longer the case when encoding data in vacuum states on patches, as the only logical operators are specific ribbon operators extending from top to bottom, left to right or corner to corner. The code distance, and hence error resilience, of any vacuum state of the patch therefore increases linearly with the width of the patch as it is scaled, and so the square root of the number $n$ of component Hilbert spaces in the patch, that is $n\sim d^2$.
\end{remark}

\subsection{Nonabelian lattice surgery}\label{sec:surgery}
Lattice surgery was invented as a method of fault-tolerant computation with the qubit, i.e. $\C\Z_2$, surface code \cite{HFDM}. The first author generalised it to qudit models using $\C\Z_d$ in \cite{Cow2}, and gave a fresh perspective on lattice surgery as `simulating' the Hopf algebras $\C\Z_d$ and $\C(\Z_d)$ on the logical space $\CH_{\rm vac}$ of a patch. In this section, we prove that lattice surgery generalises to arbitrary finite group models, and `simulates' $\C G$ and $\C(G)$ in a similar way. Throughout, we assume that the projectors $A(v)$ and $B(p)$ may be performed deterministically for simplicity. In Appendix~\ref{app:measurements} we discuss the added complication that in practice we may only perform measurements which yield projections nondeterministically. 

\begin{remark}\rm
When proving the linear maps that nonabelian lattice surgeries yield, we will use specific examples, but the arguments clearly hold generally. For convenience, we will also tend to omit normalising scalar factors, which do not impact the calculations as the maps are $\C$-linear.
\end{remark}

Let us begin with a large rectangular patch. We now remove a line of edges from left to right by projecting each one onto $e$:
\[\input{split2.tikz}\]
We call this a \textit{rough split}, as we create two new rough boundaries. We no longer apply $A(v)$ to the vertices which have had attached edges removed. If we start with a small patch in the state $|l\>_L$ for some $l\in G$ then we can explicitly calculate the linear map.
\[\input{rough_split_project.tikz}\]
where we have separated the two patches afterwards for clarity, showing that they have two separate vacuum spaces. We then have that the last expression is 
\[\input{rough_split_project2.tikz}\]
Observe the factors of $g$ in particular. The state is therefore now $\sum_g |g^{-1}\>_L\otimes |gl\>_L$, where the l.h.s. of the tensor product is the Hilbert space corresponding to the top patch, and the r.h.s. to the bottom. A change of variables gives $\sum_g |g\>_L\otimes |g^{-1}l\>_L$, the outcome of comultiplication of $\C(G)$ on the logical state $|l\>_L$ of the original patch.

Similarly, we can measure out a line of edges from bottom to top, for example
\[\input{split1.tikz}\]
We call this a \textit{smooth split}, as we create two new smooth boundaries. Each deleted edge is projected into the state ${1\over|G|}\sum_g g$. We also cease measurement of the faces which have had edges removed, and so we end up with two adjacent but disjoint patches. Working on a small example, we start with $|e\>_L$:
\[\input{smooth_split_project.tikz}\]
where in the last step we have taken $b\mapsto jc$, $g\mapsto kh$ from the $\delta$-functions and then a change of variables $j\mapsto jc^{-1}$, $k\mapsto kh^{-1}$ in the summation. Thus, we have ended with two disjoint patches, each in state $|e\>_L$. One can see that this works for any $|h\>_L$ in exactly the same way, and so the smooth split linear map is $|h\>_L \mapsto |h\>_L\otimes|h\>_L$, the comultiplication of $\C G$.

The opposite of splits are merges, whereby we take two disjoint patches and introduce edges to bring them together to a single patch. For the rough merge below, say we start with the basis states $|k\>_L$ and $|j\>_L$ on the bottom and top. First, we introduce an additional joining edge in the state $e$.
\[\input{merge1.tikz}\]
This state $|\psi\>$ automatically satisfies $B(p)|\psi\> = |\psi\>$ everywhere. But it does not satisfy the conditions on vertices, so we apply $A(v)$ to the two vertices adjacent to the newest edge. Then we have the last expression
\[\input{rough_merge_project.tikz}\]
which by performing repeated changes of variables yields
\[\input{rough_merge_project2.tikz}\]
Thus the rough merge yields the map $|j\>_L\otimes|k\>_L\mapsto|jk\>_L$, the multiplication of $\C G$, where again the tensor factors are in order from top to bottom.

Similarly, we perform a smooth merge with the states $|j\>_L, |k\>_L$ as 
\[\input{merg2.tikz}\]
We introduce a pair of edges connecting the two patches, each in the state $\sum_m m$.
\[\input{smooth_merge_project.tikz}\]
The resultant patch automatically satisfies the conditions relating to $A(v)$, but we must apply $B(p)$ to the freshly created faces to acquire a state in $\CH_{\rm vac}$, giving
\[\input{smooth_merge_project2.tikz}\]
where the $B(p)$ applications introduced the $\delta$-functions
\[\delta_{e}(bf^{-1}m^{-1}),\quad \delta_{e}(dh^{-1}n^{-1}),\quad \delta_e(dj^{-1}b^{-1}bf^{-1}fkh^{-1}hd^{-1}) = \delta_e(j^{-1}k).\]
In summary, the linear map on logical states is evidently $|j\>_L\otimes |k\>_L \mapsto \delta_{j,k}|j\>_L$, the multiplication of $\C(G)$.

The units of $\C G$ and $\C(G)$ are given by the states $|e\>_L$ and $|1;0,0\>_L$ respectively. The counits are given by the maps $|g\>_L \mapsto 1$ and $|g\>_L\mapsto \delta_{g,e}$ respectively. The logical antipode $S_L$ is given by applying the antipode to each edge individually, i.e. inverting all group elements. For example:
\[\input{antipode_1A.tikz}\]
This state is now no longer in the original $\CH_{\rm vac}$, so to compensate we must modify the lattice. We flip all arrows in the lattice -- this is only a conceptual flip, and does not require any physical modification:
\[\input{antipode_1B.tikz}\]
This amounts to exchanging left and right regular representations, and redefining the Hamiltonian accordingly. In the resultant new vacuum space, the state is now $|g^{-1}\>_L = F_\xi^{e,g^{-1}}|e\>_L$, with $\xi$ running from the bottom left corner to bottom right as previously.

\begin{remark}\rm
This trick of redefining the vacuum space is employed in \cite{HFDM} to perform logical Hadamards, although in their case the lattice is rotated by $\pi/2$, and the edges are directionless as the model is restricted to $\C\Z_2$.
\end{remark}

Thus, we have all the ingredients of the Hopf algebras $\C G$ and $\C(G)$ on the same vector space $\CH_{\rm vac}$. For applications, one should like to know which quantum computations can be performed using these algebras (ignoring the subtlety with nondeterministic projectors). Recall that a quantum computer is called approximately universal if for any target unitary $U$ and desired accuracy $\eps\in\R$, the computer can perform a unitary $V$ such that $||V-U||\leq\eps$, i.e. the operator norm error of $V$ from $U$ is no greater than $\eps$. 

We believe that when the computer is equipped with just the states $\{|h\>_L\}_{h\in G}$ and the maps from lattice surgery above then one cannot achieve approximately universal computation \cite{Stef}, but leave the proof to a further paper. If we also have access to all matrix algebra states $|\pi;i,j\>_L$ as defined in Corollary~\ref{cor:matrix_basis}, we do not know whether the model of computation is then universal for some choice of $G$, and we do not know whether these states can be prepared efficiently. In fact, how these states are defined depends on a choice of basis for each irrep, so whether it is universal may depend not only on the choice of $G$ but also choices of basis. This is an interesting question for future work.

\section{Quasi-Hopf algebra structure of $\Xi(R,K)$}\label{sec:quasi}

We now return to our boundary algebra $\Xi$. It is known that $\Xi$ has a great deal more structure, which we give more explicitly in this section than we have seen elsewhere. This structure generalises a well-known bicrossproduct Hopf algebra construction for when a finite group $G$ factorises as $G=RK$ into two subgroups $R,K$. Then each acts on the set of the other to form a {\em matched pair of actions} $\la,\ra$ and we use $\la$ to make a cross product algebra $\C K\rcross \C(R)$ (which has the same form as our algebra $\Xi$ except that we have chosen to flip the tensor factors) and $\ra$ to make a cross product coalgebra $\C K\lcocross \C(R)$. These fit together to form a bicrossproduct Hopf algebra $\C K\cobicross \C(R)$. This construction has been used in the Lie group version to construct quantum Poincar\'e groups for quantum spacetimes\cite{Ma:book}. 

In \cite{Be} was considered the more general case where we are just given a subgroup $K\subseteq G$ and a choice of transversal $R$ with the group identity $e\in R$. As we noted, we still have unique factorisation $G=RK$ but in general $R$ need not be a group. We can still follow the same steps. First of all, unique factorisation entails that $R\cap K=\{e\}$. It also implies maps 
\[\la : K\times R \rightarrow R,\quad \ra: K\times R\rightarrow K,\quad \cdot : R\times R \rightarrow R,\quad \tau: R \times R \rightarrow K\]
defined by  
\[xr = (x\la r)(x\ra r),\quad rs = r\cdot s \tau(r,s)\]
for all $x\in R, r,s\in R$, but this time these inherit the properties 
 \begin{align} (xy) \la  r  &= x \la  (y \la  r), \quad e \la  r = r,\nonumber\\ \label{lax}
 x \la  (r\cdot s)&=(x \la  r)\cdot((x\ra r)\la  s),\quad  x \la  e = e,\end{align}
 \begin{align}
(x\ra r)\ra s &= \tau\left(x\la  r, (x\ra r)\la  s)^{-1}(x\ra (r\cdot s)\right)\tau(r,s),\quad 
x \ra e = x,\nonumber\\ \label{rax}
(xy) \ra r &= (x\ra (y\la  r))(y\ra r),\quad e\ra r = e,\end{align}
\begin{align}
\tau(r, s\cdot t)\tau(s,t)& = \tau\left(r\cdot s,\tau(r,s)\la  t\right)(\tau(r,s)\ra t),\quad \tau(e,r) = \tau(r,e) = e,\nonumber\\ \label{tax}
r\cdot(s\cdot t) &= (r\cdot s)\cdot(\tau(r,s)\la  t),\quad r\cdot e=e\cdot r=r\end{align}
for all $x,y\in K$ and $r,s,t\in R$. We see from (\ref{lax}) that $\la$ is indeed an action (we have been using it in preceding sections) but $\ra$ in (\ref{rax}) is only only up to $\tau$ (termed in \cite{KM2} a `quasiaction'). Both $\la,\ra$ `act' almost by automorphisms but with a back-reaction by the other just as for a matched pair of groups. Meanwhile, we see from (\ref{tax}) that $\cdot$ is associative only up to $\tau$ and $\tau$ itself obeys a kind of cocycle condition. 

Clearly, $R$ is a subgroup via $\cdot$ if and only if $\tau(r,s)=e$ for all $r,s$, and in this case we already see that $\Xi(R,K)$ is a bicrossproduct Hopf algebra, with the only difference being that we prefer to build it on the flipped tensor factors. More generally, \cite{Be} showed that there is still a natural monoidal category associated to this data but with nontrivial associators. This corresponds by Tannaka-Krein reconstruction to a $\Xi$ as quasi-bialgebra which in some cases is a quasi-Hopf algebra\cite{Nat}. Here we will give these latter structures explicitly and in maximum generality compared to the literature (but still needing a restriction on $R$ for the antipode to be in a regular form). We will also show that the obvious $*$-algebra structure makes a $*$-quasi-Hopf algebra in an appropriate sense under restrictions on $R$. These aspects are  new, but more importantly, we give direct proofs at an algebraic level rather than categorical arguments, which we believe are essential for detailed calculations. Related works on similar algebras and coset decompositions include \cite{PS,KM1} in addition to \cite{Be,Nat,KM2}.

\begin{lemma}\cite{Be,Nat,KM2}
$(R,\cdot)$ has the same unique identity $e$ as $G$ and has the left division property, i.e. for all $t, s\in R$, there is a unique solution $r\in R$ to the equation $s\cdot r = t$ (one writes $r = s\backslash t$). In particular, we let $r^R$ denote the unique solution to $r\cdot r^R=e$, which we call a right inverse.\end{lemma}

This means that $(R,\cdot)$ is a left loop (a left quasigroup with identity). The multiplication table for $(R,\cdot)$ has one of each element of $R$ in each row, which is the left division property. In particular, there is one instance of $e$ in each row. One can recover $G$ knowing $(R,\cdot)$, $K$ and the data $\la,\ra,\tau$\cite[Prop.3.4]{KM2}. Note that a parallel property of left inverse $(\ )^L$ need not be present. 

\begin{definition} We say that $R$ is {\em regular} if $(\ )^R$ is bijective.
\end{definition}
 $R$ is regular iff it has both left and right inverses, and this is iff it satisfies $RK=KR$ by\cite[Prop.~3.5]{KM2}. If there is also right division then we have a loop (a quasigroup with identity) and under further conditions\cite[Prop.~3.6]{KM2}  we have $r^L=r^R$ and a 2-sided inverse property quasigroup. The case of regular $R$ is studied in \cite{Nat} but this excludes some interesting choices of $R$ and we do not always assume it. Throughout, we will specify when $R$ is required to be regular for results to hold. Finally, if $R$ obeys a further condition $x\la(s\cdot t)=(x\la s)\la t$ in \cite{KM2} then $\Xi$ is a Hopf quasigroup in the sense introduced in \cite{KM1}. This is even more restrictive but will apply to our octonions-related example. Here we just give the choices for our go-to cases for $S_3$.

\begin{example}\label{exS3R}\rm  $G=S_3$ with $K=\{e,u\}$ has four choices of transversal $R$ meeting our requirement that $e\in R$. Namely 
\begin{enumerate}
\item $R=\{e,uv,vu\}$ (our standard choice) {\em is a subgroup} $R=\Z_3$, so it is associative and there is 2-sided division and a 2-sided inverse. We also have $u\la(uv)=vu, u\la (vu)=uv$ but $\ra,\tau$ trivial.  
\item $R=\{e,w,v\}$ which is {\em not a subgroup} and indeed $\tau(v,w)=\tau(w,v)=u$ (and all others are necessarily $e$). There is an action $u\la v=w, u\la w=v$ but $\ra$ is still trivial. For examples
\begin{align*} vw&=wu \Rightarrow  v\cdot w=w,\ \tau(v,w)=u;\quad  wv=vu \Rightarrow  w\cdot v=v,\  \tau(w,v)=u\\ 
uv&=wu \Rightarrow  u\la v=w,\  u\ra v=u;\quad uw=vu \Rightarrow  u\la w=v,\  u\ra w=u.  \end{align*}
This has left division/right inverses as it must but {\em not right division} as $e\cdot w=v\cdot w=w$ and $e\cdot v=w\cdot v=v$. We also have $v\cdot v=w\cdot w=e$ and $(\ )^R$ is bijective so this {\em is regular}. 

\item $R=\{e,uv, v\}$ which is {\em not a subgroup} and $\tau,\la,\ra$ are all nontrivial with
\begin{align*} \tau(uv,uv)&=\tau(v,uv)=\tau(uv,v)=u,\quad \tau(v,v)=e,\\
 v\cdot v&=e,\quad v\cdot uv=uv,\quad  uv\cdot v=e,\quad uv\cdot uv=v,\\
u\la v&=uv,\quad u\la (uv)=v,\quad u\ra v=e,\quad u\ra uv=e\end{align*}
and all other cases determined from the properties of $e$. Here $v^R=v$ and $(uv)^R=v$ so this is {\em not regular}. 

\item $R=\{e,w,vu\}$ which is analogous to the preceding case, so {\em not a subgroup}, $\tau,\la,\ra$  all nontrivial and {\em not regular}.
\end{enumerate}
\end{example}

We will also need the following useful lemma in some of our proofs. 
\begin{lemma}\label{leminv}\cite{KM2} For any transversal $R$ with $e\in R$, we have
\begin{enumerate}
\item $(x\ra r)^{-1}=x^{-1}\ra(x\la r)$.
\item $(x\la r)^R=(x\ra r)\la r^R$.
\item $\tau(r,r^R)^{-1}\ra r=\tau(r^R,r^{RR})^{-1}$. 
\item $\tau(r,r{}^R)^{-1}\ra r=r^R{}^R$.
\end{enumerate}
for all $x\in K, r\in R$. 
\end{lemma}
\proof  The first two items are elementary from the matched pair axioms. For (1), we use  $e=(x^{-1}x)\ra r=(x^{-1}\ra(x\la r))(x\ra r)$ and for (2) $e=x\la(r\cdot r^R)=(x\la r)\cdot((x\ra r)\la r^R)$. The other two items are a left-right reversal of \cite[Lem.~3.2]{KM2} but given here for completeness. For (3), \begin{align*} e&=(\tau(r,r^R)\tau(r,r^R)^{-1})\ra r=(\tau(r,r^R)\ra (\tau(r,r^R)\la r))(\tau(r,r^R)^{-1}\ra r)\\
&=(\tau(r,r^R)\ra r^{RR})(\tau(r,r^R)^{-1}\ra r)\end{align*}
which we combine with 
\[ \tau(r^R,r^{RR})=\tau(r\cdot r^R,r^{RR})\tau(r^R,r^{RR})=\tau(r\cdot r^R, \tau(r,r^R)\la r^{RR})(\tau(r,r^R)\ra r^{RR})=\tau(r,r^R)\ra r^{RR}\]
by the cocycle property. For (4),  $\tau(r,r^R)\ra r^R{}^R=(r\cdot r^R) \tau(r,r^R)\ra r^R{}^R=r\cdot (r^R\cdot r^R{}^R)=r$
by one of the matched pair conditions.  \endproof

Using this lemma, it is not hard to prove cf\cite[Prop.3.3]{KM2} that 
\begin{equation}\label{leftdiv}s\backslash t=s^R\cdot\tau^{-1}(s,s^R)\la t;\quad s\cdot(s\backslash t)=s\backslash(s\cdot t)=t,\end{equation}
which can also be useful in calculations.  

\subsection{$\Xi(R,K)$ as a quasi-bialgebra} 

We recall that a quasi-bialgebra is a unital algebra $H$, a coproduct $\Delta:H\to H\tens H$ which is an algebra map but is no longer required to be coassociative, and $\eps:H\to \C$ a counit for $\Delta$ in the usual sense  $(\id\tens\eps)\Delta=(\eps\tens\id)\Delta=\id$. Instead, we have a weaker form of coassociativity\cite{Dri,Ma:book}
\[ (\id\tens\Delta)\Delta=\phi((\Delta\tens\id)\Delta(\ ))\phi^{-1}\]
for an invertible element $\phi\in H^{\tens 3}$ obeying the 3-cocycle identity
\[ (1\tens\phi)((\id\tens\Delta\tens\id)\phi)(\phi\tens 1)=((\id\tens\id\tens\Delta)\phi)(\Delta\tens\id\tens\id)\phi,\quad  (\id\tens\eps\tens\id)\phi=1\tens 1\]
(it follows that $\eps$ in the other positions also gives $1\tens 1$). 
In our case, we already know that $\Xi(R,K)$ is a unital algebra.


\begin{lemma}\label{Xibialg} $\Xi(R,K)$ is a quasi-bialgebra with 
\[ \Delta x=\sum_{s\in R}x\delta_s \tens x\ra s, \quad \Delta \delta_r = \sum_{s,t\in R} \delta_{s\cdot t,r}\delta_{s}\otimes \delta_{t},\quad \eps x=1,\quad \eps \delta_r=\delta_{r,e}\]
for all $x\in K, r\in R$, and
\[ \phi=\sum_{r,s\in R} \delta_r \otimes  \delta_s  \otimes   \tau(r,s)^{-1},\quad \phi^{-1} = \sum_{r,s\in R} \delta_r\otimes \delta_s  \otimes    \tau(r,s).\]
\end{lemma}
\proof 
This follows by reconstruction arguments, but it is useful to check directly,
\begin{align*}
(\Delta x)(\Delta y)&=\sum_{s,r}(x\delta_s\tens x\ra s)(y\delta_r\tens y\ra r)=\sum_{s,r}(x\delta_sy\delta_r\tens ( x\ra s)( y\ra r)\\
&=\sum_{r,s}xy\delta_{y^{-1}\la s}\delta_r\tens (x\ra s)(y\ra r)=\sum_r xy \delta_r\tens (x\ra(y\la r))(y\ra r)=\Delta(xy)
\end{align*}
as $s=y\la r$ and using the formula for $(xy)\ra r$ at the end. Also,
\begin{align*}
\Delta(\delta_{x\la s}x)&=(\Delta\delta_{x\la s})(\Delta x)=\sum_{r, p.t=x\la s}\delta _p x\delta_r\tens \delta_t x\ra r\\
&=\sum_{r, p.t=x\la s}x\delta_{x^{-1}\la p}\delta_r\tens x\ra r\delta_{(x\ra r)^{-1}\la t}=\sum_{(x\la r).t=x\la s}x \delta_r\tens x\ra r\delta_{(x\ra r)^{-1}\la t}\\
&=\sum_{(x\la r).((x\ra r)\la t')=x\la s}x \delta_r\tens x\ra r\delta_{t'}=\sum_{r\cdot t'=s}x\delta_r\tens (x\ra r)\delta_{t'}=(\Delta x)(\Delta \delta _s)=\Delta(x\delta_s)
\end{align*}
using the formula for $x\la(r\cdot t')$. This says that the coproducts stated are compatible with the algebra cross relations. Similarly, one can check that 
\begin{align*}
(\sum_{p,r}\delta_p\tens\delta_r\tens &\tau(p,r))((\id\tens\Delta )\Delta x)=\sum_{p,r,s,t}(\delta_p\tens\delta_r\tens \tau(p,r))(x\delta_s\tens (x\ra s)\delta_t\tens (x\ra s)\ra t)\\
&=\sum_{p,r,s,t}\delta_px\delta_s\tens\delta_r(x\ra s)\delta_t\tens \tau(p,r)((x\ra s)\ra t)\\
&=\sum_{s,t}x\delta_s\tens (x\ra s)\delta_t\tens\tau(x\la s,(x\ra s)\la t)(x\ra s)\ra t)\\
&=\sum_{s,t}x\delta_s\tens (x\ra s)\delta_t\tens(x\ra(s.t))\tau(s,t)\\
&=\sum_{p,r,s,t}(x\delta_s\tens (x\ra s)\delta_t\tens(x\ra(s.t))(\delta_p\tens\delta_r\tens\tau(p,r)\\
&=( (\Delta\tens\id)\Delta x   )  (\sum_{p,r}\delta_p\tens\delta_r\tens\tau(p,r))
\end{align*}
as $p=x\la s$ and $r=(x\ra s)\la t$ and using the formula for $(x\ra s)\ra t$. For the remaining cocycle relations, we have
\begin{align*}
(\id\tens\eps\tens\id)\phi = \sum_{r,s}\delta_{s,e}\delta_r\tens\tau(r,s)^{-1} = \sum_r\delta_r\tens 1 = 1\tens 1
\end{align*}
and 
\[ (1\tens\phi)((\id\tens\Delta\tens\id)\phi)(\phi\tens 1)=\sum_{r,s,t}\delta_r\tens\delta_s\tens \delta_t\tau(r,s)^{-1}\tens\tau(s,t)^{-1}\tau(r,s\cdot t)\]
after multiplying out $\delta$-functions and renaming variables.  Using the value of $\Delta\tau(r,s)^{-1}$ and similarly multiplying out, we obtain on the other side
\begin{align*} ((\id\tens&\id\tens\Delta)\phi)(\Delta\tens\id\tens\id)\phi=\sum_{r,s,t}\delta_r\tens\delta_s\tens\tau(r,s)^{-1}\delta_t\tens(\tau(r,s)^{-1}\ra t)\tau(r\cdot s,t)^{-1}\\
&=\sum_{r,s,t'}\delta_r\tens\delta_s\tens\delta_{t'}\tau(r,s)^{-1}\tens(\tau(r,s)^{-1}\ra (\tau(r,s)\la t'))\tau(r\cdot s,\tau(r,s)\la t')^{-1}\\
&=\sum_{r,s,t'}\delta_r\tens\delta_s\tens\delta_{t'}\tau(r,s)^{-1}\tens(\tau(r,s)\ra t')^{-1}\tau(r\cdot s,\tau(r,s)\la t')^{-1},
\end{align*}
where we change summation to $t'=\tau(r,s)\la t$ then use Lemma~\ref{leminv}. Renaming $t'$ to $t$, the two sides are equal  in view of the cocycle identity for $\tau$. Thus, we have a quasi-bialgebra with $\phi$ as stated.
\endproof

We can also write the coproduct (and the other structures) more explicitly.

\begin{remark}\rm (1) If we want to write the coproduct on $\Xi$ explicitly as a vector space, the above becomes 
\[ \Delta(\delta_r\tens x)=\sum_{s\cdot t=r}\delta_s\tens x\tens\delta_t\tens (x^{-1}\la s)^{-1},\quad \eps(\delta_r\tens x)=\delta_{r,e}\]
which is ugly due to our decision to build it on $\C(R)\tens\C K$. (2) If we built it on the other order then we could have $\Xi=\C K\rcross \C(R)$ as an algebra, where we have a right action
\[ (f\la x)(r)= f(x\ra r);\quad \delta_r\la x=\delta_{x^{-1}\ra r}\]
on $f\in \C(R)$. Now make a right handed cross product
\[ (x\tens \delta_r)(y\tens \delta_s)= xy\tens (\delta_r\la y)\delta_s=xy\tens\delta_s\delta_{r,y\ra s}\]
which has cross relations $\delta_r y=y\delta_{y^{-1}\ra r}$. These are the same relations as before. So this is the same algebra, just we prioritise a basis $\{x\delta_r\}$ instead of the other way around. This time, we have
\[ \Delta (x\tens\delta_r)=\sum_{s\cdot t=r} x\tens\delta_s\tens x\la s\tens\delta_t.\]
We do not do this in order to be compatible with the most common form of $D(G)$ as $\C(G)\lcross \C G$ as in \cite{CowMa}. 
\end{remark}

\subsection{$\Xi(R,K)$ as a quasi-Hopf algebra}

A quasi-bialgebra is a quasi-Hopf algebra if there are elements $\alpha,\beta\in H$ and an antialgebra map $S:H\to H$ such that\cite{Dri,Ma:book}
\[(S \xi_1)\alpha\xi_2=\eps(\xi)\alpha,\quad \xi_1\beta S\xi_2=\eps(\xi)\beta,\quad \phi^1\beta(S\phi^2)\alpha\phi^3=1,\quad (S\phi^{-1})\alpha\phi^{-2}\beta S\phi^{-3}=1\] 
where $\Delta\xi=\xi_1\tens\xi_2$, $\phi=\phi^1\tens\phi^2\tens\phi^3$ with inverse $\phi^{-1}\tens\phi^{-2}\tens\phi^{-3}$ is a compact notation (sums of such terms to be understood). It is usual to assume $S$ is bijective but we do not require this. The $\alpha,\beta, S$ are not unique and can be changed to $S'=U(S\ ) U^{-1}, \alpha'=U\alpha, \beta'=\beta U^{-1}$ for any invertible $U$. In particular, if $\alpha$ is invertible then  we can transform to a standard form replacing it by $1$. For the purposes of this paper, we therefore call the case of $\alpha$ invertible a (left) {\em regular antipode}.

\begin{proposition}\label{standardS} If $(\ )^R$ is bijective, $\Xi(R,K)$ is a quasi-Hopf algebra with regular antipode
\[ S(\delta_r\tens x)=\delta_{(x^{-1}\la r)^R}\tens x^{-1}\ra r,\quad \alpha=\sum_{r\in R}\delta_r\tens 1,\quad \beta=\sum_r\delta_r\tens \tau(r,r^R).\]
Equivalently in subalgebra terms, 
\[ S\delta_r=\delta_{r^R},\quad Sx=\sum_{s\in R}(x^{-1}\la s)\delta_{s^R} ,\quad \alpha=1,\quad \beta=\sum_{r\in R}\delta_r\tau(r,r^R).\]
\end{proposition}
\proof
 For the axioms involving $\phi$, we have
\begin{align*}\phi^1\beta&(S \phi^2)\alpha\phi^3=\sum_{s,t,r}(\delta_s\tens 1)(\delta_r\tens \tau(r,r^R))(\delta_{t^R}\tens\tau(s,t)^{-1})\\
&=\sum_{s,t}(\delta_s\tens\tau(s,s^R))(\delta_{t^R}\tens \tau(s,t)^{-1})=\sum_{s,t}\delta_s\delta_{s,\tau(s,s^R)\la t^R}\tens\tau(s,s^R)\tau(s,t)^{-1}\\
&=\sum_{s^R.t^R=e}\delta_s\tens \tau(s,s^R)\tau(s,t)^{-1}=1,
\end{align*}
where we used $s.(s^R.t^R)=(s.s^R).\tau(s,s^R)\la t^R=\tau(s,s^R)\la t^R$. So  $s=\tau(s,s^R)\la t^R$ holds iff  $s^R.t^R=e$ by left cancellation. In the sum, we can take $t=s^R$ which contributes $\delta_s\tens e$. Here $s^R.t^R=s^R.(s^R)^R=e$; there is a unique element $t^R$ which does this and hence a unique $t$ provided $(\ )^R$ is injective, and hence a bijection. 
\begin{align*}
S(\phi^{-1})\alpha&\phi^{-2}\beta S(\phi^{-3}) = \sum_{s,t,u,v}(\delta_{s^R}\otimes 1)(\delta_t\otimes 1)(\delta_u\otimes\tau(u,u^R))(\delta_{(\tau(s,t)^{-1}\la v)^R}\otimes (\tau(s,t)^{-1}\ra v))\\
&= \sum_{s,v}(\delta_{s^R}\otimes\tau(s^R,s^R{}^R))(\delta_{(\tau(s,s^R)^{-1}\la v)^R}\otimes \tau(s,s^R)^{-1}\ra v).
\end{align*}
Upon multiplication, we will have a $\delta$-function dictating that
\[s^R = \tau(s^R,s^R{}^R)\la (\tau(s,s^R)^{-1}\la v)^R,\]
so we can use the fact that
\begin{align*}s\cdot s^R = e &= s\cdot(\tau(s^R,s^R{}^R)\la (\tau(s,s^R)^{-1}\la v)^R)\\ &= s\cdot(s^R\cdot(s^R{}^R\cdot (\tau(s,s^R)^{-1}\la v)^R))\\
&= \tau(s,s^R)\la (s^R{}^R\cdot(\tau(s,s^R)\la v)^R),
\end{align*}
where we use similar identities to before. Therefore $s^R{}^R\cdot (\tau(s,s^R)^{-1}\la v)^R = e$, so $(\tau(s,s^R)^{-1}\la v)^R = s^R{}^R{}^R$. When $(\ )^R$ is injective, this gives us $v = \tau(s,s^R)\la s^R{}^R$. Returning to our original calculation we have that our previous expression is 
\begin{align*}
\cdots &= \sum_s \delta_{s^R}\otimes \tau(s^R,s^R{}^R)(\tau(s,s^R)^{-1}\ra (\tau(s,s^R)\la s^R{}^R))\\
&= \sum_s \delta_{s^R}\otimes \tau(s^R,s^R{}^R)(\tau(s,s^R)\ra s^R{}^R)^{-1} = \sum_s \delta_{s^R}\otimes 1 = 1
\end{align*}

We now prove the antipode axiom involving $\alpha$,
\begin{align*}
(S(\delta_s \otimes& x)_1)(\delta_s \otimes x)_2 = \sum_{r\cdot t = s}(\delta_{(x^{-1}\la r)^R}\otimes (x^{-1}\ra r))(\delta_t\otimes (x^{-1}\ra r)^{-1})\\
&= \sum_{r\cdot t = s}\delta_{(x^{-1}\la r)^R, (x^{-1}\ra r)\la t}\delta_{(x^{-1}\la r)^R}\otimes 1 = \delta_{e,s}\sum_r \delta_{(x^{-1}\la r)^R}\otimes 1 = \eps(\delta_s\otimes x)1.
\end{align*}
The condition from the $\delta$-functions is 
\[ (x^{-1}\la r)^R=(x^{-1}\ra r)\la t\]
which by uniqueness of right inverses holds iff
\[ e=(x^{-1}\la r)\cdot (x^{-1}\ra r)\la t=x^{-1}\la(r\cdot t)\]
which is iff  $r.t=e$, so $t=r^R$. As we also need $r.t=s$, this becomes $\delta_{s,e}$ as required. 

We now prove the axiom involving $\beta$, starting with
\begin{align*}(\delta_s\otimes& x)_1 \beta S((\delta_s\otimes x)_2) = \sum_{r\cdot t=s, p}(\delta_r\tens x)(\delta_p\tens\tau(p,p^R))S(\delta_t\tens (x^{-1}\ra r)^{-1})\\
&=\sum_{r\cdot t=s, p}(\delta_r\delta_{r,x\la p}\tens x\tau(p,p^R))(\delta_{((x^{-1}\ra r)\la t)^R}\tens (x^{-1}\ra r)\ra t)\\
&=\sum_{r\cdot t=s}(\delta_r\tens x\tau(x^{-1}\la r,(x^{-1}\la r)^R))(\delta_{((x^{-1}\ra r)\la t)^R}\tens (x^{-1}\ra r)\ra t).
\end{align*}
When we multiply this out, we will need from the product of $\delta$-functions that
\[ \tau(x^{-1}\la r,(x^{-1}\la r)^R)^{-1}\la (x^{-1}\la r)=((x^{-1}\ra r)\la t)^R,\]
but note that $\tau(q,q{}^R)^{-1}\la q=q^R{}^R$  from Lemma~\ref{leminv}. So the condition from the $\delta$-functions is 
\[ (x^{-1}\la r)^R{}^R=((x^{-1}\ra r)\la t)^R,\]
so
\[ (x^{-1}\la r)^R=(x^{-1}\ra r)\la t\]
when $(\ )^R$ is injective. By uniqueness of right inverses, this holds iff
\[ e=(x^{-1}\la r)\cdot ((x^{-1}\ra r)\la t)=x^{-1}\la(r\cdot t),\]
where the last equality is from the matched pair conditions. This holds iff  $r\cdot t=e$, that is, $t=r^R$. This also means in the sum that we need $s=e$.
Hence, when we multiply out our expression so far, we have 
\[\cdots=\delta_{s,e}\sum_r\delta_r\tens x\tau(x^{-1}\la r,(x^{-1}\la r)^R)(x^{-1}\ra r)\ra r^R=\delta_{s,e}\sum_r\delta_r\tens\tau(r,r^R)=\delta_{s,e}\beta,\]
as required, where we used
\[ x\tau( x^{-1}\la r,(x^{-1}\la r)^R)(x^{-1}\ra r)\ra r^R=\tau(r,r^R)\]
by the matched pair conditions. The subalgebra form of $Sx$ is the same using the commutation relations and Lemma~\ref{leminv} to reorder. 

It remains to check that 
\begin{align*}S(\delta_s&\tens y)S(\delta_r\tens x)=(\delta_{(y^{-1}\la s)^R}\tens y^{-1}\ra s)(\delta_{(x^{-1}\la x)^R}\tens x^{-1}\ra r)\\
&=\delta_{r,x\la s}\delta_{(y^{-1}\la s)^R}\tens (y^{-1}\ra s)(x^{-1}\ra r)=\delta_{r,x\la s}\delta_{(y^{-1}x^{-1}\la r)^R}\tens( y^{-1}\ra(x^{-1}\la r))(x^{-1}\ra r)\\
&=S(\delta_r\delta_{r,x\la s}\tens xy)=S((\delta_r\tens x)(\delta_s\tens y)),
\end{align*}
where the product of $\delta$-functions requires $(y^{-1}\la s)^R=( y^{-1}\ra s)\la (x^{-1}\la r)^R$, which is equivalent to $s^R=(x^{-1}\la r)^R$ using Lemma~\ref{leminv}. This imposes $\delta_{r,x\la s}$. We then replace $s=x^{-1}\la r$ and recognise the answer using the matched pair identities. 
\endproof

\subsection{$\Xi(R,K)$ as a $*$-quasi-Hopf algebra}

The correct notion of a $*$-quasi-Hopf algebra $H$ is not part of Drinfeld's theory but a natural notion is to have further structure so
as to make the monoidal category of modules a bar category in the sense of \cite{BegMa:bar}.  If $H$ is at least a quasi-bialgebra, the additional structure we need, fixing a typo in \cite[Def.~3.16]{BegMa:bar}, is the first three of:

\begin{enumerate}\item An antilinear algebra map $\theta:H\to H$.

\item An invertible element $\gamma\in H$ such that $\theta(\gamma)=\gamma$ and $\theta^2=\gamma(\ )\gamma^{-1}$.

\item An invertible element $\CG\in H\tens H$ such that
\begin{equation}\label{*GDelta}\Delta\theta =\CG^{-1}(\theta\tens\theta)(\Delta^{op}(\ ))\CG,\quad (\eps\tens\id)(\CG)=(\id\tens\eps)(\CG)=1,\end{equation}
\begin{equation}\label{*Gphi} (\theta\tens\theta\tens\theta)(\phi_{321})(1\tens\CG)((\id\tens\Delta)\CG)\phi=(\CG\tens 1)((\Delta\tens\id)\CG).\end{equation}
\item We say the $*$-quasi bialgebra is strong if
\begin{equation}\label{*Gstrong} (\gamma\tens\gamma)\Delta\gamma^{-1}=((\theta\tens\theta)(\CG_{21}))\CG.\end{equation}
\end{enumerate}

Note that if we have a quasi-Hopf algebra then $S$ is antimultiplicative and $\theta=* S$ defines an antimultiplicative antilinear map $*$. However, $S$ is not unique and it appears that specifying $\theta$ directly is more canonical.  

\begin{lemma} Let $(\ )^R$ be bijective. Then $\Xi$ has an antilinear algebra automorphism $\theta$ such that 
\[  \theta(x)=\sum_s x\ra s\, \delta_{s^R},\quad \theta(\delta_s)=\delta_{s^R},\]
\[\theta^2=\gamma(\  )\gamma^{-1};\quad \gamma=\sum_s\tau(s,s^R)^{-1}\delta_s,\quad\theta(\gamma)=\gamma.\]
\end{lemma}
\proof We compute,
\[ \theta(\delta_s\delta_t)=\delta_{s,t}\delta_{s^R}=\delta_{s^R,t^R}\delta_{s^R}=\theta(\delta_s)\theta(\delta_t)\]
\[\theta(x)\theta(y)=\sum_{s,t}x\ra s\delta_{s^R} y\ra t\delta_{t^R}=\sum_{t}(x\ra (y\la t)) y\ra t\delta_t=\sum_t (xy)\ra t\delta_{t^R}=\theta(xy)\]
where imagining commuting $\delta_t$ to the left fixes $s=(y\ra t)\la t^R=(y\la t)^R$  to obtain the 2nd equality. We also have
\[ \theta(x\delta_s)=\sum_tx\ra t\delta_{t^R}\delta_{s^R}=x\ra s\delta_{s^R}=\delta_{(x\ra s)\la s^R}x\ra s=\delta_{(x\la s)^R}x\ra s\]
\[ \theta(\delta_{x\la s}x)=\sum_t\delta_{(x\la s)^R}x\ra t\delta_{t^R}=\sum_t\delta_{(x\la s)^R}\delta_{(x\ra t)\la t^R}=\sum_t\delta_{(x\la s)^R}\delta_{(x\la t)^R}
x\ra t,\]
which is the same as it needs $t=s$. Next 
\[ \gamma^{-1}=\sum_s \tau(s,s^R)\delta_{s^{RR}}=\sum_s \delta_s \tau(s,s^R),\]
where we recall from previous calculations that $\tau(s,s^R)\la s^{RR}=s$. Then
\begin{align*}\theta^2(x)&=\sum_s\theta(x\ra s\delta_{s^R})=\sum_{s,t}(x\ra s)\ra t\delta_{t^R}\delta_{s^R}=\sum_s(x\ra s)\ra s\delta_{s^R}=\sum_s (x\ra s)\ra x^R\delta_{s^{RR}}\\
&=\sum_s \tau(x\la s,(x\la s)^R)^{-1}x\tau(s,s^R)\delta_{s^{RR}}=\sum_{s,t}\tau(t,t^R)^{-1}\delta_{t} x\tau(s,s^R)\delta_{s^{RR}}\\
&=\sum_{s,t}\delta_{t^{RR}}\tau(t,t^R)^{-1}x\tau(s,s^R)\delta_{s^{RR}}=\gamma x\gamma^{-1}&\end{align*}
where for the 6th equality if we were to commute $\delta_{s^{RR}}$ to the left, this would fix $t=x\tau(s,s^R)\la s^{RR}=x\la s$. We then use $\tau(t,t^R)^{-1}\la t=t^{RR}$ and recognise the answer. We also check that
\begin{align*}\gamma\delta_s\gamma^{-1}&= \tau(s,s^R)^{-1}\delta_s\tau(s,s^R)=\delta_{s^{RR}}=\theta^2(\delta_s),\\
 \theta(\gamma)  &= \sum_{s,t}\tau(s,s^R)^{-1}\ra t\delta_{t^R}\delta_{s^R}=\sum_s\tau(s,s^R)^{-1}\ra s\delta_{s^R}=\sum_s\tau(s^R,s^{RR})^{-1}\delta_{s^R}=\gamma\end{align*}
using Lemma~\ref{leminv}. 
\endproof

Next, we find $\CG$ obeying the conditions above.

\begin{lemma} If $(\ )^R$ is bijective then equation (\ref{*GDelta}) holds with  
\[ \CG=\sum_{s,t}  \delta_{t^R}\tau(s,t)^{-1}\tens \delta_{s^R}\tau(t,t^R)  (\tau(s,t)\ra t^R)^{-1}, \]
\[\CG^{-1}=\sum_{s,t}  \tau(s,t)\delta_{t^R}\tens  (\tau(s,t)\ra t^R)\tau(t,t^R)^{-1} \delta_{s^R}.\]
\end{lemma}
\proof The proof that $\CG,\CG^{-1}$ are indeed inverse is straightforward on matching the $\delta$-functions to fix the summation variables in $\CG^{-1}$ in terms of $\CG$. This then comes down to proving that the map
$(s,t)\to (p,q):=(\tau(s,t)\la t^R, \tau'(s,t)\la s^R)$ is injective. Indeed, the map $(p,q)\mapsto (p,p\cdot q)$ is injective by left division, so it's enough to prove that 
\[ (s,t)\mapsto (p,p\cdot q)=(\tau(s,t)\la t^R, \tau(s,t)\la(t^R\cdot\tau(t,t^R)^{-1}\la s^R))=((s\cdot t)\backslash s,(s\cdot t)^R)\]
is injective. We used $(s\cdot t)\cdot \tau(s,t)\la t^R=s\cdot(t\cdot t^R)=s$ by quasi-associativity to recognise $p$, recognised $t^R\cdot\tau(t,t^R)^{-1}\la s^R=t\backslash s^R$ from (\ref{leftdiv}) and then
\[ (s\cdot t)\cdot \tau(s,t)\la (t\backslash s^R)=s\cdot(t\cdot(t\backslash s^R))=s\cdot s^R=e\]
to recognise $p\cdot q$. That the desired map is injective is then immediate by  $(\ )^R$ injective and elementary properties of division.

We use similar methods in the other proofs. Thus, writing
\[ \tau'(s,t):=(\tau(s,t)\ra t^R)\tau(t,t^R)^{-1}=\tau(s\cdot t, \tau(s,t)\la t^R)^{-1}\]
for brevity, we have
\begin{align*}\CG^{-1}(\theta\tens\theta)(\Delta^{op} \delta_r)&=\CG^{-1}\sum_{p\cdot q=r}(\delta_{q^R}\tens\delta_{p^R})=\sum_{s\cdot t=r}\tau(s,t)\delta_{t^R}\tens\tau'(s,t)\delta_{s^R},\\
 (\Delta\theta(\delta_r))\CG^{-1}&=\sum_{p\cdot q=r^R}(\delta_p\tens\delta_q)\CG^{-1}=\sum_{p\cdot q=r^R} \tau(s,t)\delta_{t^R}\tens\tau'(s,t)\delta_{s^R},
\end{align*}
where in the second line, commuting the $\delta_{t^R}$ and $\delta_{s^R}$ to the left sets $p=\tau(s,t)\la t^R$, $q=\tau'(s,t)\la s^R$ as studied above. Hence $p\cdot q=r^R$ in the sum is the same as $s\cdot t=r$, so the two sides are equal and we have proven 
(\ref{*GDelta}) on $\delta_r$.  Similarly, 
\begin{align*}\CG^{-1}&(\theta\tens\theta)(\Delta^{op} x)\\
&=\sum_{p,q,s,t} \left(\tau(p,q)\delta_{q^R}\tens  (\tau(p,q)\ra q^R)\tau(q,q^R)^{-1} \delta_{p^R}  \right)\left((x\ra s)\ra t\, \delta_{t^R}\tens\delta_{(x\la s)^R}x\ra s\right)\\
&=\sum_{s,t}(x\ra s\cdot t)\tau(s,t)\delta_{t^R}\tens \tau(x\la(s\cdot t),(x\ra s\cdot t)\tau(s,t)\la t^R)^{-1}(x\ra s)\delta_{s^R}
 \end{align*}
 where we first note that for  the $\delta$-functions to connect, we need
 \[ p=x\la s,\quad ((x\ra s)\ra t)\la t^R=q^R,\]
 which is equivalent to $q=(x\ra s)\la t$ since $e=(x\ra s)\la (t\cdot t^R)=((x\ra s)\la t)\cdot(( (x\ra s)\ra t)\la t^R)$. In this case 
 \[ \tau(p,q)((x\ra s)\ra t)=\tau(x\la s, (x\ra s)\la t)((x\ra s)\ra t)=(x\ra s\cdot t)\tau(s,t)\]
 by the cocycle axiom. Similarly, $(x\ra s)^{-1}\la(x
 \la s)^R=s^R$ by Lemma~\ref{leminv} gives us $\delta_{s^R}$. For its coefficient, note that $p\cdot q=(x\la s)\cdot((x\ra s)\la t)=x\la(s\cdot t)$ so that, using the other form of $\tau'(p.q)$, we obtain
 \[ \tau(p\cdot q,\tau(p,q)\la q^R)^{-1}(x\ra s)=\tau(x\la(s\cdot t),\tau(p,q)((x\ra s)\ra t)\la t^R)^{-1}(x\ra s) \]
 and we use our previous calculation to put this in terms of $s,t$. On the other side, we have
 \begin{align*}
 (\Delta\theta(x))&\CG^{-1}= \sum_t\Delta(x\ra t\, \delta_{t^R}   )\CG^{-1}\\
 &=\sum_{p,q,s\cdot r=t^R}x\ra t\, \delta_s\tau(p,q)\delta_{q^R}\tens (x\ra t)\ra r\, \delta_r \tau(p\cdot q,\tau(p,q)\la q^R)^{-1}\delta_{p^R}\\
 &=\sum_{p,q}x\ra(p\cdot q)\, \tau(p,q)\delta_{q^R}\tens (x\ra p\cdot q)\ra s\, \tau(p\cdot q,s)^{-1}\delta_{p^R}, 
 \end{align*}
where, for the $\delta$-functions to connect, we need
\[ s=\tau(p,q)\la q^R,\quad r=\tau'(p,q)\la p^R.\]
The map $(p,q)\mapsto (s,r)$ has the same structure as the one we studied above but applied now to $p,q$ in place of $s,t$. It follows that $s\cdot r=(p\cdot q)^R$ and hence this being equal $t^R$ is equivalent to $p\cdot q=t$. Taking this for the value of $t$, we obtain the second expression for $(\Delta\theta(x))\CG^{-1}$. 

We now use the identity for $(x\ra p\cdot q)\ra s $ and $(p\cdot q)\cdot \tau(p,q)\la q^R=p\cdot(q\cdot q^R)=p$ to obtain the same as we obtained for $\CG^{-1}(\theta\tens\theta)(\Delta^{op} x)$ on $x$, upon renaming $s,t$ there to $p,q$.  The proofs of (\ref{*Gphi}), (\ref{*Gstrong}) are similarly quite involved, but omitted given that it is known that the category of modules is a strong bar category. \endproof


\begin{corollary}\label{corstar} For $(\ )^R$ bijective and the standard antipode in Proposition~\ref{standardS}, we have a $*$-quasi Hopf algebra with $\theta=* S$, where $x^*=x^{-1},\delta_s^*=\delta_s$ is the standard $*$-algebra structure on $\Xi$ as a cross product and $\gamma,\CG$ are as above. 
\end{corollary}\proof We check that
\[*Sx=*(\sum_s \delta_{(x^{-1}\la s)^R}x^{-1}\ra s)=\sum_s (x^{-1}\ra s)^{-1}\delta_{(x^{-1}\la s)^R}=\sum_{s'}x\ra s'\delta_{s'{}^R}=\theta(x),\] where $s'=x^{-1}\la s$ and we used Lemma~\ref{leminv}.  \endproof 

The key property of any quasi-bialgebra is that its category of modules is monoidal with associator $\phi_{V,W,U}: (V\tens W)\tens U\to V\tens (W\tens U)$ given by the action of $\phi$. In the $*$-quasi case, this becomes a bar category as follows\cite{BegMa:bar}. First, there is a functor ${\rm bar}$ from the category to itself which sends a module $V$ to a `conjugate', $\bar V$.  In our case, this has the same set and abelian group structure as $V$ but $\lambda.\bar v=\overline{\bar\lambda v}$ for all $\lambda\in \C$, i.e. a conjugate action of the field, where we write $v\in V$ as $\bar v$ when viewed in $\bar V$. Similarly,
\[ \xi.\bar v=\overline{\theta(\xi).v}\]
for all $\xi\in \Xi(R,K)$. On morphisms $\psi:V\to W$, we define $\bar\psi:\bar V\to \bar W$ by $\bar \psi(\bar v)=\overline{\psi(v)}$.  Next, there is a natural isomorphism $\Upsilon: {\rm bar}\circ\tens \Rightarrow \tens^{op}\circ({\rm bar}\times{\rm bar})$, given in our case for all modules $V,W$ by
\[ \Upsilon_{V,W}:\overline{V\tens W}\isom \bar W\tens \bar V, \quad \Upsilon_{V,W}(\overline{v\tens w})=\overline{ \CG^2.w}\tens\overline{\CG^1.v}\]
and making a hexagon identity with the associator, namely
\[ (\id\tens\Upsilon_{V,W})\circ\Upsilon_{V\tens W, U}=\phi_{\bar U,\bar V,\bar W}\circ(\Upsilon_{W,U}\tens\id)\circ\Upsilon_{V,W\tens U}\circ \overline{\phi_{V,W,U}}.\]
 We also have a natural isomorphism ${\rm bb}:\id\Rightarrow {\rm bar}\circ{\rm bar}$, given in our case for all modules $V$ by
\[ {\rm bb}_V:V\to \overline{\overline V},\quad {\rm bb}_V(v)=\overline{\overline{\gamma.v}}\]
and obeying $\overline{{\rm bb}_V}={\rm bb}_{\bar V}$. In our case, we have a strong bar category, which means also
\[ \Upsilon_{\bar W,\bar V}\circ\overline{\Upsilon_{V,W}}\circ {\rm bb}_{V\tens W}={\rm bb}_V\tens{\rm bb}_W.\]
Finally, a bar category has some  conditions on the unit object $\underline 1$, which in our case is the trivial representation with these automatic. That $G=RK$ leads to a strong bar category is in \cite[Prop.~3.21]{BegMa:bar} but without the underlying $*$-quasi-Hopf algebra structure as found above.  

\begin{example}\label{exS3quasi} \rm {\sl (i) $\Xi(R,K)$ for $S_2\subset S_3$ with its standard transversal.} As an algebra, this is generated by $\Z_2$, which means by an element $u$ with $u^2=e$, and by $\delta_{0},\delta_{1},\delta_{2}$
for $\delta$-functions as the points of $R=\{e,uv,vu\}$. The relations are $\delta_i$ orthogonal and add to $1$, and cross relations
\[ \delta_0u=u\delta_0,\quad \delta_1u=u\delta_2,\quad \delta_2u=u\delta_1.\]
The dot product is the additive group $\Z_3$, i.e. addition mod 3. The coproducts etc are
\[ \Delta \delta_i=\sum_{j+k=i}\delta_j\tens\delta_k,\quad \Delta u=u\tens u,\quad \phi=1\tens 1\tens 1\]
with addition mod 3. The cocycle and right action are trivial and the dot product is that of $\Z_3$ as a subgroup generated by $uv$. This gives an ordinary cross product Hopf algebra $\Xi=\C(\Z_3)\lcross\C \Z_2$. Here $S\delta_i=\delta_{-i}$ and $S u=u$.  For the $*$-structure, the cocycle is trivial  so $\gamma=1$ and $\CG=1\tens 1$ and we have an ordinary Hopf $*$-algebra.

 {\sl (ii) $\Xi(R,K)$ for $S_2\subset S_3$ with its second transversal.} For this $R$, the dot product is specified by $e$ the identity and $v\cdot w=w$, $w\cdot v=v$.  The algebra has relations
 \[ \delta_e u=u\delta_e,\quad \delta_v u=u\delta_w,\quad \delta_w u=u\delta_v\]
and the quasi-Hopf algebra coproducts etc. are
\[ \Delta \delta_e=\delta_e\tens \delta_e+\delta_v\tens\delta_v+\delta_w\tens\delta_w,\quad 
\Delta \delta_v=\delta_e\tens \delta_v+\delta_v\tens\delta_e+\delta_w\tens\delta_v,\]
\[ 
\Delta \delta_w=\delta_e\tens \delta_w+\delta_w\tens\delta_e+\delta_v\tens\delta_w,\quad  \Delta u=u\tens u,\]
\[ \phi=1\tens 1\tens  1+(\delta_v \tens\delta_w+\delta_w\tens\delta_v    )\tens (u-1)=\phi^{-1}.\]
The antipode is 
\[  S\delta_s=\delta_{s^R}=\delta_s,\quad  S u=\sum_{s}\delta_{(u\la s)^R}u=u,\quad \alpha=1,\quad \beta=\sum_s \delta_s\tens\tau(s,s)=1\]
from the antipode lemma, since the map $(\ )^R$ happens to be injective and indeed acts as the identity. In this case, we see that $\Xi(R,K)$ is nontrivially a quasi-Hopf algebra. Only $\tau(v,w)=\tau(w,v)=u$ are nontrivial, hence for the $*$-quasi Hopf algebra structure, we have
\[ \gamma=1,\quad \CG=1\tens 1+(\delta_v\tens\delta_w+\delta_w\tens\delta_v)(u\tens u-1\tens 1)\]
with $\theta=*S$ acting as the identity on our basis, $\theta(x)=x$ and $\theta(\delta_s)=\delta_s$. \end{example}

We also note that the algebras $\Xi(R,K)$ here are manifestly isomorphic for the two $R$, but the coproducts are different, so the tensor products of representations is different, although they turn out isomorphic. The set of irreps does not change either, but how we construct them can look different.  We will see in the next that this is part of a monoidal equivalence of categories. 

\begin{example}\rm $S_2\subset S_3$ with its 2nd transversal. Here $R$ has two orbits: (a) $\CC=\{e\}$ with $r_0=e, K^{r_0}=K$ with two 1-diml irreps $V_\rho$ as $\rho$=trivial and $\rho={\rm sign}$, and hence two irreps of $\Xi(R,K)$;  (b)  $\CC=\{w,v\}$ with $r_0=v$ or $r_0=w$, both  with $K^{r_0}=\{e\}$ and hence only $\rho$ trivial, leading to one 2-dimensional irrep of $\Xi(R,K)$. So, altogether, there are again three irreps of $\Xi(R,K)$:
\begin{align*} V_{(\{e\},\rho)}:& \quad \delta_r.1 =\delta_{r,e},\quad u.1 =\pm 1,\\
V_{(\{w,v\}),1)}:& \quad \delta_r. v=\delta_{r,v}v,\quad \delta_r. w=\delta_{r,w}w,\quad u.v= w,\quad u.w=v
\end{align*}
acting on $\C$ and on the span of $v,w$ respectively. These irreps are equivalent to what we had in Example~\ref{exS3n} when computing irreps from the standard $R$. 
\end{example}

\section{Categorical justification and twisting theorem}\label{sec:cat_just}

We have shown that the boundaries can be defined using the action of the algebra $\Xi(R,K)$ and that one can perform novel methods of fault-tolerant quantum computation using these boundaries. The full story, however, involves the quasi-Hopf algebra structure verified in the preceding section and now we would like to  connect back up to the category theory behind this. 

\subsection{$G$-graded $K$-bimodules.} We start by proving the equivalence ${}_{\Xi(R,K)}\CM \simeq {}_K\CM_K^G$ explicitly and use it to derive the coproduct studied in Section~\ref{sec:quasi}. Although this equivalence is known\cite{PS}, we believe this to be a new and more direct derivation. 

\begin{lemma} If $V_\rho$ is a $K^{r_0}$-module and $V_{\CO,\rho}$ the associated $\Xi(R,K)$ irrep, then
\[ \tilde V_{\CO,\rho}= V_{\CO,\rho}\tens \C K,\quad x.(r\tens v\tens z).y=x\la r\tens\zeta_r(x).v\tens (x\ra r)zy,\quad |r\tens v\tens z|=rz\]
is a $G$-graded $K$-bimodule. Here $r\in \CO$ and $v\in V_\rho$ in the construction of $V_{\CO,\rho}$. 
\end{lemma}
\proof That this is a $G$-graded right $K$-module commuting with the left action of $K$ is trivial. That the left action  works and is $G$-graded is
\begin{align*}x.(y.(r\tens v\tens z))&=x.(y\la r\tens \zeta_r(y).v\tens (y\ra r)z)= xy\la r\tens \zeta_r(xy).v\tens (x\ra(y\la r))(y\ra r)z\\
&=xy\la r\tens \zeta_r(xy).v\tens ((xy)\ra r)z\end{align*}
and
\[ |x.(r\tens v\tens z).y|=(x\la r) (x\ra r)zy= xrzy=x|r\tens v \tens z|y.\]
\endproof

\begin{remark}\rm Recall that we can also think more abstractly of $\Xi=\C(G/K)\lcross \C K$ rather than using a transversal. In these terms, a representation of $\Xi(R,K)$ as an $R$-graded $K$-module $V$ such that $|x.v|=x\ra |v|$ now becomes  a $G/K$-graded $K$-module such that $|x.v|=x|v|$, where $|v|\in G/K$ and we multiply from the left by $x\in K$. Moreover, the role of an orbit $\CO$ above is played by a double coset $T=\CO K\in {}_KG_K$. In these terms, the role of the isometry group $K^{r_0}$ is played by 
\[  K^{r_T}:=K\cap r_T K r_T^{-1}, \] 
where $r_T$ is any representative of the same double coset. One can take $r_T=r_0$ but we can also chose it more freely. 
Then an irrep is given by a double coset $T$ and an irreducible representation $\rho_T$ of $K^{r_T}$. If we denote by $V_{\rho_T}$ the carrier space for this then the associated irrep of $\C(G/K)\lcross\C K$ is  $V_{T,\rho_T}=\C K\tens_{K^{r_T}}V_{\rho_T}$ which is manifestly a $K$-module and we give it the $G/K$-grading by $|x\tens_{K^{r_T}} v|=xK$. The construction in the last lemma is then equivalent to 
\[ \tilde V_{T,\rho_T}=\C K\tens_{K^{r_T}} V_{\rho_T}\tens\C K,\quad |x\tens_{K^{r_T}} v\tens z|=xz\]
as  manifestly a $G$-graded $K$-bimodule. This is an equivalent point of view, but we prefer our more explicit one based on $R$, hence details are omitted. 
\end{remark}

Also note that the category ${}_K\CM_K^G$ of $G$-graded $K$-bimodules has an obvious monoidal structure inherited from that of $K$-bimodules, where we tensor product over $\C K$. Here $|w\tens_{\C K} w'|=|w||w'|$ in $G$ is well-defined and $x.(w\tens_{\C K}w').y=x.w\tens_{\C K} w'.y$ has degree $x|w||w'|y=x|w\tens_{\C K}w'|y$ as required. 

\begin{proposition} \label{prop:mon_equiv} 
We let $R$ be a transversal and $W=V\tens \C K$ made into a $G$-graded $K$-bimodule by
\[ x.(v\tens z).y=x.v\tens (x\ra|v|)zy, \quad  |v\tens z|=  |v|z\in G,\]
where now we view $|v|\in R$ as the chosen representative of $|v|\in G/K$. This gives a functor $F:{}_\Xi\CM\to {}_K\CM_K^G$ which is a monoidal equivalence for a suitable quasibialgebra structure on $\Xi(R,K)$. The latter depends on $R$ since $F$ depends on $R$.
\end{proposition}
\proof We define $F(V)$ as stated, which is clearly a right module that commutes with the left action, and the latter is a module structure as 
\[ x.(y.(v\tens z))=x.(y.v\tens (y\ra |v|)z)=xy.v\tens (x\ra (y\la |v|))(y\ra |v|)z=(xy).(v\tens z)\]
using the matched pair axiom for $(xy)\ra |v|$.  We also check that $|x.(v\tens z).y|=|x.v|zy=(x\la |v|)(x\ra |v|)zy=x|v|zy=x|v\tens z|y$. Hence, we have a $G$-graded $K$-bimodule. Conversely, if $W$ is a $G$-graded $K$-bimodule, we let
\[ V=\{w\in W\ |\ |w|\in R\},\quad  x.v=xv(x\ra |v|)^{-1},\quad \delta_r.v=\delta_{r,|v|}v,\]
where $v$ on the right is viewed in $W$ and we use the $K$-bimodule structure. This is arranged so that $x.v$ on the left lives in $V$. Indeed, $|x.v|=x|v|(x\ra |v|)^{-1}=x\la |v|$ and $x.(y.v)=xyv(y\ra |v|)^{-1}(x\ra(y\la |v|))^{-1}=xyv((xy)\ra |v|)^{-1}$ by the matched pair condition, as required for a representation of $\Xi(R,K)$.  One can check that this is inverse to the other direction. Thus, given $W=\oplus_{rx\in G}W_{rx}=\oplus_{x\in K}  W_{Rx}$, where we let $W_{Rx}=\oplus_{r\in R}W_{rx}$, the right action by $x\in K$ gives an isomorphism $W_{Rx}\isom V\tens x$ as vector spaces and hence recovers $W=V\tens\C K$. This clearly has the correct right $K$-action and from the left $x.(v\tens z)=xv(x\ra|v|)^{-1}\tens (x\ra|v|)z$, which under the identification maps to $xv(x\ra|v|)^{-1} (x\ra|v|)z=xvz\in W$ as required given that $v\tens z$ maps to $vz$ in $W$. 

Now, if $V,V'$ are $\Xi(R,K)$ modules then as vector spaces, 
\[ F(V)\tens_{\C K}F(V')=(V\tens \C K)\tens_{\C K} (V'\tens \C K)=V\tens V'\tens \C K{\buildrel f_{V,V'}\over\isom}F(V\tens V')\]
 by the obvious identifications except that in the last step we allow ourselves the possibility of a nontrivial isomorphism as vector spaces. For the actions on the two sides, 
\[ x.(v\tens v'\tens z).y=x.(v\tens v')\tens (x\ra |v\tens v'|)zy= x.v\tens (x\ra |v|).v'\tens ((x\ra|v|)\ra|v'|)zy,\]
where on the right, we have $x.(v\tens 1)=x.v \tens x\ra|v|$ and then take $x\ra|v|$ via the $\tens_{\C K}$ to act on $v'\tens z$ as per our identification. Comparing the $x$ action on the $V\tens V'$ factor, we need
\[\Delta x=\sum_{r\in R}x\delta_r\tens  x\ra r= \sum_{r\in R}\delta_{x\la r}\tens x \tens 1\tens x\ra r\]
as a modified coproduct without requiring a nontrivial $f_{V,V'}$ for this to work. The first expression is viewed in $\Xi(R,K)^{\tens 2}$ and the second is on the underlying vector space.  Likewise, looking at the grading of $F(V\tens V')$ and comparing with the grading of $F(V)\tens_{\C K}F(V')$, we need to define $|v\tens v'|=|v|\cdot|v'|\in R$ and use $|v|\cdot|v'|\tau(|v|,|v'|)=|v||v'|$ to match the degree on the left hand side. This amounts to the coproduct of $\delta_r$ in $\Xi(R,K)$,
\[ \Delta\delta_r=\sum_{s\cdot t=r}\delta_s\tens\delta_t=\sum_{s\cdot t=r} \delta_s\tens 1\tens \delta_t \tens 1\]
{\em and} a further isomorphism 
\[ f_{V,V'}(v\tens v'\tens z)= v\tens v'\tens\tau(|v|,|v'|)z\] on the underlying vector space.
After applying this, the degree of this element is $|v\tens v'|\tau(|v|,|v'|)z=|v||v'|z=|v\tens 1||v'\tens z|$, which is the degree on the original $F(V)\tens_{\C K}F(V')$ side. Now we show that $f_{V,V'}$ respects associators on each side of $F$. Taking the associator on the $\Xi(R,K)$-module side as
\[ \phi_{V,V',V''}:(V\tens V')\tens V''\to V\tens(V'\tens V''),\quad \phi_{V,V',V''}((v\tens v')\tens v'')=\phi^1.v\tens (\phi^2.v'\tens\phi^3.v'')\]
and $\phi$ trivial on the $G$-graded $K$-bimodule side, for $F$ to be monoidal with the stated $f_{V,V'}$ etc, we need
\begin{align*}
F(\phi_{V,V,V'})&f_{V\tens V',V''}f_{V,V'}(v\tens v'\tens z)\\
&=F(\phi_{V,V,V'})f_{V\tens V',V''}(v\tens v'\tens \tau(|v|,|v'|).v''\tens (\tau(|v|,|v'|)\ra|v''|)z)\\
&=F(\phi_{V,V,V'})(v\tens v'\tens \tau(|v|,|v'|).v''\tens\tau(|v|.|v'|,\tau(|v|,|v'|)\la |v''|)(\tau(|v|,|v'|)\ra|v''|)z)\\
&=F(\phi_{V,V,V'})(v\tens v'\tens \tau(|v|,|v'|).v''\tens \tau(|v|,|v'|.|v''|)\tau(|v'|,|v''|)z,\\
f_{V,V'\tens V''}&f_{V',V''}(v\tens v'\tens v''\tens z)=f_{V,V'\tens V''}(v\tens v'\tens v''\tens \tau(|v'|,|v''|)z) \\
&=v\tens v'\tens v''\tens\tau(|v|,|v'\tens v''|)\tau(|v'|,|v''|)z =v\tens v'\tens v''\tens\tau(|v|,|v'|.|v''|)\tau(|v'|,|v''|)z,\end{align*}
where for the first equality we moved $\tau(|v|,|v'|)$ in the output of $f_{V,V'}$ via $\tens_{\C K}$ to act on the $v''$. We used 
the cocycle property of $\tau$ for the 3rd equality. Comparing results, we need
\[ \phi_{V,V',V''}((v\tens v')\tens v'')=v\tens( v'\tens \tau(|v|,|v'|)^{-1}.v''),\quad \phi=\sum_{s,t\in R}(\delta_s\tens 1)\tens(\delta_s\tens1)\tens (1\tens \tau(s,t)^{-1}).\]
Note that we can write
\[ f_{V,V'}(v\tens v'\tens z)=(\sum_{s,t\in R}(\delta_s\tens 1)\tens(\delta_t\tens 1)\tens \tau(s,t)).(v\tens v'\tens z)\]
but we are not saying that $\phi$ is a coboundary since this is not given by the action of an element of $\Xi(R,K)^{\tens 2}$. 
\endproof

This derives the quasibialgebra structure on $\Xi(R,K)$ used in Section~\ref{sec:quasi} but now so as to obtain an 
equivalence of categories. 

\subsection{Drinfeld twists induced by change of transversal} 

 We recall that if $H$ is a quasiHopf algebra and $\chi\in H\tens H$ is a {\em cochain} in the sense of invertible and $(\id\tens
\eps)\chi=(\eps\tens\id)\chi=1$, then its {\em Drinfeld twist} $\bar H$ is another quasi-Hopf algebra
\[ \bar\Delta=\chi^{-1}\Delta(\ )\chi,\quad \bar\phi=\chi_{23}^{-1}((\id\tens\Delta)\chi^{-1})\phi ((\Delta\tens\id)\chi)\chi_{12},\quad \bar\eps=\eps\]
\[ S=S,\quad\bar\alpha=(S\chi^1)\alpha\chi^2,\quad \bar\beta=(\chi^{-1})^1\beta S(\chi^{-1})^2\]
where $\chi=\chi^1\tens\chi^2$ with a sum of such terms understood and we use same notation for $\chi^{-1}$, see \cite[Thm.~2.4.2]{Ma:book} but note that our $\chi$ is denoted $F^{-1}$ there. In categorical terms, this twist 
corresponds to a monoidal equivalence $G:{}_{H}\CM\to {}_{H^\chi}\CM$ which is the identity on objects and morphisms but has a nontrivial natural transformation 
\[ g_{V,V'}:G(V)\bar\tens G(V')\isom G(V\tens V'),\quad  g_{V,V'}(v\tens v')= \chi^1.v\tens\chi^2.v'.\]
The next theorem follows by the above reconstruction arguments, but here we check it directly. The logic is that for different $R,R'$ the category of modules are both monoidally equivalent to ${}_K\CM_K^G$ and hence monoidally equivalent  but not in a manner that is compatible with the forgetful functor to Vect. Hence these should be related by a cochain twist.

\begin{theorem}\label{thmtwist} Let $R,\bar R$ be two transversals with $\bar r\in\bar R$ representing the same coset as $r\in R$. Then  $\Xi(\bar R,K)$ is a cochain twist of  $\Xi(R,K)$  at least as quasi-bialgebras (and as quasi-Hopf algebras if one of them is). The Drinfeld cochain is $\chi=\sum_{r\in R}(\delta_r\tens 1)\tens (1\tens r^{-1}\bar r)$. \end{theorem}
\proof  Let $R,\bar R$ be two transversals. Then for each $r\in R$, the class $rK$ has a unique  representative $\bar rK$ with $\bar r\in R'$. Hence  $\bar r= r c_r$ for some function $c:R\to K$ determined by the two transversals as $c_r=r^{-1}\bar r$ in $G$. One can show that the cocycle matched pairs are related by
\[ x\bar\la \bar r=(x\la r)c_{x\la r},\quad x\bar\ra \bar r= c_{x\la r}^{-1}(x\ra r)c_r\]
among other identities. On using 
\begin{align*} \bar s\bar t=sc_s tc_t=s (c_s\la t)(c_s\ra t)c_t&= (s\cdot c_s\la t)\tau(s, c_s\la t)(c_s\ra t)c_t\\&=\overline{ s\cdot (c_s\la t)}c_{s\cdot c_s\la t}^{-1}\tau(s, c_s\la t)(c_s\ra t)c_t\end{align*}
and factorising using $\bar R$, we see that 
\begin{equation}\label{taucond} \bar s\, \bar\cdot\,  \bar t= \overline{s\cdot c_s\la t},\quad\bar\tau(\bar s,\bar t)=c_{s\cdot c_s\la t}^{-1}\tau(s, c_s\la t)(c_s\ra t)c_t.\end{equation} 

We will construct a monoidal functor $G:{}_{\Xi(R,K)}\CM\to {}_{\Xi(\bar R,K)}\CM$ with $g_{V,V'}(v\tens v')= \chi^1.v\tens\chi^2.v'$ for a suitable $\chi\in \Xi(R,K)^{\tens 2}$. First, let $F:{}_{\Xi(R,K)}\CM\to {}_K\CM_K^G$ be the monoidal functor above with natural isomorphism $f_{V,V'}$ and $\bar F:{}_{\Xi(\bar R,K)}\CM\to {}_K\CM_K^G$ the parallel for $\Xi(\bar R,K)$ with isomorphism $\bar f_{V,V'}$. Then
\[ C:F\to \bar F\circ G,\quad C_V:F(V)=V\tens\C K\to V\tens \C K=\bar FG(V),\quad C_V(v\tens z)=v\tens c_{|v|}^{-1}z\]
is a natural isomorphism. Check on the right we have, denoting the $\bar R$ grading by $||\ ||$, the $G$-grading and $K$-bimodule structure 
\begin{align*} |C_V(v\tens z)|&= |v\tens c_{|v|}^{-1}z|= ||v||c_{|v|}^{-1}z=|v|z=|v\tens z|,\\ 
x.C_V(v\tens z).y&=x.(v\tens c_{|v|}^{-1}z).y=x.v\tens (x\bar\ra ||v||)c_{|v|}^{-1}zy=x.v \tens c_{x\la |v|}^{-1} (x\ra |v|)zy\\
&= C_V(x.(v\tens z).y).\end{align*}
We want these two functors to not only be naturally isomorphic but for this to respect that they are both monoidal functors. Here $\bar F\circ G$ has the natural isomorphism
\[ \bar f^g_{V,V'}= \bar F(g_{V,V'})\circ \bar f_{G(V),G(V')}\]
by which it is a monoidal functor. 
 
The natural condition on a natural isomorphism $C$ between monoidal functors is  that $C$ behaves in the obvious way on 
tensor product objects via the natural isomorphisms associated to each monoidal functor.  In our case, this means
\[ \bar f^g_{V,V'}\circ (C_{V}\tens C_{V'}) = C_{V\tens V'}\circ f_{V,V'}: F(V)\tens F(V')\to \bar F G(V\tens V').\]
Putting in the specific form of these maps, the right hand side is
\[C_{V\tens V'}\circ f_{V,V'}(v\tens 1\tens_K v'\tens z)=C_{V\tens V'}(v\tens v'\tens \tau(|v|,|v'|)z)=v\tens v'\tens c^{-1}_{|v\tens v'|}\tau(|v|,|v'|)z,\]
while the left hand side is
\begin{align*}\bar f^g_{V,V'}\circ (C_{V}\tens C_{V'})&(v\tens 1\tens_K v'\tens z)=\bar f^g_{V,V'}(v\tens c^{-1}_{|v|}\tens_K v'\tens c^{-1}_{|v'|}z)\\
&=\bar f^g_{V,V'}(v\tens 1\tens_K c^{-1}_{|v|}.v'\tens (c^{-1}_{|v|}\bar\la ||v'||)c^{-1}_{|v'|}z)\\
&=\bar F(g_{V,V'})(v\tens c^{-1}_{|v|}.v'\tens \bar\tau(||v||,||c^{-1}_{|v|}.v'||)(c^{-1}_{|v|}\bar\la||v'||)c^{-1}_{|v'|}z)\\
&=\bar F(g_{V,V'})(v\tens c^{-1}_{|v|}.v'\tens c^{-1}_{|v\tens v'|}\tau(|v|,|v'|)z,
\end{align*}
using the second of (\ref{taucond}) and $|v\tens v'|=|v|\cdot|v'|$. We also used  $\bar f^g_{V,V'}=\bar F(g_{V,V'})\bar f_{G(V),G(V')}:\bar FG(V)\tens \bar FG(V')\to \bar FG(V\tens V')$. Comparing, we need $\bar F(g_{V,V'})$ to be the action of the element 
\[ \chi=\sum_{r\in R} \delta_r\tens c_r\in \Xi(R,K)^{\tens 2}.\]
It follows from the arguments, but one can also check directly, that $\phi$ indeed twists as stated to $\bar\phi$ when these are given by Lemma~\ref{Xibialg}, again using (\ref{taucond}).  \endproof

The twisting of a quasi-Hopf algebra is again one. Hence, we have:

\begin{corollary}\label{twistant} If $R$ has $(\ )^R$ bijective giving a quasi-Hopf algebra with regular antipode $S,\alpha=1,\beta$ as in Proposition~\ref{standardS} and $\bar R$ is another transversal then $\Xi(\bar R,K)$ in the twisting form of Theorem~\ref{thmtwist} has an antipode
\[ \bar S=S,\quad \bar \alpha=\sum_r \delta_{r^R} c_r ,\quad \bar \beta =\sum_r \delta_r \tau(r,r^R)(c_r^{-1}\ra r^R)^{-1} . \]
This is a regular antipode if $(\ )^R$ for $\bar R$ is also bijective (i.e. $\bar\alpha$ is then invertible and can be transformed back to standard form to make it 1).\end{corollary}
\proof  We work with the initial quasi-Hopf algebra $\Xi(R,K)$ and $\la,\ra,\tau$ refer to this but note that $\Xi(\bar R,K)$ is the same algebra when $\delta_r$ is identified with the corresponding $\delta_{\bar r}$. Then
\begin{align*}\bar \alpha&=(S\chi^{1})\chi^{2}=\sum_r S\delta_r\tens c_r=\delta_{r^R}c_r\end{align*}
using the formula for $S\delta_r=\delta_{r^R}$ in Proposition~\ref{standardS}. Similarly,
$\chi^{-1}=\sum_r \delta_r\tens c_r^{-1}$ and we use $S,\beta$ from the above lemma, where
\[ S (1\tens x)= \sum_s \delta_{(x^{-1}\la s)^R}\tens x^{-1}\ra s=\sum_t\delta_{t^R}\tens x^{-1}\ra(x\la t)=\sum_t\delta_{t^R}\tens (x\ra t)^{-1}.\]
Then
\begin{align*} \bar \beta &=\chi^{-1}\beta S\chi^{-2}=\sum_{r,s,t}\delta_r\delta_s\tau(s,s^R)\delta_{t^R}(c_r^{-1}\ra t)^{-1}\\
&=\sum_{r,t} \delta_r\tau(r,r^R)\delta_{t^R}(c_r^{-1}\ra t)^{-1}=\sum_{r,t}\delta_r\delta_{\tau(r,r^R)\la t^R}\tau(r,r^R)  (c_r^{-1}\ra t)^{-1}.\end{align*}
 Commuting the $\delta$-functions to the left requires $r=\tau(r,r^R)\la t^R$ or $r^{RR}=\tau(r,r^R)^{-1}\la r= t^R$ so $t=r^R$ under our assumptions, giving the answer stated. 

If $(\ )^R$ is bijective then $\bar\alpha^{-1}=\sum_r c_r^{-1}\delta_{r^R}=\sum_r \delta_{c_r^{-1}\la r^R}c_r^{-1}$ provides the left inverse.  On the other side, we need $c_r^{-1}\la r^R= c_s^{-1}\la s^R$ iff $r=s$. This is true  if $(\ )^{R}$ for $\bar R$ is also bijective. That is because, if we write $(\ )^{\bar R}$ for the right inverse with respect to $\bar R$, one can show by comparing the factorisations that
\[ \bar s^{\bar R}=\overline{c_s^{-1}\la s^R},\quad \overline{s^R}=c_s\bar\la \bar s^{\bar R}\]
and we use the first of these. \endproof

\begin{example}\rm With reference to the list of transversals for $S_2\subset S_3$, we have four quasi-Hopf algebras of which two were already computed in Example~\ref{exS3quasi}. 

{\sl (i) 2nd transversal as twist of the first.} Here $\bar\Xi$ is generated by $\Z_2$ as $u$ again and $\delta_{\bar r}$ with $\bar R=\{e,w,v\}$. We have the same cosets represented by these with $\bar e=e$, $\overline{uv}=w$ and  $\overline{vu}=v$, which means $c_e=e, c_{vu}=u, c_{uv}=u$.  To compare the algebras in the two cases, we identify $\delta_0=\delta_e,\delta_1=\delta_w, \delta_2=\delta_v$  as delta-functions on $G/K$ (rather than on $G$) in order to identify the algebras of $\bar\Xi$ and $\Xi$. The cochain from Theorem~\ref{thmtwist} is \[ \chi=\delta_e\tens e+(\delta_{vu}+\delta_{uv})\tens u=\delta_0\tens 1+ (\delta_1+\delta_2)\tens u=\delta_0\tens 1+ (1-\delta_0)\tens u \]
as an element of $\Xi\tens\Xi$. One can check that this conjugates the two coproducts as claimed. We also have
\[ \chi^2=1\tens 1,\quad (\eps\tens\id)\chi=(\id\tens\eps)\chi=1.\]
We spot check (\ref{taucond}), for example $v\bar\cdot w=\overline{vu}\, \bar\cdot\, \overline{uv}=\overline{uv}=\overline{vuvu}=\overline{vu( u\la (uv))}$, as it had to be. We should therefore find that 
\[((\Delta\tens\id)\chi)\chi_{12}=((\id\tens\Delta)\chi)\chi_{23}\bar\phi. \]
We have checked directly that this indeed holds. Next, the antipode of the first transversal should twist to 
\[ \bar S=S,\quad \bar\alpha=\delta_e c_e+\delta_{uv}c_{vu}+\delta_{vu}c_{uv}=\delta_e(e-u)+u=\delta_e c_e+\delta_{vu}c_{vu}+\delta_{uv}c_{uv}=\bar\beta\]
by Corollary~\ref{twistant} for twisting the antipode. Here, $U=\bar\alpha^{-1}=\bar\beta = U^{-1}$ and $\bar S'=U(S\ )U^{-1}$ with $\bar\alpha'=\bar\beta'=1$ should also be an antipode. We can check this:
\[U u = (\delta_0(e-u)+u)u = \delta_0(u-e)+e = u(\delta_{u^{-1}\la 0}(e-u)+u) = u U\]
so $\bar S' u = UuU^{-1} = u$, and 
\[\bar S' \delta_1 = U(S\delta_1)U= U\delta_2 U = (\delta_0(e-u)+u)\delta_2(\delta_0(e-u)+u) = \delta_1.\]
\bigskip

{\sl (ii) 3rd transversal as a twist of the first.} A mixed up choice is $\bar R=\{e,uv,v\}$ which is not a subgroup so $\tau$ is nontrivial. One has 
\[ \tau(uv,uv)=\tau(v,uv)=\tau(uv,v)=u,\quad \tau(v,v)=e,\quad v.v=e,\quad v.uv=uv,\quad  uv.v=e,\quad uv.uv=v,\]
\[ u\la v=uv,\quad u\la (uv)=v,\quad u\ra v=e,\quad u\ra uv=e\]
and all other cases implied from the properties of $e$. Here $v^R=v$ and $(uv)^R=v$. These are with respect to $\bar R$, but note that twisting calculations will take place with respect to $R$.

Writing $\delta_0=\delta_e,\delta_1=\delta_{uv},\delta_2=\delta_v$ we have the same algebra as before (as we had to) and now the coproduct etc., 
\[ \bar\Delta u=u\tens 1+\delta_0u\tens (u-1),\quad \bar\Delta\delta_0=\delta_0\tens\delta_0+\delta_2\tens\delta_2+\delta_1\tens\delta_2  \]
\[  
\bar\Delta\delta_1=\delta_0\tens\delta_1+\delta_1\tens\delta_0+\delta_2\tens\delta_1,\quad \bar\Delta\delta_2=\delta_0\tens\delta_2+\delta_2\tens\delta_0+\delta_1\tens\delta_1,\]
\[ \bar\phi=1\tens 1\tens 1+ (\delta_1\tens\delta_2+\delta_2\tens\delta_1+\delta_1\tens\delta_1)(u-1)=\bar\phi^{-1}\]
for the quasibialgebra.  We used the $\tau,\la,\ra,\cdot$ for $\bar R$ for these direct calculations. 

Now we consider twisting with
\[ c_0=e,\quad c_1=(uv)^{-1}uv=1,\quad c_2=v^{-1}vu=u,\quad \chi=1\tens 1+ \delta_2\tens (u-1)=\chi^{-1}\]
and check twisting the coproducts 
\[ (1\tens 1+\delta_2\tens (u-1))(u\tens u)(1\tens 1+\delta_2u\tens (u-1))=u\tens 1+\delta_0\tens(u-1)=\bar\Delta u, \]
\[  (1\tens 1+\delta_2\tens (u-1))(\delta_0\tens\delta_0+\delta_1\tens\delta_2+\delta_2\tens\delta_1)(1\tens 1+\delta_2\tens (u-1))=\bar\Delta\delta_0,\]
\[  (1\tens 1+\delta_2\tens (u-1))(\delta_0\tens\delta_1+\delta_1\tens\delta_0+\delta_2\tens\delta_2)(1\tens 1+\delta_2\tens (u-1))=\bar\Delta\delta_1,\]
\[  (1\tens 1+\delta_2\tens (u-1))(\delta_0\tens\delta_2+\delta_2\tens\delta_0+\delta_1\tens\delta_1)(1\tens 1+\delta_2\tens (u-1))=\bar\Delta\delta_2.\]
One can also check that (\ref{taucond}) hold, e.g. for the first half,
\[  \bar 2=\bar 1\bar\cdot\bar 1=\overline{1+c_1\la 1}=\overline{1+1},\quad \bar 0=\bar 1\bar\cdot\bar 2=\overline{1+c_1\la 2}=\overline{1+2},\]
\[ \bar 1=\bar2\bar\cdot\bar 1=\overline{2+c_2\la 1}=\overline{2+2},\quad  \bar 0=\bar2\bar\cdot\bar 2=\overline{2+c_2\la 2}=\overline{2+1}\]
as it must.

Now we apply the twisting of antipodes in Corollary~\ref{twistant}, remembering to do calculations now with  $R$ where $\tau,\ra$ are trivial, to get
\[ \bar S=S,\quad \bar\alpha=\delta_0+\delta_1c_2+\delta_2c_1=1+\delta_1(u-1),\quad \bar\beta=\delta_0+\delta_2c_2+\delta_1c_1=1+\delta_2(u-1),\]
which obey $\bar\alpha^2=\bar\alpha$ and $\bar\beta^2=\bar\beta$ and are therefore not (left or right) invertible. Hence, we cannot set either equal to 1 by $U$ and there is an antipode, but it is not regular. One can check the antipode indeed works:
\begin{align*}(Su)\alpha+ (Su) (S\delta_0)\alpha(u-1)&=u(1+\delta_1(u-1))+\delta_0 u(1+\delta_1(u-1))(u-1)\\
&=u+\delta_2(1-u)+\delta_0(1-u)=u+(1-\delta_1)(1-u)=\alpha\\
u\beta+\delta_0u\beta S(u-1)&=u(1+\delta_2(u-1))+\delta_0 u(1+\delta_2(u-1))(u-1)\\
&=u+\delta_1(1-u)+\delta_0(1-u)=u+(1-\delta_2)(1-u)=\beta	
\end{align*}
\begin{align*} (S\delta_0)\alpha\delta_0&+(S\delta_2)\alpha\delta_2+(S\delta_1)\alpha\delta_2=\delta_0(1+\delta_1(u-1))\delta_0+(1-\delta_0)(1+\delta_1(u-1))\delta_2\\
&=\delta_0+(1-\delta_0)\delta_2+\delta_1(\delta_1 u-\delta_2)=\delta_0+\delta_2+\delta_1u=\alpha\\
\delta_0\beta S\delta_0&+\delta_2\beta S\delta_2+\delta_1\beta S\delta_2=\delta_0(1+\delta_2(u-1))\delta_0+(1-\delta_0)(1+\delta_2(u-1))\delta_1\\
&=\delta_0+(1-\delta_0)\delta_1+(1-\delta_0)\delta_2(u-1)\delta_1=\delta_0+\delta_1+\delta_2(\delta_2u-\delta_1)=\beta
\end{align*}
and more simply on $\delta_1,\delta_2$. 

The fourth transversal has a similar pattern to the 3rd, so we do not list its coproduct etc. explicitly. 
\end{example}

In general, there will be many different choices of transversal. For  $S_{n-1}\subset S_n$, the first  two transversals for $S_2\subset S_3$ generalise as follows, giving a  Hopf algebra and a strictly quasi-Hopf algebra respectively. 

\begin{example}\rm {\sl (i) First transversal.} Here  $R=\Z_n$ is a subgroup with $i=0,1,\cdots,n-1$ mod $n$ corresponding to the elements  $(12\cdots n)^i$. Neither subgroup
is normal for $n\ge 4$, so both actions are nontrivial but $\tau$ is trivial. This expresses $S_n$ as a double cross product $\Z_n\dcross S_{n-1}$ (with trivial $\tau$) and the matched pair of actions
\[ \sigma\la i=\sigma(i),\quad (\sigma\ra i)(j)=\sigma(i+j)-\sigma(i)\]
for $i,j=1,\cdots,n-1$, where we add and subtract mod $n$ but view the results in the range $1,\cdots, n$. This was actually found by twisting from the 2nd transversal below, but we can check it directly as follows. First.
\[\sigma (1\cdots n)^i= (\sigma\la i)(\sigma\ra i)=(12\cdots n)^{\sigma(i)}\left((1\cdots n)^{-\sigma(i)}\sigma(12\cdots n)^i\right)\]
and we check that the second factor sends $n\to i\to \sigma(i) \to n$, hence lies in $S_n$. It follows by the known fact of unique factorisation into these subgroups that this factor is $\sigma\ra i$. Its action on  $j=1,\cdots, n-1$ is 
\[ (\sigma\la i)(j)=(12\cdots n)^{-\sigma(i)}\sigma(12\cdots n)^i(j)=\begin{cases} n-\sigma(i) & i+j=n\\ \sigma(i+j)-\sigma(i) & i+j\ne n\end{cases}=\sigma(i+j)-\sigma(i),\]
where $\sigma(i+j)\ne \sigma(i)$ as $i+j\ne i$ and $\sigma(n)=n$ as $\sigma\in S_{n-1}$. It also follows since the two factors are subgroups that these are indeed a matched pair of actions. We can also check  the matched pair axioms directly. Clearly, $\la$ is an action and
\[ \sigma(i)+ (\sigma\ra i)(j)=\sigma(i)+\sigma(i+j)-\sigma(i)=\sigma\la(i+j)\] for $i,j\in\Z_n$. On the other side, 
\begin{align*}( (\sigma\ra i)\ra j)(k)&=(\sigma\ra i)(j+k)-(\sigma\ra i)(j)=\sigma(i+(j+k))-\sigma(i)-\sigma(i+j)+\sigma(i)\\
&=\sigma((i+j)+k)-\sigma(i+j)=(\sigma\ra(i+j))(k),\\
((\sigma\ra(\tau\la i))(\tau\ra i))(j)&=(\sigma\ra\tau(i))(\tau(i+j))-\tau(i))=\sigma(\tau(i)+\tau(i+j)-\tau(i))  -\sigma(\tau(i))\\
&=  \sigma(\tau(i+j))-\sigma(\tau(i))=((\sigma\tau)\ra i)(j)\end{align*}
for $i,j\in \Z_n$ and $k\in 1,\cdots,n-1$.  

This gives $ \C S_{n-1}\cobicross\C(\Z_n)$ as a natural bicrossproduct Hopf algebra which we identify with $\Xi$ (which we prefer to build  on the other tensor product order). From Lemma~\ref{Xibialg} and Proposition~\ref{standardS}, this is spanned by products of $\delta_i$ for $i=0,\cdots n-1$ as our labelling of $R=\Z_n$ and  $\sigma\in S_{n-1}=K$, with cross relations $\sigma\delta_i=\delta_{\sigma(i)}\sigma$, $\sigma\delta_0=\delta_0\sigma$, and coproduct etc., 
\[ \Delta \delta_i=\sum_{j\in \Z_n}\delta_j\tens\delta_{i-j},\quad \Delta\sigma=\sigma\delta_0+\sum_{i=1}^{n-1}(\sigma\ra i),\quad \eps\delta_i=\delta_{i,0},\quad\eps\sigma=1,\]
\[ S\delta_i=\delta_{-i},\quad  S\sigma=\sigma^{-1}\delta_0+(\sigma^{-1}\ra i)\delta_{-i},\]
where $\sigma\ra i$ is as above for $i=1,\cdots,n-1$. This is a usual Hopf $*$-algebra with $\delta_i^*=\delta_i$ and $\sigma^*=\sigma^{-1}$ according to Corollary~\ref{corstar}. 

\medskip
{\sl  (ii) 2nd transversal.} Here $R=\{e, (1\, n),(2\, n),\cdots,(n-1\, n)\}$,  which has nontrivial $\la$ in which $S_{n-1}$ permutes the 2-cycles according to the $i$ label, but again trivial $\ra$ since
\[ \sigma(i\, n)=(\sigma(i)\, n)\sigma,\quad \sigma\la (i\ n)=(\sigma(i)\, n)\]
for all $i=1,\cdots,n-1$ and $\sigma\in S_{n-1}$.  It has nontrivial $\tau$ as 
\[ (i\, n )(j\, n)=(j\, n)(i\, j)\Rightarrow (i\, n)\cdot (j\, n)=(j\, n),\quad \tau((i\, n),(j\, n))=(ij)\]
for $i\ne j$ and we see that $\cdot$ has right but not left division or left but not right cancellation. We also have $(in)\cdot(in)=e$ and $\tau((in),(in))=e$ so that $(\ )^R$ is the identity map, hence $R$ is regular.   

This transversal gives a cross-product quasiHopf algebra $\Xi=\C S_{n-1}\rcross_\tau \C(R)$ where $R$ is a left quasigroup (i.e. unital and with left cancellation) except that we prefer to write it with the tensor factors in the other order. From Lemma~\ref{Xibialg} and Proposition~\ref{standardS}, this is 
spanned by products of $\delta_i$ and $\sigma\in S_{n-1}$, where $\delta_0$ is the delta function at $e\in R$ and $\delta_i$ at $(i,n)$ for $i=1,\cdots,n-1$. The cross relations have the same algebra $\sigma\delta_i=\delta_{\sigma(i)}\sigma$ for $i=1,\cdots,n-1$ as before but now
the tensor coproduct etc., and nontrivial associator
\[\Delta\delta_0=\sum_{i=0}^{n-1}\delta_i\tens\delta_i,\quad \Delta\delta_i=1\tens\delta_i+\delta_i\tens\delta_0,\quad \Delta \sigma=\sigma\tens\sigma,\quad \eps\delta_i=\delta_{i,0},\quad\eps\sigma=1,\]
\[ S\delta_i=\delta_{i},\quad S\sigma=\sigma^{-1},\quad \alpha=\beta=1,\]
\[\phi=(1\tens\delta_0+\delta_0\tens(1-\delta_0)+\sum_{i=1}^{n-1}\delta_i\tens\delta_i)\tens 1+ \sum_{i,j=1\atop i\ne j}^{n-1}\delta_i\tens\delta_j\tens (ij).\]
This is a $*$-quasi Hopf algebra with the same $*$ as before but now nontrivial
\[ \gamma=1,\quad \CG=1\tens\delta_0+\delta_0\tens(1-\delta_0)+\sum_{i=1}^{n-1}\delta_i\tens\delta_i+ \sum_{i,j=1\atop i\ne j}^{n-1}\delta_i(ij)\tens\delta_j(ij)\]
from Corollary~\ref{corstar}.

\medskip{\sl (iii) Twisting between the above two transversals.}  We denote the first transversal $R=\Z_n$, where $i$ is identified with $(12\cdots n)^i$, and we denote the 2nd transversal by $\bar R$ with corresponding elements $\bar i=(i\ n)$. Then 
\[ c_i=(12\cdots n)^{-i}(i\ n)\in S_{n-1},\quad c_i(j)=\begin{cases} n-i & j=i\\ j-i & else \end{cases}\]
for $i,j=1,\cdots,n-1$. If we use the stated $\la$ for the first transversal then one can check that the first half of (\ref{taucond}) holds,
\[  \overline{i+c_i\la i}=\overline{i+n-i}=e=\bar i\bar\cdot \bar i,\quad \overline{i+c_i\la j}=\overline{i+j-i}=\bar j=\bar i\bar\cdot \bar j\]
as it must. We can also check that the actions are indeed related by twisting. Thus,
\[ \sigma\ra\bar i=c_{\sigma\la i}^{-1}(\sigma\ra i)c_i=(\sigma(i),n)(12\cdots n)^{\sigma(i)}(\sigma\ra i)(12\cdots n)^{-i}(i,n)=(\sigma(i),n)\sigma(i,n)=\sigma\]
\[ \sigma\bar\la \bar i=(\sigma\la i)c_{\sigma\la i}=(12\cdots n)^{\sigma(i)}(12\cdots n)^{-\sigma(i)}(\sigma(i),n)=(\sigma(i),n),\]
where we did the computation with $\Z_n$ viewed in $S_n$. 

It follows that the Hopf algebra from case  (i)  cochain twists to a simpler quasihopf algebra in case (ii). The required cochain from Theorem~\ref{thmtwist} is
\[ \chi=\delta_0\tens 1+ \sum_{i=1}^{n-1}\delta_i\tens (12\cdots n)^{-i}(in).\]
\end{example}
The above example is a little similar to the Drinfeld $U_q(g)$ as Hopf algebras which are cochain twists of $U(g)$ viewed as a quasi-Hopf algebra. We conclude with the promised example related to the octonions. This is a version of \cite[Example~4.6]{KM2}, but with left and right swapped and some cleaned up conventions. 

\begin{example}\rm 
We let $G=Cl_3\lcross \Z_2^3$, where  $Cl_3$ is generated by $1,-1$ and $e_{i}$, $i=1,2,3$, with relations 
\[ (-1)^2=1,\quad (-1)e_i=e_i(-1),\quad e_i^2=-1,\quad e_i e_j=-e_j e_i  \]
for $i\ne j$ and the usual combination rules for the product of signs.  Its elements can be enumerated as $\pm e_{\vec a}$ where $\vec{a}\in \Z_2^3$ is viewed in the additive group of 3-vectors with entries in the field $\F_2=\{0,1\}$ of order 2 
and
\[ e_{\vec a}=e_1^{a_1}e_2^{a_2}e_3^{a_3},\quad e_{\vec a} e_{\vec b}=e_{\vec a+\vec b}(-1)^{\sum_{i\ge j}a_ib_j}. \]
This is the twisted group ring description of the 3-dimensional Clifford algebra over $\R$ in \cite{AlbMa}, but now restricted to coefficients $0,\pm1$ to give a group of order 16. For an example,
\[ e_{110}e_{101}=e_2e_3 e_1e_3=e_1e_2e_3^2=-e_1e_2=-e_{011}=-e_{110+101}\]
with the sign given by the formula.

We similarly write the elements of $K=\Z_2^3$ multiplicatively as $g^{\vec a}=g_1^{a_1}g_1^{a_2}g_3^{a_3}$ labelled by 3-vectors with values in $\F_2$. The generators $g_i$ commute and obey $g_i^2=e$. The general group product becomes the vector addition, and the cross relations are 
\[ (-1)g_i=g_i(-1),\quad e_i g_i= -g_i e_i,\quad  e_i g_j=g_j e_i\]
for $i\ne j$. This implies that $G$ has order 128.  

(i) If we take $R=Cl_3$ itself then this will be a subgroup and we will have for $\Xi(R,K)$ an ordinary Hopf $*$-algebra as a semidirect product  $\C \Z_2^3\rcross \C(Cl_3)$ except that we build it on the opposite tensor product. 

(ii) Instead, we take as representatives the eight elements again labelled by 3-vectors over $\F_2$, 
\[ r_{000}=1,\quad r_{001}=e_3,\quad r_{010}=e_2,\quad r_{011}=e_2e_3g_1\]
\[ r_{100}=e_1,\quad r_{101}=e_1e_3 g_2,\quad r_{110}=e_1e_2g_3,\quad r_{111}=e_1e_2e_3  g_1g_2g_3 \]
and their negations, as a version of \cite[Example~4.6]{KM2}. This can be written compactly as
\[ r_{\vec a}=e_{\vec a}g_1^{a_2 a_3}g_2^{a_1a_3}g_3^{a_1a_2}\]

\begin{proposition}\cite{KM2} This choice of transversal makes $(R,\cdot)$ the octonion two sided inverse property quasigroup $G_{\O}$ in the Albuquerque-Majid description of the octonions\cite{AlbMa},
\[ r_{\vec a}\cdot r_{\vec b}=(-1)^{f(\vec a,\vec b)} r_{\vec a+\vec b},\quad f(\vec a,\vec b)=\sum_{i\ge j}a_ib_j+ a_1a_2b_3+ a_1b_2a_3+b_1a_2a_3 \]
with the product on signed elements behaving as if bilinear. The action $\ra$ is trivial, and left action and  cocycle $\tau$ are 
\[ g^{\vec a}\la r_{\vec b}=(-1)^{\vec a\cdot \vec b}r_{\vec b},\quad \tau(r_{\vec a},r_{\vec b})=g^{\vec a\times\vec b}=g_1^{a_2 b_3+a_3 b_2}g_2^{a_3 b_1+a_1b_3} g_3^{a_1b_2+a_2b_1}\]
with the action  extended with signs as if linearly and $\tau$  independent of signs in either argument. 
\end{proposition}
\proof  We check in the group
\begin{align*} r_{\vec a}r_{\vec b}&=e_{\vec a}g_1^{a_2 a_3}g_2^{a_1a_3}g_3^{a_1a_2}e_{\vec b}g_1^{b_2 b_3}g_2^{b_1b_3}g_3^{b_1b_2}\\
&=e_{\vec a}e_{\vec b}(-1)^{b_1a_2a_3+b_2a_1a_3+b_3a_1a_2} g_1^{a_2a_3+b_2b_3}g_2^{a_1a_3+b_1b_3}g_3^{a_1a_2+b_1b_2}\\
&=(-1)^{f(a,b)}r_{\vec a+\vec b}g_1^{a_2a_3+b_2b_3-(a_2+b_2)(a_3+b_3)}g_2^{a_1a_3+b_1b_3-(a_1+b_1)(a_3+b_3)}g_3^{a_1a_2+b_1b_2-(a_1+b_1)(a_2+b_2)}\\
&=(-1)^{f(a,b)}r_{\vec a+\vec b}g_1^{a_2b_3+b_2a_3} g_2^{a_1b_3+b_1a_3}g_3^{a_1b_2+b_1a_2},
\end{align*}
from which we read off $\cdot$ and $\tau$. For the second equality, we moved the $g_i$ to the right using the commutation rules in $G$. For the third equality we used the product in $Cl_3$ in our description above and then converted $e_{\vec a+\vec b}$ to $r_{\vec a+\vec b}$. \endproof

The product of the quasigroup $G_\O$ here is the same as the octonions product as an algebra over $\R$ in the description of \cite{AlbMa}, restricted to elements of the form $\pm r_{\vec a}$. The cocycle-associativity property of $(R,\cdot)$ says
\[ r_{\vec a}\cdot(r_{\vec b}\cdot r_{\vec c})=(r_{\vec a}\cdot r_{\vec b})\cdot\tau(\vec a,\vec b)\la r_{\vec c}=(r_{\vec a}\cdot r_{\vec b})\cdot  r_{\vec c} (-1)^{(\vec a\times\vec b)\cdot\vec c}\]
giving -1 exactly when the 3 vectors are linearly independent as 3-vectors over $\F_2$. One also has $r_{\vec a}\cdot r_{\vec b}=\pm r_{\vec b}\cdot r_{\vec a}$ with $-1$ exactly when the two vectors are linearly independent, which means both nonzero and not equal, and $r_{\vec a} \cdot r_{\vec a}=\pm1 $ with $-1$ exactly when the one vector is linearly independent, i.e. not zero. (These are exactly the quasiassociativity, quasicommutativity and norm properties of the octonions algebra in the description of \cite{AlbMa}.)  The 2-sided inverse is
\[ r_{\vec a}^{-1}=(-1)^{n(\vec a)}r_{\vec a},\quad n(0)=0,\quad n(\vec a)=1,\quad \forall \vec a\ne 0\]
with the inversion operation extended as usual with respect to signs. 

The quasi-Hopf algebra $\Xi(R,K)$ is spanned by $\delta_{(\pm,\vec a)}$ labelled by the points of $R$ and products of the $g_i$ with the relations $g^{\vec a}\delta_{(\pm, \vec b)}=\delta_{(\pm (-1)^{\vec a\cdot\vec b},\vec b)} g^{\vec a}$ and tensor coproduct etc.,
\[ \Delta \delta_{(\pm, \vec a)}=\sum_{(\pm', \vec b)}\delta_{(\pm' ,\vec b)}\tens\delta_{(\pm\pm'(-1)^{n(\vec b)},\vec a+\vec b)},\quad \Delta g^{\vec a}=g^{\vec a}\tens g^{\vec a},\quad \eps\delta_{(\pm,\vec a)}=\delta_{\vec a,0}\delta_{\pm,+},\quad \eps g^{\vec a}=1,\]
\[S\delta_{(\pm,\vec a)}=\delta_{(\pm(-1)^{n(\vec a)},\vec a},\quad S g^{\vec a}=g^{\vec a},\quad\alpha=\beta=1,\quad \phi=\sum_{(\pm, \vec a),(\pm',\vec{b})} \delta_{(\pm,\vec a)}\tens\delta_{(\pm',\vec{b})}\tens g^{\vec a\times\vec b}\]
and from Corollary~\ref{corstar} is a $*$-quasi-Hopf algebra with $*$ the identity on $\delta_{(\pm,\vec a)},g^{\vec a}$ and 
\[ \gamma=1,\quad \CG=\sum_{(\pm, \vec a),(\pm',\vec{b})} \delta_{(\pm,\vec a)}g^{\vec a\times\vec b}
\tens\delta_{(\pm',\vec{b})}g^{\vec a\times\vec b}.\]
The general form here is not unlike our $S_n$ example. 
\end{example}

\subsection{Module categories context}

This section does not contain anything new beyond \cite{Os2,EGNO}, but completes the categorical picture that connects our algebra $\Xi(R,K)$ to the more general context of module categories, adapted to our notations. 

Our first observation is that if $\tens: \CC\times \CV\to \CV$ is a left action of a monoidal category $\CC$ on a category $\CV$ (one says that $\CV$ is a left $\CC$-module) then one can  check that this is the same thing as a monoidal functor $F:\CC\to \End(\CV)$ where the set ${\rm End}(\CV)$ of endofunctors can  be viewed as a strict monoidal category with monoidal product the endofunctor composition $\circ$. Here ${\rm End}(\CV)$ has monoidal unit   $\id_\CV$ and its morphisms are natural transformations between endofunctors. $F$ just sends an object $X\in \CC$ to $X\tens(\ )$ as a monoidal functor from $\CV$ to $\CV$. A monoidal functor comes with natural isomorphisms $\{f_{X,Y}\}$ and these are given tautologically by
\[  f_{X,Y}(V): F(X)\circ F(Y)(V)=X\tens (Y\tens V)\cong  (X\tens Y)\tens V=   F(X\tens Y)(V)\]
as part of the monoidal action. Conversely, if given a functor $F$, we define $X\tens V=F(X)V$ and extend the monoidal associativity of $\CC$ to  mixed objects using $f_{X,Y}$ to define $X\tens (Y\tens V)= F(X)\circ F(Y)V\isom F(X\tens Y)V= (X\tens Y)\tens V$.  The notion of a left module category is a categorification of the bijection between an algebra action $\cdot: A \tens V\rightarrow V$ and a representation as an algebra map $A \rightarrow {\rm End}(V)$. There is an equally good notion of a right $\CC$-module category extending $\tens$ to $\CV\times\CC\to \CV$. In the same way as one uses $\cdot$ for both the algebra product and the module action, it is convenient to use $\tens$ for both in the categorified version. Similarly for the right module version.

Another general observation is that if $\CV$ is a $\CC$-module category for a monoidal category $\CC$ then ${\rm Fun}_\CC(\CV,\CV)$, the (left exact) functors from $\CV$ to itself that are compatible with the action of $\CC$, is another monoidal category. This is denoted $\CC^*_{\CV}$ in \cite{EGNO}, but should not be confused with the dual of a monoidal functor which was one of the origins\cite{Ma:rep} of the  centre $\CZ(\CC)$ construction as a special case. Also note that if $A\in \CC$ is an algebra in the category then $\CV={}_A\CC$, the left modules of $A$ in the category, is a {\em right} $\CC$-module category. If $V$ is an $A$-module then we define $V\tens X$ as the tensor product in $\CC$ equipped with an $A$-action from the left on the first factor. Moreover, for certain `nice' right module categories $\CV$, there exists a suitable algebra $A\in \CC$ such that $\CV\simeq {}_A\CC$, see \cite{Os2}\cite[Thm~7.10.1]{EGNO} in other conventions. For such module categories, ${\rm Fun}_\CC(\CV,\CV)\simeq {}_A\CC_A$ the category of $A$-$A$-bimodules in $\CC$. Here, if given an  $A$-$A$-bimodule $E$ in $\CC$, the corresponding endofunctor is given by $E\tens_A(\ )$, where we require $\CC$ to be Abelian so that we can define $\tens_A$. This turns $V\in {}_A\CC$ into another $A$-module in $\CC$ and $E\tens_A(V\tens X)\isom (E\tens_A V)\tens X$, so the construction commutes with the right $\CC$-action.

Before we explain how these abstract ideas lead to ${}_K\CM^G_K$, a more `obvious' case is the study of  left module categories for $\CC = {}_G\CM$. If $K\subseteq G$ is a subgroup, we set  $\CV = {}_K\CM$ for $i: K\subseteq G$. The functor $\CC\to \End(\CV)$ just sends $X\in \CC$ to $i^*(X)\tens(\ )$ as a functor on $\CV$, or more simply $\CV$ is a left $\CC$-module by $X\tens V=i^*(X)\tens V$. More generally\cite{Os2}\cite[Example~7..4.9]{EGNO}, one can include a cocycle $\alpha\in H^2(K,\C^\times)$  since we are only interested in monoidal equivalence, and this data $(K,\alpha)$ parametrises all indecomposable left ${}_G\CM$-module categories. Moreover, here $\End(\CV)\simeq {}_K\CM_K$, the category of $K$-bimodules, where a bimodule $E$ acts by $E\tens_{\C K}(\ )$. So the data we need for a ${}_G\CM$-module category is a monoidal functor ${}_G\CM\to {}_K\CM_K$. This is of potential interest but is not the construction we were looking for. 

Rather, we are interested in right module categories of $\CC=\CM^G$, the category of $G$-graded vector spaces. It turns out that these are classified by the exact same data $(K,\alpha)$ (this is related to the fact that the $\CM^G,{}_G\CM$ have the same centre) but the construction is different. Thus, if $K\subseteq G$ is a subgroup, we consider $A=\C K$  regarded as an algebra in $\CC=\CM^G$ by $|x|=x$ viewed in $G$. One can also twist this by a cocycle $\alpha$, but here we stick to the trivial case. Then $\CV={}_A\CC={}_K\CM^G$, the category of $G$-graded left $K$-modules, is a right $\CC$-module category. Explicitly, if $X\in \CC$ is a $G$-graded vector space and $V\in\CV$ a $G$-graded left $K$-module then 
\[  V\tens X,\quad  x.(v\tens w)=v.x\tens w,\quad |v\tens w|=|v||w|,\quad \forall\ v\in V,\ w\in X\]
is another $G$-graded left $K$-module. Finally, by the general theory, there is an associated monoidal category 
\[ \CC^*_\CV:={\rm Fun}_{\CC}(\CV,\CV)\simeq {}_K\CM^G_K\simeq {}_{\Xi(R,K)}\CM.\]
which is the desired category to describe quasiparticles on boundaries in \cite{KK}. Conversely, if $\CV$ is an indecomposable right $\CC$-module category for $\CC=\CM^G$, it is explained in \cite{Os2}\cite[Example~7.4.10]{EGNO} (in other conventions) that the set of indecomposable objects has a transitive action of $G$ and hence can be identified with $G/K$ for some subgroup $K\subseteq G$. This can be used to put the module category up to equivalence in the above form (with some cocycle $\alpha$).

\section{Concluding remarks}\label{sec:rem}
We have given a detailed account of the algebra behind the treatment of boundaries in the Kitaev model based on subgroups $K$ of a finite group $G$. New results include the quasi-bialgebra $\Xi(R,K)$ in full generality, a more direct derivation from the category ${}_K\CM^G_K$ that connects to the module category point of view, a theorem that $\Xi(R,K)$ changes by a Drinfeld twist as $R$ changes, and a $*$-quasi-Hopf algebra structure that ensures a nice properties for the category of representations (these form a strong bar category). On the computer science side, we edged towards how one might use these ideas in quantum computations and detect quasiparticles across ribbons where one end is on a boundary. We also gave new decomposition formulae relating representations of $D(G)$ in the bulk to those of $\Xi(R,K)$ in the boundary. 

Both the algebraic and the computer science aspects can be taken much further. The case treated here of trivial cocycle $\alpha$ is already complicated enough but the ideas do extend to include these and should similarly be worked out. Whereas most of the abstract literature on
such matters  is at the conceptual level where we work up to categorical equivalence, we set out to give constructions more explicitly, which we we believe is essential for concrete calculations and should also be relevant to the physics. For example, much of the literature on anyons is devoted to so-called $F$-moves which express the associativity isomorphisms even though, by Mac Lane's theorem, monoidal categories are equivalent to strict ones. On the physics side, the covariance properties of ribbon operators also involve the coproduct and hence how they are realised depends on the choice of $R$. The same applies to how $*$ interacts with tensor products, which would be relevant to the unitarity properties of composite systems. Of interest, for example, should be the case of a lattice divided into two parts $A,B$ with a boundary between them and how the entropy of states in the total space relate to those in the subsystem. This is an idea of considerable interest in quantum gravity, but the latter has certain parallels with quantum computing and could be explored concretely using the results of the paper. We also would like to expand further the concrete use of patches and lattice surgery, as we considered only the cases of boundaries with $K=\{e\}$ and $K=G$, and only a square geometry. Additionally, it would be useful to know under what conditions the model gives universal quantum computation. While there are broadly similar such ideas in the physics literature, e.g., \cite{CCW}, we believe our fully explicit treatment will help to take these forward. 

Further on the algebra side, the Kitaev model generalises easily to replace $G$ by a finite-dimensional semi-simple Hopf algebra, with some aspects also in the non-semisimple case\cite{CowMa}. The same applies easily enough to at least a quasi-bialgebra associated to an inclusion $L\subseteq H$ of finite-dimensional Hopf algebras\cite{PS3} and to the corresponding module category picture. Ultimately here, it is the nonsemisimple case that is of interest as such Hopf algebras (e.g. of the form of reduced quantum groups $u_q(g)$) generate the categories where anyons as well as TQFT topological invariants live. It is also known that by promoting the finite group input of the Kitaev model to a more general weak Hopf algebra, one can obtain any unitary fusion category in the role of $\CC$\cite{Chang}. There remains a lot of work, therefore, to properly connect these theories to computer science and in particular to established methods for quantum circuits. A step here could be braided ZX-calculus\cite{Ma:fro}, although precisely how remains to be developed. These are some directions for further work.

\section*{Data availability statement}

Data sharing is not applicable to this article as no new data were created or analysed in this study.

\appendix

\section{Boundary ribbon operators with $\Xi(R,K)^\star$}\label{app:ribbon_ops}

\begin{definition}\rm\label{def:Y_ribbon}
Let $\xi$ be a ribbon, $r \in R$ and $k \in K$. Then $Y^{r \otimes \delta_k}_{\xi}$ acts on a direct triangle $\tau$ as
\[\input{Y_action_direct.tikz},\]
and on a dual triangle $\tau^*$ as
\[\input{Y_action_dual.tikz}.\]
Concatenation of ribbons is given by
\[Y^{r \otimes \delta_k}_{\xi'\circ\xi} = Y^{(r \otimes \delta_k)_2}_{\xi'}\circ Y^{(r \otimes \delta_k)_1}_{\xi} = \sum_{x\in K} Y^{(x^{-1}\rightharpoonup r) \otimes \delta_{x^{-1}k}}_{\xi'}\circ Y^{r\otimes\delta_x}_{\xi},\]
where we see the comultiplication $\Delta(r \otimes \delta_k)$ of $\Xi(R,K)^*$. Here, $\Xi(R,K)^*$ is a coquasi-Hopf algebra, and so has coassociative comultiplication (it is the multiplication which is only quasi-associative). Therefore, we can concatenate the triangles making up the ribbon in any order, and the concatenation above uniquely defines $Y^{r\otimes\delta_k}_{\xi}$ for any ribbon $\xi$.
\end{definition}

Let $s_0 = (v_0,p_0)$ and $s_1 = (v_1,p_1)$ be the sites at the start and end of a triangle. The direct triangle operators satisfy
\[k'\la_{v_0}\circ Y^{r\otimes \delta_k}_{\tau} =Y^{r\otimes \delta_{k'k}}_{\tau}\circ k'\la_{v_0},\quad k'\la_{v_1}\circ Y^{r\otimes\delta_k}_\tau = Y^{r\otimes\delta_{k'k^{-1}}}_\tau\circ k'\la_{v_1}\]
and
\[[\delta_{r'}\la_{s_i},Y^{r\otimes\delta_k}_{\tau}]= 0\]
for $i\in \{1,2\}$.
For the dual triangle operators, we have
\[k'\la_{v_i}\circ \sum_k Y^{r\otimes\delta_k}_{\tau^*} = Y^{(k'\la r)\otimes\delta_k}_{\tau^*}\circ k'\la_{v_i}\]
again for $i\in \{1,2\}$. However, there do not appear to be similar commutation relations for the actions of $\C(R)$ on faces of dual triangle operators. In addition, in the bulk, one can reconstruct the vertex and face actions using suitable ribbons \cite{Bom,CowMa} because of the duality between $\C(G)$ and $\C G$; this is not true in general for $\C(R)$ and $\C K$.

\begin{example}\label{ex:Yrib}\rm
Given the ribbon $\xi$ on the lattice below, we see that $Y^{r\otimes \delta_k}_{\xi}$ acts only along the ribbon and trivially elsewhere. We have
\[\input{Y_action_ribbon.tikz}\]
if $g^2,g^4,g^6(g^7)^{-1}\in K$, and $0$ otherwise, and
\begin{align*}
&y^1 = (rx^1)^{-1}\\
&y^2 = ((g^2)^{-1}rx^2)^{-1}\\
&y^3 = ((g^2g^4)^{-1}rx^3)^{-1}\\
&y^4 = ((g^2g^4g^6(g^7)^{-1})^{-1}rx^3)^{-1}
\end{align*}
One can check this using Definition~\ref{def:Y_ribbon}.
\end{example}

It is claimed in \cite{CCW} that these ribbon operators obey similar equivariance properties with the site actions of $\Xi(R,K)$
as the bulk ribbon operators, but we could not reproduce these properties. Precisely, we find that when such ribbons are `open' in the sense of \cite{Kit, Bom, CowMa} then an intermediate site $s_2$ on a ribbon $\xi$ between either endpoints $s_0,s_1$ does \textit{not} satisfy
\[\Lambda_{\C K}\la_{s_2}\circ Y^{r\otimes \delta_k}_{\xi} = Y^{r\otimes \delta_k}_{\xi}\circ \Lambda_{\C K}\la_{s_2}.\]
in general, nor the corresponding relation for $\Lambda_{\C(R)}\la_{s_2}$.

\section{Measurements and nonabelian lattice surgery}\label{app:measurements}
In Section~\ref{sec:surgery}, we described nonabelian lattice surgery for a general underlying group algebra $\C G$, but for simplicity of exposition we assumed that the projectors $A(v)$ and $B(p)$ could be applied deterministically. In practice, we can only make a measurement, which will only sometimes yield the desired projectors. As the splits are easier, we discuss how to handle these first, beginning with the rough split. We demonstrate on the same example as previously:
\[\input{rough_split_calc.tikz}\]
\[\input{rough_split_calc2.tikz}\]
where we have measured the edge to be deleted in the $\C G$ basis. The measurement outcome $n$ informs which corrections to make. The last arrow implies corrections made using ribbon operators. These corrections are all unitary, and if the measurement outcome is $e$ then no corrections are required at all. The generalisation to larger patches is straightforward, but requires keeping track of multiple different outcomes.

Next, we discuss how to handle the smooth split. In this case, we measure the edges to be deleted in the Fourier basis, that is we measure the self-adjoint operator $\sum_{\pi} p_{\pi} P_{\pi}\la$ at a particular edge, where 
\[P_{\pi} := P_{e,\pi} = {{\rm dim}(W_\pi)\over |G|}\sum_{g\in G} {\rm Tr}_\pi(g^{-1}) g\]
from Section~\ref{sec:lattice} acts by the left regular representation. Thus, for a smooth split, we have the initial state $|e\>_L$:
\[\input{smooth_split_calc1.tikz}\]
\[\input{smooth_split_calc2.tikz}\]
\[\input{smooth_split_calc3.tikz}\]
and afterwards we still have coefficients from the irreps of $\C G$. In the case when $\pi = 1$, we are done. Otherwise, we have detected quasiparticles of type $(e,\pi)$ and $(e,\pi')$ at two vertices. In this case, we appeal to e.g. \cite{BKKK, Cirac}, which claim that one can modify these quasiparticles deterministically using ribbon operators and quantum circuitry. The procedure should be similar to initialising a fresh patch in the zero logical state, but we do not give any details ourselves. Then we have the desired result.

For merges, we start with a smooth merge, as again all outcomes are in the group basis. Recall that after generating fresh copies of $\C G$ in the states $\sum_{m\in G} m$, we have
\[\input{smooth_merge_project.tikz}\]
we then measure at sites which include the top and bottom faces, giving:
\[\input{smooth_merge_measure_1.tikz}\]
for some conjugacy classes $\CC, \CC'$. There are no factors of $\pi$ as the edges around each vertex already satisfy $A(v)|\psi\> = |\psi\>$. When $\CC = \CC' = \{e\}$, we may proceed, but otherwise we require a way of deterministically eliminating the quasiparticles detected at the top and bottom faces. Appealing to e.g. \cite{BKKK, Cirac} as earlier, we assume that this may be done, but do not give details. Alternatively one could try to `switch reference frames' in the manner of Pauli frames with qubit codes \cite{HFDM}, and redefine the Hamiltonian. The former method gives
\[\input{smooth_merge_measure_2.tikz}\]
Lastly, we measure the inner face, yielding
\[\input{smooth_merge_measure_3.tikz}\]
so $|j\>_L\otimes |k\>_L \mapsto \sum_{s\in \CC''} \delta_{js,k} |js\>_L$, which is a direct generalisation of the result for when $G = \Z_n$ in \cite{Cow2}, where now we sum over the conjugacy class $\CC''$ which in the $\Z_n$ case are all singletons.

The rough merge works similarly, where instead of having quasiparticles of type $(\CC,1)$ appearing at faces, we have quasiparticles of type $(e,\pi)$ at vertices.

\end{document}